\documentclass[11pt]{article}
\usepackage{amssymb}
\usepackage{amsmath}
\usepackage{epsfig}
\usepackage{xspace}
\usepackage{cite}
\newcommand{\ie}{i.e.\xspace}
\newcommand{\eg}{e.g.\xspace}
\newcommand{\GeV}{\,\,\mathrm{GeV}}
\newcommand{\NLLB}{NLL$_\mathrm{B}$\xspace}

\newcommand{\kbar}{{\bar k}}
\newcommand{\CGO}{{\cal G}_{\omega}}
\newcommand{\CKO}{{\cal K}_{\omega}}
\newcommand{\CFO}{{\cal F}_{\omega}}
\newcommand{\CK}{{\cal K}}
\newcommand{\CKT}{\tilde{\cal K}}
\newcommand{\CG}{{\cal G}}
\newcommand{\CGT}{\tilde{\cal G}}
\newcommand{\CF}{{\cal F}}
\newcommand{\CGOT}{\tilde{\CG}_{\omega}}
\newcommand{\ci}{{\mathrm{c}}}

\newcommand{\as}{\alpha_s}              % coupling constant
\newcommand{\asb}{\bar{\alpha}_s}       % alpha_s bar
\newcommand{\az}{\alpha_0}              % frozen coupling constant
\newcommand{\eff}{{\mathrm{eff}}}
\newcommand{\fc}{M}                     % combination of 4 psi functions
\newcommand{\gres}{\gamma_{\mathrm{res}}}
\newcommand{\gs}{\gamma^{*}}            % virtual photon
\newcommand{\half}{{\textstyle \frac12}}% small 1/2
\newcommand{\ik}{{\cal H}}              % impact kernel
\newcommand{\ikc}{H}                    % perturb. coefficient of impact kernel
\newcommand{\kc}{\mathrm{kc}}           % kinematical constraint
\newcommand{\krn}{H_\lambda}            % K_0 regularised a la Marcello
\newcommand{\kt}{{\boldsymbol k}}       % transverse momentum of exchanged gluon
\newcommand{\Li}{{\mathrm{Li}_2}}       % polylogarithm
\renewcommand{\ln}{\log}
\newcommand{\nf}{n_f}
\newcommand{\om}{\omega}
\newcommand{\omhalf}{{\textstyle\frac{\omega}{2}}}
\newcommand{\omc}{\omega_c}
\newcommand{\omp}{\omega_\mathbb{P}}
\newcommand{\oms}{\omega_s}
\newcommand{\omsb}{\bar{\omega}_s}
\newcommand{\ord}{{\cal O}}             % of order
\newcommand{\pa}{\text{\sf a}}          % particle a
\newcommand{\pb}{\text{\sf b}}          % particle b
\newcommand{\pg}{\mathsf{g}}            % gluon
\newcommand{\pq}{\mathsf{q}}            % quark
\newcommand{\qt}{\boldsymbol{q}}        % transverse momentum of emitted gluon
\newcommand{\appb}[3]{{\it Acta Phys.~Polon.~}{\bf B #1} (#2) #3}
\newcommand{\epjc}[3]{{\it Eur.~Phys.~J.~}{\bf C #1} (#2) #3}
\newcommand{\hep}[1]{{\tt hep-ph/#1}}
\newcommand{\jhep}[3]{{\it JHEP }{\bf #1} (#2) #3}
\newcommand{\jetp}[3]{{\it Sov.~Phys.~JETP }{\bf #1} (#2) #3}
\newcommand{\jetpl}[3]{{\it JETP Lett.~}{\bf #1} (#2) #3}
\newcommand{\jpg}[3]{{\it J.~Phys.~}{\bf G#1} (#2) #3}
\newcommand{\npb}[3]{{\it Nucl.~Phys.~}{\bf B #1} (#2) #3}
\newcommand{\npbps}[3]{{\it Nucl.~Phys.~B (Proc.~Suppl.) }{\bf #1} (#2) #3}
\newcommand{\plb}[3]{{\it Phys.~Lett.~}{\bf B #1} (#2) #3}
\newcommand{\prd}[3]{{\it Phys.~Rev.~}{\bf D #1} (#2) #3}
\newcommand{\prl}[3]{{\it Phys.~Rev.~Lett.~}{\bf #1} (#2) #3}
\newcommand{\prep}[3]{{\it Phys.~Rep.~}{\bf #1} (#2) #3}
\newcommand{\sjnp}[3]{{\it Sov.~J.~Nucl.~Phys.~}{\bf #1} (#2) #3}
\newcommand{\zpc}[3]{{\it Z.~Phys.~}{\bf C #1} (#2) #3}
\begin{document}

\titlepage
\begin{flushright}
DESY 03--060 \\ DFF 404/05/03\\  LPTHE--03--20 \\ hep-ph/0307188 \\
July 2003
%Draft $Revision: 1.74 $
\end{flushright}

\vspace*{1in}
\begin{center}
{\Large \bf
Renormalisation group improved small-$\boldsymbol{x}$ Green's function}\\
\vspace*{0.4in}
M.~Ciafaloni$^{(a)}$,
D.~Colferai$^{(a)}$,
G.P.~Salam$^{(b)}$
and A.M.~Sta\'sto$^{(c)}$ \\
{\small
\vspace*{0.5cm}
$^{(a)}$ {\it  Dipartimento di Fisica, Universit\`a di Firenze,
 50019 Sesto Fiorentino (FI), Italy}; \\
\vskip 2mm
{\it  INFN Sezione di Firenze,  50019 Sesto Fiorentino (FI), Italy}\\
\vskip 2mm
$^{(b)}$ {\it LPTHE, Universities of Paris VI \& VII and CNRS,
75252 Paris 75005, France}\\
\vskip 2mm
$^{(c)}$ {\it Theory Division, DESY, D22603 Hamburg};\\
\vskip 2mm
{\it H.~Niewodnicza\'nski Institute of Nuclear Physics, Krak\'ow, Poland}\\
\vskip 2mm}
\end{center}
%\vspace*{1cm}
%\centerline{(\today)}

\vskip1cm
\begin{abstract}
  
  We investigate the basic features of the gluon density predicted by
  a renormalisation group improved small-$x$ equation which
  incorporates both the gluon splitting function at leading collinear
  level and the exact BFKL kernel at next-to-leading level.  We
  provide resummed results for the Green's function and its hard
  Pomeron exponent $\oms(\as)$, and for the splitting function and its
  critical exponent $\omc(\as)$. We find that non-linear resummation
  effects considerably extend the validity of the hard Pomeron regime
  by decreasing diffusion corrections to the Green's function exponent
  and by slowing down the drift towards the non-perturbative Pomeron
  regime. As in previous analyses, the resummed exponents are reduced
  to phenomenologically interesting values. Furthermore, significant
  preasymptotic effects are observed. In particular, the resummed
  splitting function departs from the DGLAP result in the moderate
  small-$x$ region, showing a shallow dip followed by the expected
  power increase in the very small-$x$ region.  Finally, we outline
  the extension of the resummation procedure to include the photon
  impact factors.

\end{abstract}

%%%%%%%%%%%%%%%%%%%%%%%%%%%%%%%%%%%%%%%%%%%%%%%%%%%%%%%%%%%%%%%%%%%%%%%%%%%%%%
\newpage
\section{Introduction}

Progress in understanding small-$x$ physics has been characterized by quite a
number of steps: first the BFKL evolution equation~\cite{BFKL} and its early
prediction of the small-$x$ rise of hard cross sections, leading to the notion
of hard Pomeron in perturbative QCD; then, the qualitative confirmation of such
a rise at HERA~\cite{HERA}, showing however a somewhat milder effect and, at the
same time, good agreement with DGLAP evolution~\cite{DGLAP} at two-loop
level; then the parallel calculation of the next-to-leading (NL) BFKL
kernel~\cite{NLLFL,NLLCC}, leading to a dramatic decrease of the effect and to
possible instabilities~\cite{BLUMVOGT,ROSS98,LEVIN} of the leading $\log s$
series; finally, the proposal of various resummation
approaches~\cite{Salam1998,CC,CCS1,ABF2000,THORNE,SCHMIDT,BFKLP} and recipes to
stabilize the series, in order to provide reliable predictions for
processes with two hard scales and
DIS-type processes.

The resummation approach proposed by some of us~\cite{Salam1998,CC,CCS1} and
summarized in Sec.~\ref{s:rgia} identifies a few physical QCD effects that
lead to large corrections. Firstly, the cross section dependence on the ratio
of the hard scales of the problem, which is constrained by the renormalisation
group (RG) requirement of single-logarithmic scaling violations in the relevant
Bjorken variables. Secondly, the occurrence, at NL level, of the non-singular
part (in moment space) of the anomalous dimension, yielding a sizable negative
contribution. Finally, the running coupling effects which modify and make
ambiguous the very notion of a hard Pomeron.

A key effect of the running coupling is that the BFKL evolution drifts towards
smaller momentum scales, which are more strongly coupled, thus making
non-perturbative physics more important at high energies. This means that the
asymptotically leading Pomeron $\omp$ \cite{JKCOL} is actually a
non-perturbative strong coupling quantity~\cite{Lipatov86,CC97}. This feature
can be taken into account by the initial condition in the DGLAP evolution of
structure functions, but may be a problem in the processes with two hard scales
(like Mueller-Navelet jets \cite{MUELLERNAVELET}, $\gs \gs$ scattering
\cite{GAMSTAR} etc.) where the perturbative hard Pomeron behavior can be
observed at intermediate energies only.

Recently, it has been noticed that the transition to the Pomeron regime is
driven, in some small-$x$ models, by a sudden tunneling effect~\cite{CCS2,CCSS2}
at moderate values of $\as(t)\log 1/x$, so that the $b$-expansion~\cite{CCSS1}
may be needed to suppress the Pomeron and to identify the hard Pomeron exponent
$\om_s(t)$ and its diffusion corrections~\cite{KOVMUELLER,ABB,LEVIN,CMT}
(here $t=\ln \kt^2/\Lambda^2$, where $\kt$ is the transverse momentum of the
hard probe, and $\Lambda = \Lambda_{\mathrm{QCD}}$). 
Furthermore, the gluon splitting function is expected to be power behaved in
the small-$x$ region too, but with a different exponent $\omc(t)$,
due to running coupling effects. Therefore, in a resummed
approach with running coupling one has to investigate various high-energy
exponents: the hard Pomeron index $\oms(t)$ just mentioned, the resummed
anomalous dimension singularity $\omc(t)$, which are generally different and
perturbatively calculable, finally the asymptotic Pomeron $\omp$ which is determined by  the strong coupling behavior of the model.

The calculation of $\om_s$ and $\om_c$ was performed in the renormalisation
group improved (RGI) approach of~\cite{CCS1}. The result was that $\om_s(t)$
carries important non-linear effects, leading to a stable and sizable decrease
with respect to its LL BFKL value, and that $\om_c(t)$ is sizably smaller than
$\om_s(t)$ also. However, the method of solution of the RGI equation used
in~\cite{CCS1} was best suited for the homogeneous equation, rather than the
Green's function (cfr.~Sec.~\ref{s:rgia}). Therefore, no real estimate of hard
small-$x$ cross sections was really possible.

The purpose of the present paper is to further investigate the RGI approach by
providing a numerical calculation in $\kt$ and rapidity space of the Green's
function and of the corresponding splitting function. By then using
$\kt$-factorization~\cite{KTFAC} and the corresponding impact
factors~\cite{KTSUB,FM,Bartels02,BaCoGiKy}, this sets the ground for a full
cross section calculation. Here we also provide the high energy exponents and a
semi-analytical treatment of the diffusion corrections.  Part of the results of
this paper have been summarized elsewhere~\cite{CCSSletter}.

In order to perform such an analysis, we introduce a resummation scheme slightly
different from that proposed in~\cite{CCS1}, which turns out to be more
convenient for numerical implementation, and belongs to a class of schemes
that are identical modulo NNL$x$ (and NLO in $Q^2$)
ambiguities intrinsic in the resummation approach.
Recall, that --- as summarized in the introductory Sec.~\ref{s:rgia} --- the
RGI approach incorporates leading and next-to-leading kernel information
exactly, with some extra $\om$-dependence ($\om$ is the Mellin variable conjugated
to $Y\sim\log1/x$) so as to implement the RG constraints and the resummation of
leading log collinear singularities mentioned before. Such requirements fix the
form of the $\om$-dependence of the kernel, apart from NNL terms, which remain
and allow some freedom in the choice of the resummation scheme.

The exact definition of the kernel and of the resummation scheme is provided in
Sec.~\ref{s:reskernel}. Stated in words, the main difference of the present
formulation with respect to that of Ref.~\cite{CCS1} is that the resummation of
the collinear behavior quoted before is  obtained here by the $\om$-dependence
of the leading kernel, rather than by a string of subleading ones. 
This allows us to include the full
$\om$-dependence of the one-loop anomalous dimension in a more direct way, 
while of course, leading plus NL kernel
information is correctly incorporated, as in all such schemes.

The detailed investigation of the gluon Green's function with its hard Pomeron
behavior and its diffusion corrections is performed in Sec.~\ref{s:resGGF},
by analytical and numerical methods. 
The full numerical evaluation relies on the method introduced in
Ref.~\cite{BoMaSaSc97}. Through the numerical study we are able to
analyze the border between perturbative and non-perturbative Pomeron
behavior, at realistic values of $Y$ and $\as$, and to extract the
leading terms ($\sim b Y$, $\sim b^2 Y^3$) in the exponent of the
perturbative part. Such terms can also be calculated analytically by
the $b$-expansion method \cite{KOVMUELLER,CCSS1}. We are thus able to
identify both the hard Pomeron exponent at order $\ord(b)$ and its
diffusion corrections, and we notice sizable non-linear effects which
stabilize the intercept, decrease the diffusion effects and slow down
the drift towards the non-perturbative Pomeron regime.

We also provide in Sec.~\ref{s:resanomdim} the resummed splitting function. At
the analytical level, we notice that the $\om$-expansion method~\cite{CC,CCS1}
allows one to define a resummed characteristic function which, in the
saddle-point 
approximation, can be related to the ``duality'' approach of
Ref.~\cite{ABF2000}, depending on the choice of the intercept in the
latter. Beyond the saddle-point estimate, the resummed splitting function is
evaluated numerically by the method of Ref.~\cite{CCS3}, and shows a
power increase $\sim x^{-\omc(\as)}$ in the very small-$x$ region, together
with a shallow dip (compared to the DGLAP result) at moderately small
$x$ values.

A preliminary discussion of the off-shell photon impact factors is provided in
Sec.~\ref{s:iif}. Here we show how the resummation scheme incorporating
collinear leading logs can be extended to the impact factor, and how the latter
can be extracted from the result obtained in the recent
literature~\cite{BaCoGiKy}. We finally summarize and discuss our results in
Sec.~\ref{s:discussion}.

%%%%%%%%%%%%%%%%%%%%%%%%%%%%%%%%%%%%%%%%%%%%%%%%%%%%%%%%%%%%%%%%%%%%%%
\section{Renormalisation Group improved approach\label{s:rgia}}

The size of subleading corrections~\cite{NLLFL,NLLCC} to the BFKL kernel
$\CK(\kt,\kt')$ and the ensuing instabilities~\cite{BLUMVOGT,ROSS98,LEVIN}
make it mandatory to understand the physical origin of the large terms and
possibly resum them. In a series of papers~\cite{Salam1998,CC,CCS1} (for a
review see~\cite{Salam1999}) it was argued that most of the large corrections
were due to collinear contributions, so as to achieve consistency of high-energy
factorization~\cite{KTFAC} at subleading level~\cite{KTSUB} with the
renormalisation group. This requires resummation~\cite{Salam1998} of both the
energy scale-dependent terms of the kernel~\cite{NLLCC} and of the leading-log
collinear logarithms~\cite{CC} for both $Q \gg Q_0$ and $Q \ll Q_0$, with $Q$,
$Q_0$ being the hard scales of the process. In the following we summarize the
approach of~\cite{CCS1}, which incorporates both the renormalisation group
requirements and the known exact forms of the leading~\cite{BFKL} and
next-to-leading~\cite{NLLFL,NLLCC} BFKL kernel.  A resummation for anomalous
dimensions within a single collinear regime $Q \gg Q_0$ has been proposed
in~\cite{ABF2000}, and alternative resummations in~\cite{THORNE,SCHMIDT,BFKLP}.

\subsection{$\kt$-factorization and high-energy exponents\label{s:kf}}

We consider a general process of scattering of two hard probes $A$ and $B$ with
scales $Q$ and $Q_0$ at high center-of-mass energy $\sqrt{s}$.  We assume that
the cross section can be written in the following $\kt$-factorized
form~\cite{KTFAC}:
\begin{equation}
 \sigma_{AB}(s;Q,Q_0) = \int \frac{d \om}{2 \pi i}\,
 \frac{d^2 \kt}{\kt^2} \, \frac{d^2 \kt_0}{\kt_0^2}
 \left(\frac{s}{Q Q_0}\right)^{\om} h_{\om}^{A}(Q,\kt) \;
 \CGO(\kt,\kt_0) \; h_{\om}^{B}(Q_0, \kt_0)
\label{eq:sigma}
\end{equation}
where $h^{A}$ and $h^{B}$ are dimensionless impact factors which characterize
the probes and ensure that $|\kt|$ ($|\kt_0|$) is of order $Q$ ($Q_0$), and the
gluon Green's function is defined by
\begin{equation}
 \CGO(\kt,\kt_0) = \langle \kt | [\om - \CKO]^{-1} | \kt_0 \rangle \;.
\label{eq:gluongreen}
\end{equation}
The function $\CKO$ is the kernel of the small-$x$ equation of the general form
\begin{equation}
\om \CGO(\kt,\kt_0) = \delta^2(\kt-\kt_0) + \int \frac{d^2 \kt'}{\pi} \;
\CKO (\kt,\kt') \; \CGO(\kt',\kt_0) \; .
\label{eq:bfklequation}
\end{equation}
The factorization formula (\ref{eq:sigma}) involving two-(Regge)gluon exchange,
has been justified up to NL $\log s$ level in Refs.~\cite{KTSUB} for initial
partons and in~\cite{FM,Bartels02} for physical probes. At further subleading
levels, many (Regge)gluon Green's functions contribute to the cross section as
well, due to the s-channel iteration. However, our purpose here is to
incorporate leading-twist collinear behavior, and at that level the two-gluon
contribution is dominant, so that we shall consider only the contribution
(\ref{eq:sigma}) in the following.

While $\kt$-factorization is supposed to be valid for $\as \lesssim \om \ll 1$,
we shall sometimes extra\-po\-la\-te Eq.~(\ref{eq:sigma}) to sizable values of
$\om = \ord(1)$ and moderate values of $s$, encouraged by the stability of our
resummation, and by the possibility of incorporating phase space thresholds in
Eq.~(\ref{eq:sigma}) (cfr.~Sec.~\ref{s:iif}). It should be kept in mind that
such a region lies outside the validity range of Eq.~(\ref{eq:sigma}), so that
the extrapolated Green's function loses --- most probably --- its original
meaning as two-(Regge)gluon propagator.

In writing Eq.~(\ref{eq:sigma}), we have performed the choice of energy scale
$s_0 = Q Q_0$, in terms of which the high energy kinematics shows a simpler
phase space, as explained in more detail in Sec.~\ref{s:iif}. Actually, for
intermediate subenergies it is more convenient to introduce as energy variables
the scalar products of type $\nu = 2k_\mu k_0^\mu$, which have $|\kt||\kt_0|$ as
threshold, so that $|\kt||\kt_0|/\nu$ is a good Mellin variable. Correspondingly,
the energy dependence of the Green's function and of the impact factors is
defined by ($k \equiv |\kt|, k_0 \equiv |\kt_0|$)
\begin{align}
 \CG( \nu, \kt, \kt_0 ) &= \int\frac{d \om}{2 \pi i}
  \left(\frac{\nu}{k k_0}\right)^{\om} \CGO( \kt, \kt_0 ) \label{eq:CG} \\
 &\equiv \frac1{k k_0} G( Y; t, t_0 ) \;,
  \qquad \left( Y \equiv \log \frac{\nu}{k k_0}
  \;, \qquad t \equiv \log\frac{k^2}{\Lambda^2} \right) \nonumber
\end{align}
and
\begin{equation}\label{eq:if}
 h(\nu,Q,\kt) = \int\frac{d \om}{2 \pi i}
 \left( \frac{\nu}{Q k} \right)^{\om} h_{\om}(Q,\kt)\;.
\end{equation}

In this paper, we are mostly interested in the properties of the two-scale
Green's function and of its high-energy exponents.  It was pointed out
in~\cite{CCS1} that, in the improved approach with running coupling, the high
energy limits of the Green's function and of the collinear splitting functions
are regulated by different indices, which both originate from the frozen
coupling hard Pomeron exponent. We shall define the index $\oms(t)$ by
(cfr.~Sec.~\ref{s:eidc})
\begin{equation}
\label{eq:oms}
 G(Y;t,t_0) \simeq \frac{1}{\sqrt{2\pi\asb\chi'' Y}}\exp[\oms(\frac{t+t_0}{2})Y
 + \text{diffusion corrections}] \;, \qquad \asb \equiv \as \frac{N_c}{\pi}
\end{equation}
in the limit $\oms(t) Y \gg 1$ and $t\simeq t_0 \gg 1$, and the index $\omc(t)$
by
\begin{equation}\label{eq:Plimit}
 x P(\asb(k^2),x) \;  \stackrel{x \rightarrow 0}{\longrightarrow} \;
 x^{-\omc(t)} p(\asb)\;,
\end{equation}
where $P(\asb(k^2),x)$ is the resummed gluon-gluon splitting function
(Sec.~\ref{s:resanomdim}).  The exponent $\oms$ in Eq.~(\ref{eq:oms}) used to be
defined as the location of the anomalous dimension singularity in the saddle
point approximation. It is now understood~\cite{CCS1}, see also~\cite{THORNE},
that this singularity is actually an artefact of the saddle point approximation,
and that the true anomalous dimension singularity, located at $\om=\omc(t)$,
causes the power behavior of the effective splitting function. This result has
then been confirmed in the alternative resummation procedures
of~\cite{ABF2001,ABF2003,THORNE}. 

Even the definition in Eq.~(\ref{eq:oms}) is not free of ambiguities, due to the
occurrence of diffusion corrections to the
exponent~\cite{KOVMUELLER,ABB,LEVIN,CMT} which rapidly increase with $Y$, and
to the contamination of the non-perturbative Pomeron, which dominates above some
critical rapidity~\cite{CCSS2,CCSS1}.

In the following, both regimes $t \simeq t_0$ and $t \gg t_0$ will be discussed
in detail in the RG-improved approach, by emphasizing our perturbative
predictions and their range of validity.

\subsection{Scale changing transformations}

Let us note that the symmetrical scale choice $\nu_0 = k k_0$ performed in
Eq.~(\ref{eq:CG}) is not the only possible one, and is physically justified only
in the case $k \sim k_0$. This configuration occurs for example in the
process of $\gs \gs$ scattering at high energy with comparable
virtualities of both photons~\cite{GAMSTAR}, forward jet/$\pi^0$ production in
DIS~\cite{FORWARDJET} or production of 2 hard jets at hadron
colliders~\cite{MUELLERNAVELET}.  However, in the typical deep inelastic
situation, when one of the scales is much larger, $k \gg k_0$ ($k_0 \gg k$) the
correct Bjorken variable is rather $k^2/s$ ($k_0^2/s$). In order to switch to
this asymmetric case one should perform a similarity transformation on the gluon
Green's function of the form
\begin{equation}
\CGO \rightarrow \bigg({k_{>} \over k_{<}}\bigg)^{\om} \CGO \; ,
\label{eq:similarity}
\end{equation}
where $k_{>}=\max(k,k_0)$ and $k_{<}=\min(k,k_0)$.  The transformation
(\ref{eq:similarity}) implies the following change of kernel $\CKO$
\begin{subequations}\label{eq:kertransform}
\begin{align}
 \CKO(k,k')& \; \rightarrow \;\CK^{u}_{\om}(k,k') \,=\, \CKO(k,k') \bigg({k
   \over k'}\bigg)^{\om} 
 \;, \qquad  \nu_0 = k^2 \;, \\
 \CKO(k,k')& \; \rightarrow \;\CK^{l}_{\om}(k,k') \,=\, \CKO(k,k')
 \bigg({k' \over k}\bigg)^{\om} 
 \;, \qquad  \nu_0 = k'{}^2 \;,
\end{align}
\end{subequations}
where now $\CK^{u}_{\om}$ ($\CK^{l}_{\om}$) means the kernel for the
upper-$k^2$ (lower-$k'{}^2$) energy scale choice.

Our goal is to find a resummed prescription for $\CKO(k,k^{\prime})$ which takes
into account the large $Y$ terms and is consistent with renormalisation
group equations.  The kernel $\CKO(k,k')$ is not scale invariant,
and it can be expanded in powers of the coupling constant as follows
\begin{equation}
\CKO(k,k') = \sum_{n=0}^{\infty} [\asb(k^2)]^{n+1}
\; \CK_{n}^\om(k,k') \;.
\label{eq:kernexp}
\end{equation}
where
\begin{equation}
 \asb(k^2)={1 \over b\ln(k^2/\Lambda^2)} \; , \qquad
 b={11 \over 12}-{N_f \over 6 N_c} \;,
\label{eq:alphas}
\end{equation}
and the coefficient kernels $\CK_{n}^\om(k,k^{\prime})$ are now scale invariant,
and additionally carry some $\om$-dependence. We shall now see how the
renormalisation group constraints on $\CK_{\om}^u$ and $\CK_{\om}^l$ determine
the collinear behavior of $\CK_{\om}$ .

%%%%%%%%%%%%%%%%%%%%%%%%%%%%%%%%%%%%%%%%%%%%%%%%%%%%%%%%%%%%%%%%%%%%%%%%
\subsection{Renormalisation group constraints and shift of $\gamma$ poles}

It is important to notice that the $\om$-dependence of the scale
invariant kernels $\CK_n^{\om}$, present in Eq.~(\ref{eq:kernexp}), is
not negligible (even for the small $\om$ values being considered) and
follows from the requirement that collinear singularities have to be
single logarithmic in both regimes $k \gg k_0$ and $k_0 \gg k$. If $k
\gg k_0$, it is simplest to discuss the kernel in its form
$\CK_{\om}^{u}$, Eq.~(\ref{eq:kertransform}a). A leading-$\log k^2$
analysis for $k \gg k'$ shows that its collinear singularities are
determined by the non-singular part (in $\om$ space), $A_1(\om)$, of
the gluon anomalous dimension,
\begin{equation}
\bar{\alpha}_s A_1(\om) = \gamma_{gg}(\om) - {\bar{\alpha}_s \over \om},
\label{eq:a1omdef}
\end{equation}
and
\begin{equation}
 A_1(\om)  = -{11 \over 12} + {\cal O}(\om), \hspace*{1cm}  (N_f=0) \; ,
 \label{eq:a1om}
\end{equation}
In contrast the singular part $\asb/\om$ is accounted for by the
iteration of the BFKL equation itself.

To be precise, one has
\begin{equation}
 \label{eq:collinearkerup}
 \CK_{\om}^u(k,k') \simeq \frac{\asb(k^2)}{k^2}
 \exp \int_{t'}^t d(\log\kappa^2) \; A_1(\om) \asb(\kappa^2)
 = \frac{\asb(k^2)}{k^2}
 \left(1-b \asb(k^2) \log \frac{k^2}{k'{}^2} \right)^{-\frac{A_1(\om)}{b}} \;,
\end{equation}
where $t = \ln k^2/\Lambda_{QCD}^2$, indeed showing single
logarithmic scaling violations.  A similar reasoning, yields the
collinear behavior of $\CK_{\om}^l$ from Eq.~(\ref{eq:kertransform}b)
with the opposite strong ordering behavior $k' \gg k$, which is
relevant in the regime $k_0 \gg k$.

But $\CK_{\om}^u$ and $\CK_{\om}^l$ are related to $\CK_{\om}$ by the
$\om$-dependent similarity transformations
(\ref{eq:kertransform}a,\ref{eq:kertransform}b), so that the latter must
have the following collinear structure
\begin{multline}
 \label{eq:collinearker}
 \CK_{\om}(k,k') \simeq \asb(k^2) \left[\frac{1}{k^2} \left(\frac{k'}{k}\right)^{\om}
 \left(\frac{\asb(k^2)}{\asb(k'{}^2)}\right)^{-\frac{A_1(\om)}{b}} \Theta(k-k')
\;  + \right. \\
\left. + \; \frac{1}{k'{}^2} \left(\frac{k}{k'}\right)^{\om}
 \left(\frac{\asb(k^2)}{\asb(k'{}^2)}\right)^{\frac{A_1(\om)}{b}-1} \Theta(k'-k) \right] \,.
\end{multline}
In this expression one can see that the $\om$-dependence provided by
%$( k_< / k_> )^{\om}$
$\left(\frac{k_<}{k_>}\right)^{\om}$
is essential, because $k_>/k_<$ can be a
large parameter. We also keep the $\om$-dependence in $A_1(\om)$, in order to
take into account the full one-loop anomalous dimension.

By expanding in $b\asb$ the renormalisation group logarithms present in the
collinear behavior of Eqs.~(\ref{eq:collinearkerup},\ref{eq:collinearker}), we
obtain the leading collinear singularities of the coefficient kernels
$\CK^{\om}_n$ in Eq.~(\ref{eq:kernexp}).  This implies that, in $\gamma$-space,
the corresponding eigenvalues have the following structure
\begin{equation}
\chi_n^{\om}(\gamma)
={1\!\cdot\! A_1 (A_1+b) \cdots [A_1 + (n-1) b] \over( \gamma + \frac{\om}{2})^{n+1}}
+{1\!\cdot\!(A_1-b) A_1 \cdots [A_1 - n b] \over (1-\gamma+\frac{\om}{2})^{n+1}} \;,
\label{eq:chicoll}
\end{equation}
where the $\om$ dependence of $A_1$ is left implicit.
Therefore the position of the $\gamma \rightarrow 0$ ($\gamma \rightarrow 1$)
poles is shifted by $-\omhalf$ ($+\omhalf$) for the kernel
(\ref{eq:collinearker}) with symmetrical scale choice $\nu_0=k k_0$.
Through this shift one
is able to resum~\cite{Salam1998} the higher order $\gamma$-poles of the
kernel that are due to scale changing effects.

In fact, the leading and next-to-leading eigenvalues corresponding to this
symmetrical choice of scale have the collinear behavior
\begin{eqnarray}
\chi_0^{\om}(\gamma) \simeq {1 \over \gamma+  \omhalf}
+ {1 \over 1-\gamma+\omhalf}\;, \nonumber \\
\chi_1^{\om}(\gamma) \simeq {A_1(\om) \over (\gamma+  \omhalf)^2}
+ {A_1(\om) -b \over (1-\gamma+\omhalf)^2} \;.
\label{eq:chicollsym}
\end{eqnarray}

Now, in order to obtain the NLL coefficient~\cite{CCS1} in the $\bar{\alpha}_s$
expansion one has to expand in $\om$ the term $\chi_0^{\om}(\gamma)$ to first
order with subsequent identification
$\om \rightarrow \bar{\alpha}_s \chi_0^{\om=0}$, and add the $\chi_1^{\om=0}$
terms.  The result for the NLL eigenvalue in the collinear approximation then
reads
\begin{eqnarray}
\chi_1^\mathrm{coll}(\gamma) = \left[ \bar{\alpha}_s \chi_0^\om(\gamma)
 {\partial \chi_0^{\om} \over \partial \om} + \chi_1^\om \right]_{\om=0}
= - {1 \over 2 \gamma^3} - {1 \over 2 (1-\gamma)^3 }+ {A_1(0) \over \gamma^2}
+ {A_1(0) -b \over (1-\gamma)^2} + \dots \; .
\label{eq:chicollnll}
\end{eqnarray}
We note that the $\om$-dependent shift has generated cubic poles
$\frac{1}{\gamma^3},\frac{1}{(1-\gamma)^3}$ which seem to imply double logs
$\ln^2 \frac{k_<^2}{k_>^2}$, but are actually needed with the choice of scale
$k k_0$ in order to recover the correct Bjorken variable $k_>^2/s$. The collinear
terms with $A_1(\om)$ have instead generated double poles
$\frac{1}{\gamma^2},\frac{1}{(1-\gamma)^2}$ which correspond to single logs,
$\ln\frac{k_<^2}{k_>^2}$.

The double and cubic poles at $\gamma=0$ and $\gamma=1$ so obtained
are precisely those of the full NLL BFKL kernel eigenvalue. In fact
Eq.~(\ref{eq:chicollnll}) is 
a collinear approximation to the full NLL BFKL kernel
eigenvalue~\cite{NLLFL,NLLCC} which has the following form
\begin{align}\nonumber
 \chi_1(\gamma) &= -\frac{b}{2} [\chi^2_0(\gamma) + \chi'_0(\gamma)]
  -\frac{1}{4} \chi_0''(\gamma)
  -\frac{1}{4} \left(\frac{\pi}{\sin \pi \gamma} \right)^2
  \frac{\cos \pi \gamma}{3 (1-2\gamma)}
  \left(11+\frac{\gamma (1-\gamma )}{(1+2\gamma)(3-2\gamma)}\right) \\
 &\quad +\left(\frac{67}{36}-\frac{\pi^2}{12} \right) \chi_0(\gamma)
  +\frac{3}{2} \zeta(3) + \frac{\pi^3}{4\sin \pi\gamma}  \nonumber\\
 &\quad - \sum_{n=0}^{\infty} (-1)^n
 \left[ \frac{\psi(n+1+\gamma)-\psi(1)}{(n+\gamma)^2}
 +\frac{\psi(n+2-\gamma)-\psi(1)}{(n+1-\gamma)^2} \right] \; .
\label{eq:nllorg}
\end{align}

It turns out that the collinear approximation (\ref{eq:chicollnll}) above
reproduces the exact eigenvalue (\ref{eq:nllorg}) up to
7\%~\cite{CCS1,Salam1999} accuracy when $\gamma\in ]0,1[$. This suggests that
the collinear terms are the dominant contributions in the NLL kernel.

In the following, we shall normally incorporate the shift of $\gamma$-poles in
the form
\begin{equation}
 \label{eq:symmshift}
 \chi_n^{\om}(\gamma) = \chi_{nL}^{\om}(\gamma + \omhalf) 
 + \chi_{nR}^{\om}(1-\gamma+\omhalf) \;,
\end{equation}
where $\chi_{nL}^{\om}$ ($\chi_{nR}^{\om}$) have only
$\gamma \rightarrow -\omhalf$ ($\gamma\rightarrow 1 + \omhalf $) singularities
of the type in Eq.~(\ref{eq:chicoll}).  In this way the collinear singularities
are single logarithmic in both limits $k \gg k_0$ and $k_0 \gg k$, and the
energy scale dependent terms are automatically resummed.  The modified
leading-order eigenvalue that we adopt has the following structure (compare
(\ref{eq:chicollsym})):
\begin{equation}
\chi_0^{\om} = 2 \psi(1) - \psi(\gamma+\omhalf) - \psi(1-\gamma+\omhalf) \;,
\label{eq:chiomll}
\end{equation}
in the case of symmetric choice of energy scale $\nu_0 =k k_0$.  This form of
the kernel was considered previously in~\cite{AGS,KMS}.  It is obtained from the
leading order BFKL kernel by imposing the so-called kinematical (or consistency)
constraint~\cite{Ciaf88,LDC,KMS1997} which limits the virtualities of the transverse
momenta of the gluons in the real emission part of the kernel.  The origin of
this constraint is the requirement that in the multi-Regge kinematics the
virtualities of the exchanged gluons be dominated by their transverse parts.
The NLL contribution of the resummed kernel, $\chi_1^{\om}$ was then~\cite{CCS1}
constructed by the requirement that the collinear limit in
Eq.~(\ref{eq:chicollsym}) should be correctly reproduced, and the exact form of
the NL kernel (\ref{eq:nllorg}) should be obtained also.

The final NLL eigenvalue function proposed in~\cite{CC,CCS1} reads
\begin{eqnarray}
\chi_1^{\om}(\gamma) & = & \chi_1(\gamma) + {1 \over 2} \chi_0(\gamma)
{\pi^2 \over \sin^2 \pi \gamma}  \nonumber \\
& & -A_1(0) \psi^{\prime}(\gamma) -[A_1(0)-b] \psi^{\prime}(1-\gamma) \nonumber \\
& & +A_1(\om) \psi^{\prime}(\gamma+\omhalf)
+[A_1(\om)-b] \psi^{\prime}(1-\gamma+\omhalf) \nonumber \\
& &  - {\pi^2 \over 6} [\chi_0(\gamma) - \chi_0^{\om}(\gamma)] \; .
\label{eq:chiomnll}
\end{eqnarray}
The first line is the original NLL term $\chi_1(\gamma)$ with the subtraction of
the cubic poles which come from the changes of the energy scale and which are
resummed by the leading order $\om$-dependent kernel (\ref{eq:chiomll}).  The
second and third lines contain shifted collinear double poles, and finally the
last line contains the shifted single poles which additionally appear as an
artefact of the resummation procedure.

%%%%%%%%%%%%%%%%%%%%%%%%%%%%%%%%%%%%%%%%%%%%%%%%%%%%%%%%%%%%%%%%%%
\subsection{$\boldsymbol\om$-expansion and collinear resummation\label{s:omegaexp}}

In the present paper we choose a form of the improved kernel that differs
somewhat from that of Ref.~\cite{CCS1} --- quoted in
Eqs.~(\ref{eq:chiomll},\ref{eq:chiomnll}) --- by using the possibility of
translating part of the $\as$-dependence in Eq.~(\ref{eq:kernexp}) into
additional $\om$-dependence.  Actually, it was pointed out in~\cite{CC,CCS1}
that, at high energies, $\om$ is a more useful expansion parameter than
$\as(k^2)$, the relation being given roughly by $\om\simeq\asb\chi_0$, as
noticed already in connection with Eq.~(\ref{eq:chicollnll}).

The $\om$-expansion is a systematic way of solving the homogeneous equation
\begin{equation}
\label{eq:homo}
(\om-{\CK_{\om}}) \CF_{\om}(k) = 0 \; ,
\end{equation}
where $\CK_{\om}$ is given by Eq.~(\ref{eq:kernexp}), by the
$\gamma$-representation
%~\cite{JK,COLKWIE}
\begin{equation}
\label{eq:gamrep}
\CF_\om(t) = \int\frac{d\gamma}{2\pi i}\; e^{\gamma t-\frac{1}{b\om}X_\om(\gamma)}\;,
\end{equation}
in which $\chi_\om(\gamma) = X'_\om(\gamma)$ satisfies a non-linear
integro-differential equation equivalent to Eq.~(\ref{eq:homo}). The latter is
derived by using the representation $t\to-\partial_{\gamma}$ in
Eq.~(\ref{eq:kernexp}), and is given by \cite{CCS1}
\begin{equation}
\label{eq:nonlinear}
 \chi_{\om}(\gamma)=\chi^{\om}_0(\gamma)
 +\left(\frac{\om}{\chi_{\om}-b\om\partial_{\gamma}}\right)\chi^{\om}_1
 +\left(\frac{\om}{\chi_{\om}-b\om\partial_{\gamma}}\right)^2\chi^{\om}_2
 +\cdots\;.
\end{equation}
Approximate solutions to Eq.~(\ref{eq:nonlinear}) can be obtained either by
truncating at, say, NL level (i.e., setting $\chi_2 = \chi_3 = \cdots = 0$), or
by expanding in the $\om$-parameter to all orders.  The latter procedure yields
the solution \cite{CC,CCS1}
\begin{equation}
\label{eq:omexp}
\chi_{\om}(\gamma) = \chi_0^{\om}(\gamma)
+ \om \frac{\chi_1^\om(\gamma)}{\chi_0^\om(\gamma)}
+ \om^2 \frac{1}{\chi_0^\om(\gamma)} \left[\frac{\chi_2^\om(\gamma)}{\chi_0^\om(\gamma)}
+ b \left(\frac{\chi_1^\om}{\chi_0^\om} \right)'
- \left(\frac{\chi_1^\om}{\chi_0^\om} \right)^2\right] + \dots\;,
\end{equation}
and amounts to replacing the kernel $\CK_\om$ by an effective kernel
$\asb(k^2) \CK_\om^{\eff}$, where $\CK_\om^{\eff}$ is scale invariant. The
corresponding characteristic function $\chi_\om(\gamma)$ in Eq.~(\ref{eq:omexp})
is --- very roughly --- obtained by the replacement
$\asb \rightarrow \om/\chi_\om$ in Eq.~(\ref{eq:kernexp}), so that indeed $\om$
plays the role of a new expansion parameter.
A virtue of the expansion (\ref{eq:omexp}) is that it contains simple
(leading) collinear poles only, because the double-poles left in $\chi_1^\om$
after the $\om$-shift are canceled by the denominators. 

The $\om$-expansion is particularly useful for the resummation of the leading
collinear singularities of Eqs.~(\ref{eq:collinearker}) and (\ref{eq:chicoll}).
Suppose we first take $\asb$ frozen (limit $b=0$). Then, the leading poles of
Eq.~(\ref{eq:chicoll}) have approximately the factorized form
\begin{equation}
\label{eq:collfac}
 \chi_n^\om \simeq \chi_0^\om (\chi_{\ci}^\om)^n \;, \qquad \chi_{\ci}^\om
 = \frac{A_1(\om)}{\gamma+\omhalf} + \frac{A_1(\om)}{1-\gamma+\omhalf} \;,
\end{equation}
(valid for $\gamma +\omhalf \simeq 0$ or $1-\gamma+\omhalf \simeq 0$), so that
the resummed behavior (\ref{eq:collinearker}) reads
\begin{equation}
 \label{eq:kerresum}
 \CKO \simeq \sum_{n=0}^\infty \asb K_0^\om [\asb K_{\ci}^\om]^n =
 \asb K_0^\om (1-\asb K_{\ci}^\om)^{-1} \;.
\end{equation}
Exactly the same result can be obtained by the $\om$-expansion~(\ref{eq:omexp})
truncated at the NL level, by setting
$\chi_1^{\om}/\chi_0^{\om} \simeq \chi_{\ci}^{\om}$, and thus considering the
kernel
\begin{equation}
\label{eq:kerresum1}
 \CKT_\om = \asb(K^\om_0 + \om K^\om_\ci) \;.
\end{equation}
In fact, the resolvent of the latter is given by
\begin{equation}\label{d:GGFeff}
 \CGOT \equiv [\om-\CKT_\om]^{-1}
 \simeq (1-\asb K^\om_\ci)^{-1}
 \left[ \om-\asb K^\om_0 (1-\asb K^\om_\ci)^{-1} \right]^{-1} \;,
\end{equation}
and is then proportional to the Green's function of the resummed
kernel~(\ref{eq:kerresum}).

In other words, leading-log collinear singularities are equivalently
incorporated by a string of subleading kernels (as in Eq.~(\ref{eq:kerresum})),
or by a NL contribution of order $\asb \om$ (as in Eq.~(\ref{eq:kerresum1})) ---
apart from a redefinition of the impact factors. In the realistic case with
running coupling it is straightforward to check that $b$-dependence
only remains in the 
first term of the $\om$-expansion (\ref{eq:omexp})
\begin{equation}
\label{eq:omexpcol}
\chi_\om(\gamma) \simeq \chi_0^\om + \om \left( \frac{A_1}{\gamma+\omhalf}
+ \frac{A_1-b}{1-\gamma+\omhalf}\right) + \dots \;,
\end{equation}
whereas it cancels out in all remaining subleading terms.  Therefore, in order
to incorporate  the leading log collinear
behavior in the form~(\ref{eq:omexpcol}) we can set, for instance,
\begin{equation}
\label{eq:kerresum2}
\CKT_\om = \asb(\qt^2) K^\om_0 + \om \asb(k_>^2) K^\om_\ci + \mathrm{NLL} \;,
\end{equation}
as an improved leading kernel.  Here we assume that the scale for $\asb$ in the
leading BFKL part is provided by the momentum of the emitted gluon
$\qt = \kt - \kt'$, as suggested by the $b$-dependent part of the NLL eigenvalue
in Eq.~(\ref{eq:nllorg}), which corresponds to the kernel
$b \frac{1}{\qt^2} \log \frac{\qt^2}{\kt^2}\big|_\mathrm{Reg}$
(see~\cite{NLLCC}), and --- via $\om$-expansion --- to the $b$-term in
Eq.~(\ref{eq:omexpcol}).  A simplified version of Eq.~(\ref{eq:kerresum2})
without the NLL term and with one collinear term (for $\gamma \rightarrow 0$)
was used in~\cite{KMS1997} for a phenomenological analysis of the structure
functions.

Note that, if we take literally the $\om$-expansion (\ref{eq:omexp}) with the
choice of NLL term~(\ref{eq:chiomnll}), then $\chi_1^\om / \chi_0^\om$ would
coincide with $\chi_{\ci}^\om$ close to the collinear poles, but would be
different in detail away from them, and would actually contain spurious poles at
complex values of $\gamma$ due to the zeroes of $\chi_0^{\om}(\gamma)$. Such
poles cancel out if the full $\om$-expansion series (\ref{eq:omexp}) is summed
up, but are present at any finite truncation of the series, thus implying poor
convergence of the solution whenever $\gamma$-values close to the spurious poles
become important.  For this reason in this paper we prefer to resum collinear
singularities by the improved kernel (\ref{eq:kerresum2}), which contains only
collinear poles.  Furthermore, the NLL term needed to complete
Eq.~(\ref{eq:kerresum2}) --- to be detailed in the next section --- turns out to
have only simple (leading) collinear poles, because the running coupling terms
have been already included in the $q^2$-scale dependence of the running
coupling.  Therefore, the full kernel has the same virtues as
Eq.~(\ref{eq:omexp}) in the collinear limit and, lacking spurious poles, is more
suitable for numerical iteration.

%%%%%%%%%%%%%%%%%%%%%%%%%%%%%%%%%%%%%%%%%%%%%%%%%%%%%%%%%%%%%%%%%%%%%%%
\section{Form of the resummed kernel\label{s:reskernel}}

\subsection{Next-to-leading  coefficient kernel}

We have still to incorporate in our improved kernel the exact form of
the NLL result~\cite{NLLFL,NLLCC} in the scheme of the $\asb$ expansion,
i.e. (\ref{eq:kerresum2}). We choose to start from the leading kernel
in Eq.~(\ref{eq:kerresum2}) which incorporates both the collinear
resummation and the running coupling effects due to the choice of
scale $\qt^2$.  The full improved kernel then has the form
\begin{equation}
\label{eq:kerresumf}
 \CKT_{\om} = \asb(\qt^2) K^\om_0 + \om \asb(k_>^2) K^\om_\ci
 + \asb^2(k_>^2) \tilde{K}^\om_1 \;,
\end{equation}
where $k_> = \max(k,k')$, $k_< = \min(k,k')$, and $\tilde{K}^\om_1$ is
determined below.

We recall that the Mellin transform of the collinear part $\CK_{\ci}^{\om}$,
defined by
\begin{equation}
 \chi_{\ci}^{\om}(\gamma)  =   { A_1(\om) \over \gamma + \omhalf} +
 {A_1(\om)   \over 1 - \gamma + \omhalf} \;,
\label{eq:coll}
\end{equation}
leads to the expression
\begin{equation}
\label{eq:colexp}
K_{\ci}^{\om}(k,k') = \frac{A_1(\om)}{k_>^2} \left(\frac{k_<}{k_>}\right)^{\om} \; .
\end{equation}
One can match the above prescription to the standard kernel at NLL order by
expanding in $\om$ and in $b\asb$ to first order
\begin{equation}
 \CKT_{\om} \simeq \asb(k^2) (K_0^0 + \om K_0^1 + \om K_{\ci}^0)
 + \bar{\alpha}_s^2 (\tilde{K}_1^{0} + K_0^\mathrm{run}),
\label{eq:kerfirst}
\end{equation}
where we have defined
\begin{equation}
K_\ci^0 \equiv K_\ci^{\om=0} \;, \qquad
 K_j^0 \equiv K_j^{\om = 0} \;, \qquad
 K_j^1 \equiv \left.{\partial K_j^{\om} \over \partial \om}\right|_{\om= 0}
 \;, \qquad \chi_0^\mathrm{run}(\gamma)  = -\frac{b}{2}(\chi'_0+\chi_0^2 ) \;,
\end{equation}
by noting that the running coupling term has the form [see
Eqs.~(\ref{K0Regularised},\ref{chiRegularised}) and App.~\ref{a:krn}]
\begin{equation}
\label{eq:runterm}
 K_0^\mathrm{run}(k,k')
 = -b \left[\log \frac{\qt^2}{\kt^2} K_0(\kt,\kt') \right]_\mathrm{Reg} \; .
\end{equation}
By replacing the expression (\ref{eq:kerfirst}) into Eq.~(\ref{eq:sigma}) we
obtain the relationship with the customary BFKL Green's function
\begin{equation}
 [\om - \CKT_{\om}]^{-1} =
 \left( 1-\bar{\alpha}_s (K_0^1+K_{\ci}^0) \right)^{-1}
 \left[ \om - \bar{\alpha}_s \left(K_0+\bar{\alpha}_s K_1 +
 {\cal O}(\bar{\alpha}_s^2) \right) \right]^{-1} \;,
\label{impact}
\end{equation}
where $K_0$ and $K_1$ are LL and NLL $\om$-independent kernels.
The two expressions will match provided we identify
\begin{align}
 K_0 &= K_0^0 \nonumber \\
 \tilde{K}_1^0 &= K_1 - K_0^0 ( K_0^1 + K_{\ci}^0) - K_0^\mathrm{run} \;,
\label{eq:matching}
\end{align}
and we properly redefine the (so far unspecified) impact factors (see
Sec.~\ref{s:iif}).  Thus the term $\tilde{K}_1^0$ in (\ref{eq:matching})
corresponds to the customary NLL expression (\ref{eq:nllorg}) with subtractions.

In $\gamma$-space the subtracted NLL eigenvalue function which corresponds to
the $\tilde{K}_1^{\om}$ has the following form:
\begin{align}
 \tilde{\chi}_1(\gamma) &= \chi_1(\gamma)-\chi_0^0(\gamma)
 [ \chi_0^1(\gamma)+\chi_\ci^0(\gamma) ] -\chi_0^{\mathrm{run}}(\gamma)
\nonumber \\
 &=\chi_1(\gamma)
 + {1 \over 2} \chi_0(\gamma) {\pi^2 \over \sin^2(\pi\gamma)}
 - \chi_0(\gamma) {A_1(0) \over \gamma (1-\gamma) }
 + \frac{b}{2}(\chi'_0+\chi_0^2) \;.
\label{eq:nllgam}
\end{align}
The subtractions cancel the triple poles (due to change of energy
scales) and the double poles (from the non-singular part of the
anomalous dimension).  Therefore the resulting kernel $\tilde{\chi}_1$
contains at most single poles at $\gamma = 0, 1$.  Eq.~(\ref{eq:kerresum2})
together with the eigenvalues
(\ref{eq:chiomll}), (\ref{eq:coll}) and (\ref{eq:nllgam}) gives a
complete prescription for the resummed model. This new formulation is
identical to the previous $\om$-expansion \cite{CC,CCS1} near the
collinear poles.  It has the advantage that it can be easily
transformed into the $(x,k^2)$ space (it is free of ratios in
$\gamma$-space, such as $\chi_1/\chi_0$) and avoids the spurious poles
that were present in (\ref{eq:omexp}).

Note that the choice of scale in $\asb$ in the first term in
Eq.~(\ref{eq:kerresumf}) is determined by the form of the NLL part.  Any change
of scale in this term would correspond to the change of NLL terms proportional
to $b$. The scale for the collinear parts is chosen to match the standard DGLAP
formulation whereas in the NLL part is purely conventional, and its change would
be of the NNLL order.  In the following, in order to study the dependence on
renormalisation scale uncertainties, we introduce the quantity $x_\mu$ and
generalize eq.~\eqref{eq:kerresumf} as follows
\begin{equation}\label{xmukernel}
 \CKO = \left(\asb(x_\mu^2 q^2) + b\asb^2 \ln x_\mu^2 \right) K^\om_0
 + \om \left(\asb(x_\mu^2 k_>^2) + b\asb^2 \ln x_\mu^2 \right) K^\om_\ci
 + \asb^2(x_\mu^2 k_>^2) \tilde{K}^\om_1\,.
\end{equation}

%----------------------------------------------------------------------
\subsection{Form of the kernel in ($\boldsymbol{x,\kt^2}$) space
\label{s:fkxks}}

We define the resummed kernel in ($x,k^2$) space as the (integrated) inverse
Mellin transform of $\CKT_\om$:
\begin{equation}\label{defKz}
 \CKT(z;k,k') \equiv \int \frac{d \om}{2 \pi i} \; z^{-\om} \,
 \frac1{\om} \CKT_{\om}(k,k')
\end{equation}
where the real variable $z$ can assume values between $x$ and 1.

The subtractions of (\ref{eq:nllgam}) are translated into $(x,k^2)$ space to give
\begin{eqnarray}
{1 \over 2} \chi_0(\gamma) {\pi^2 \over \sin^2(\pi\gamma)} & \rightarrow &
{1 \over 4 |k^2-k^{\prime 2}|} \left[\log^2 {k^{\prime 2} \over k^2}
 + 4 \Li \Big( 1-{k_<^2 \over k_>^2} \Big) \right] \nonumber \\
- \chi_0(\gamma) {A_1(0) \over \gamma (1-\gamma) }  & \rightarrow &
-A_1(0) \;\mathrm{sign}(k^2-k^{\prime 2})
\left[ {1 \over k^2} \log {|k^2-k^{\prime 2}| \over k^{\prime 2}}
- {1 \over k^{\prime 2}} \log{|k^{\prime 2}-k^2| \over k^2 } \right] \nonumber \\
 \frac12 [ \chi_0^2(\gamma)+\chi_0'(\gamma) ] & \rightarrow &
 \left[ {1 \over q^2} \log{q^2 \over k^2} \right]_\mathrm{Reg}\;,
\label{eq:SubKspace}
\end{eqnarray}
where the dilogarithm function is defined to be
\begin{equation}
 \Li(w) \ := \; -\int_0^w {dt \over t}\log(1-t) \;, \quad
 \Li(1) = {\pi^2 \over 6}\;.
\end{equation}

In $(x,k^2)$ space the symmetric shift is translated into the symmetric
kinematical constraint which has to be imposed onto the real emission part of
the BFKL and also into the collinear non-singular DGLAP terms:
\begin{equation}
kz < k' < \frac{k}{z}
\end{equation}
(in the following we denote the imposition of the kinematical constraint onto
the appropriate parts of the kernel by the superscript (kc),
i.e. $K_0^{\kc}(k,k')$).

The final resummed kernel $\CKT(z;k,k')$ is the sum of three contributions:
\begin{multline}
 \int_x^1\frac{dz}{z}\int d k'^2 \; \CKT(z;k,k') f(\frac{x}{z},k')\\
 = \int_x^1\frac{dz}{z}\int d k'^2
 \left[\bar{\alpha}_s(\qt^2) K_0^{\kc}(z;\kt,\kt') +
 \bar{\alpha}_s(k_{>}^2) K_{\ci}^{\kc}(z;k,k')+
 \bar{\alpha}^2_s({k}^2_{>}) \tilde{K}_1(k,k') \right ] f(\frac{x}{z},k') \;.
 \label{eq:kernelxk}
\end{multline}
The different terms are as follows:\\
$\bullet$ LO BFKL with running coupling and consistency constraint
($\qt = \kt-\kt'$)
\begin{multline}
\label{eq:LOBFKLkc}
 \int_x^1\frac{dz}{z}\int dk'^2 \; \left[ \bar{\alpha}_s(\qt^2)
  K_0^{\kc}(z;\kt,\kt') \right] f(\frac{x}{z},k') \\
 = \int_x^1 \frac{dz}{z} \int {d^2 \qt \over \pi \qt^2} \;
  \bar{\alpha}_s(\qt^2) \left[ f(\frac{x}{z},|\kt+\qt|)
 \Theta(\frac{k}{z}-k')\Theta(k'-kz)-\Theta(k-q) f(\frac{x}{z},k) \right]\;,
\end{multline}
$\bullet$ non-singular DGLAP terms with consistency constraint
\begin{align}
 \int_x^1\frac{dz}{z}\int d k'{}^2 \;& \bar{\alpha}_s(k_{>}^2)
  K_{\ci}^{\kc}(z;k,k') f(\frac{x}{z},k')  \nonumber \\
 &= \int_x^1 { dz \over z} \int_{(kz)^2}^{k^2} \frac{{dk'}^2}{k^2} \;
  \bar{\alpha}_s(k^2) z{k \over k'} \tilde{P}_{gg}(z{k \over k'})
  f(\frac{x}{z},k')  \nonumber \\
 &+ \int_x^1 { dz \over z} \int_{k^2}^{(k/z)^2} \frac{{dk'}^2}{k'{}^2} \;
  \bar{\alpha}_s({k'}^2) z{k' \over k} \tilde{P}_{gg}(z{k' \over k})
  f(\frac{x}{z},k') \;,  \label{eq:dglapterms}
\end{align}
$\bullet$ NLL part of the BFKL with subtractions included
\begin{multline}
 \int_x^1\frac{dz}{z}\int d k'{}^2 \; \bar{\alpha}^2_s({k}^2_{>})
  \tilde{K}_1(k,k') f(\frac{x}{z},k')  \\
 = {1\over 4}\int_x^1 \frac{dz}{z} \int d{k'}^2 \; \bar{\alpha}^2_s({k}^2_{>})
  \bigg\{  \\
 {\left({67 \over 9} - {\pi^2 \over 3}\right) {1\over |{k'}^2-k^2|}
  \left [f(\frac{x}{z},{k'}^2) - {2 k_{<}^2 \over ({k'}^2 + k^2)}
  f(\frac{x}{z},k^2)\right] + } \\
 {\bigg[ - {1 \over 32} \left({2 \over {k'}^2} + {2 \over k^2} +
  \left({ 1\over {k'}^2 } - {1\over k^2} \right)
  \log\left({k^2 \over {k'}^2}\right)\right)
  + {4 \Li(1-k_{<}^2/k_{>}^2) \over |{k'}^2 - k^2|}} \\
 {-4 A_1(0){\rm sgn}({k}^2-{k'}^2)
  \left( {1 \over k^2} \log{|{k'}^2-k^2| \over {k'}^2} -
  {1 \over {k'}^2} \log{|{k'}^2-k^2| \over {k}^2}\right)} \\
 - \left(3 + \left({3 \over 4} - {({k'}^2+k^2)^2 \over 32{k'}^2 k^2}\right)
  \right) \int_0^{\infty} {dy \over k^2 + y^2 {k'}^2 }
  \log|{1+y \over 1-y}| \\
 + {1 \over {k'}^2 + k^2} \left( {\pi^2 \over 3} +
  4 \Li( {k_{<}^2 \over k_{>}^2})\right) \bigg]
  f(\frac{x}{z},k') \bigg\} \\
 + {1\over 4} 6 \zeta(3) \int_x^1 \frac{dz}{z} \;
  \bar{\alpha}^2_s(k^2)  f(\frac{x}{z},k) \;. \hspace{40mm}
\end{multline}

The non-singular splitting function in the DGLAP terms is defined as follows:
\begin{equation}
 \tilde{P}_{gg}  =  P_{gg} - { 1 \over z} \;,
\end{equation}
where we take
\begin{equation}
 \label{eq:pgg}
 P_{gg} = \frac{1-z}{z} + z (1-z) + \frac{z}{(1-z)_+} +
 \frac{11}{12} \delta(1-z)\;,
\end{equation}
(we only consider purely gluonic channel, $n_f=0$).
Also we note that the argument of the splitting function ${\tilde P}$ has to be
shifted in (\ref{eq:dglapterms}) in order to reproduce the correct collinear
limit when the kinematic constraint ($kz < k' < \frac{k}{z}$) is included.
This follows from the inverse Mellin transform of Eq.~(\ref{eq:colexp})
\begin{equation}
 \label{eq:argsplit}
 K_{\ci}^{\kc}(z;k,k') = \int \frac{d\om}{2\pi i} \; \frac{A_1(\om)}{k_>^2}
 \left(\frac{k_<}{k_>} \right)^{\om} \, z^{-\om} =\frac{1}{k_>^2}
 \left(z\frac{k_>}{k_<} \right) \tilde{P} \left(z\frac{k_>}{k_<}\right) \; .
\end{equation}
In other words the correct variable in the splitting function is modified by the
ratio of two virtualities in the case when the kinematical constraint is included
\begin{eqnarray}
 z & \rightarrow & z { k \over k'} < 1 \, \, {\rm for} \, \, k' < k \nonumber \\
 z & \rightarrow & z { k' \over k} < 1 \, \, {\rm for} \, \, k < k'  \; .
\end{eqnarray}

%%%%%%%%%%%%%%%%%%%%%%%%%%%%%%%%%%%%%%%%%%%%%%%%%%%%%%%%%%%%%%%
\subsection{Choice of scheme\label{s:cos}}

The prescription formulated above for the kernel eigenvalue
(\ref{eq:nllgam}) is free of double and cubic poles in $\gamma=0$ (and
$\gamma=1$), however there are still some residual single poles.
These poles come from the constant terms from the expansion of
subtraction $\chi_0^1+\chi_{\ci}^0$ around $\gamma=0$ ($\gamma=1$).
Expanding this subtraction around $\gamma=0$ one obtains
\begin{equation}
 -\chi_0^0 (\chi_0^1+\chi_{\ci}^0) = -\chi_0^0 \; \bigg [ -{1 \over 2\gamma^2}
 - {\pi^2 \over 6}+{A_1(0) \over \gamma} + A_1(0) + {\cal O}(\gamma) \; \bigg] \;,
\end{equation}
therefore there appear additional singular terms,
\begin{equation}
 \left[ {\pi^2 \over 6} - A_1(0)  \right] {1 \over \gamma} \;,
 \label{eq:sub}
\end{equation}
in the subtracted kernel $\tilde{\chi}{}_1^0$ which are not shifted. Furthermore,
the term~(\ref{eq:sub}) contributes to the 2-loop anomalous dimension, together
with the constant term arising from the leading kernel as follows:
\begin{equation} \label{eq:subBis}
 \chi_0^{\om} + \om \chi_{\ci}^{\om} \simeq \frac{1 + \om A_1}{\gamma +\omhalf}
 - \om \, C(\om) +\ord( \gamma +\omhalf ) \;,
\end{equation}
where
\begin{eqnarray}
 C(\om) & = & -\frac{A_1(\om)}{\om+1}
  +\frac{\psi(1+\om)-\psi(1)}{\om} \nonumber \\
 C(0) & = & \frac{\pi^2 }{6}-A_1(0) \; .
 \label{eq:sub1}
\end{eqnarray}
By combining~(\ref{eq:sub}) with~(\ref{eq:subBis}) we would get the
contribution
\begin{equation}\label{eq:anomfalse}
 \Delta \gamma^{(2)} = \frac{\asb^2}{\om} C(0) -\asb C(\om) \gamma^{(1)}
 \simeq \frac{\asb^2}{\om} \left[C(0)-C(\om) \big(1+\om A_1(\om)\big)\right]\;,
\end{equation}
where $\gamma^{(1)}=\asb(1+\om A_1(\om))/\om$ is the DGLAP anomalous
dimension in the leading order.
The expression (\ref{eq:anomfalse}) violates the momentum sum rule
$\Delta \gamma^{(2)}(\om=1)=0$.

We thus consider two possible forms of subtraction.
In the first scheme {\bf A}
we add and subtract from the NLL part the term proportional to
$C(0)$ in the following way,
\begin{equation}
 \tilde{\chi}_1(\gamma) \to \tilde{\chi}_1^{\om}(\gamma) =
 \tilde{\chi}_1(\gamma) -C(0) \chi_0(\gamma)+ C(0) \chi_0^{\om}(\gamma) \;,
 \label{eq:subver1}
\end{equation}
which leads to the following modification of the kernel in $(x,k^2)$ space
\begin{multline}\label{eq:kerver1}
 \int_x^1\frac{dz}{z}\int d k'^2
  \left \{ \bar{\alpha}_s(\qt^2)
  K_0^{\kc}(z;k,k')  + \bar{\alpha}_s(k_{>}^2)
  K_{\ci}^{\kc}(z;k,k') + \right. \\
 \left. + \bar{\alpha}^2_s({k}^2_{>})
  \left[ \tilde{K}_1(k,k') + C(0) K_0^{\kc}(z;k,k') - C(0)
   K_0(k,k') \right ] \right \} f(\frac{x}{z},k') \,.
\end{multline}
This scheme satisfies general RG constraints, but contains the
anomalous dimension (\ref{eq:anomfalse}) and violates the momentum sum
rule.

In the second scheme {\bf B}
we shall consider a modification which adds the shifted pole to the NLL kernel
with the $\om$-dependent coefficient $(1+\om A_1(\om))$
\begin{equation} \label{eq:subver3}
 \tilde{\chi}_1(\gamma) \to \tilde{\chi}_1^{\om}(\gamma) =
 \tilde{\chi}_1(\gamma) -
 \left( \frac1{\gamma} + \frac1{1-\gamma} \right) C(0) +
 \left( \frac1{\gamma+\omhalf} + \frac1{1+\omhalf-\gamma} \right)
 C(\om) [ 1+\om A_1(\om)] \;.
\end{equation}
It is straightforward to check that in this case the 2-loop anomalous dimension
vanishes%
\footnote{We use here a generalization of the $Q_0$-scheme~\cite{Ci95}. We do
  not try to include the known 2-loop expression in the
  $\overline{\mathrm{MS}}$ scheme because it is subject to a scheme change and
  to kernel ambiguities which are not fully understood yet.
},
due to a cancellation between the pole term (\ref{eq:subver3}) and the
constant term in (\ref{eq:subBis}). Therefore, scheme {\bf B}\ satisfies
energy-momentum conservation.

The change in the resummed kernel in $(x,k^2)$ space corresponding to scheme
{\bf B} is obtained by inverse Mellin transform of (\ref{eq:subver3}) and is
given by
\begin{multline}\label{eq:kerver3}
 \int_x^1\frac{dz}{z}\int d k'^2
  \left \{ \bar{\alpha}_s(\qt^2)
  K_0^{\kc}(z;k,k')  + \bar{\alpha}_s(k_{>}^2)
  K_{\ci}^{\kc}(z;k,k') + \bar{\alpha}^2_s({k}^2_{>})\tilde{K}_1(k,k')  \right\}
 f(\frac{x}{z},k') - \\
 -\int_x^1 { dz \over z} \left\{ C(0)\left[ \int_0^{k^2}
 \frac{{dk'}^2}{k^2}
  \bar{\alpha}^2_s(k^2)  f(\frac{x}{z},k')
+ \int_{k^2}^\infty  \frac{{dk'}^2}{k'{}^2}
 \bar{\alpha}^2_s({k'}^2)  f(\frac{x}{z},k') \right] - \right. \\
\left. - \left[ \int_{(kz)^2}^{k^2}
\frac{{dk'}^2}{k^2}
 \bar{\alpha}^2_s(k^2)   z{k \over k'} S_{2}(z{k \over k'})  f(\frac{x}{z},k')
+ \int_{k^2}^{(k/z)^2} \frac{{dk'}^2}{k'{}^2}
 \bar{\alpha}^2_s({k'}^2)  z{k' \over k} S_{2}(z{k' \over k}) f(\frac{x}{z},k') \right] \right\} \;,
\end{multline}
with the function $S_2(z)$ given by
\begin{multline}\label{d:S2}
 S_2(z) =\frac{1}{144 z} \left\{ 132 + 24 {\pi}^2 + z
  [-541 + 24 {\pi}^2 + 72 z (1 + 3 z)]
  -  144 \ln(-1 + \frac{1}{z}) \ln(\frac{1}{z}) \right. \\
 + 12 \bigg( \ln(1 - z) [-1 - 2 z (23 + z (-15 + 8 z))
  - 12 (1 + z) \ln(1 - z)]  + 12 z \ln(-1 + \frac{1}{z}) \ln(\frac{1}{z}) \\
  + 2 z [1 + z (-21 + 5 z) - 6 \ln(1 - z)] \ln(z) -
  6 (-1 + 2 z) \ln^2(z)  \bigg) \\
   + 144 (-1 + z) \left. [ \Li(z)+\frac{1}{2} \ln(\frac{1}{z})
  \ln \bigg[ \frac{z}{(1-z)^2}
  \bigg] -\frac{{\pi}^2}{6} ] - 144 (1 + 2 z) \Li(1 - z) \right\}\;.
\end{multline}

Note that whatever scheme we choose, ${\tilde K}_1$ contains
higher-twist poles (at $\gamma=-1,-2,\ldots$ and $\gamma=2,3,\ldots$),
which are not shifted. In the calculations that follow we keep these
poles unshifted independently of the choice of energy-scale. This means
that calculations of the Green's function carried out with different
energy-scale choices will formally differ at NNLL level. In practice
however we find that this energy-scale dependence is very small.

%%%%%%%%%%%%%%%%%%%%%%%%%%%%%%%%%%%%%%%%%%%%%%%%%%%%%%%%%%%%%%%
\section{Characteristic features of the resummed Green's function
\label{s:resGGF}}

We shall first investigate the features of the two-scale Green's
function%
\footnote{In Secs.~\ref{s:resGGF} and \ref{s:resanomdim} we remove for
  simplicity the~~$\widetilde{}$~~symbols used before to denote RGI quantities
  in our present scheme.
}
$G(Y;k^2,k_0^2)$ based on the form of the resummed kernel just proposed.  In the
perturbative regime $k^2,k_0^2 \gg \Lambda_{QCD}^2$ with $\oms(k^2) Y$
large we have both 
perturbative contributions, leading to the hard Pomeron exponent, and
non-perturbative ones, due to the asymptotic Pomeron, which is sensitive to the
strong coupling region. It was noticed in~\cite{CCS2,CCSS2} that the hard
Pomeron dominates for energies below a certain threshold
$\asb(k^2) Y < 1 / b \omp$ beyond which there is a tunneling transition to the
non-perturbative regime.
It has also been noticed~\cite{CCSS1}, that in the formal limit
$b\rightarrow 0$ with $\asb(k^2)$ fixed the Pomeron is suppressed as
$\exp(-1/b\asb)$, so that one can define a purely perturbative Green's
functions and investigate the diffusion corrections to the hard
Pomeron exponent. In the following, we use the $b$-expansion up to
second order, so as to obtain the exponent $\oms(t)$ and the
additional parameters occurring in the diffusion corrections predicted
by our improved small-$x$ equation. Furthermore, we analyze the
perturbative non-perturbative interface numerically so as to estimate,
as a function of $\log Q^2$, the critical rapidity beyond which the
non-perturbative Pomeron takes over.

Since the perturbative rapidity range turns out to be considerably
extended with respect to LL expectations, we shall be able to extract
numerically the full perturbative Green's function and among other
things its high-energy exponent and diffusion corrections to it.

%%%%%%%%%%%%%%%%%%%%%%%%%%%%%%%%%%%%%%%%%%%%%%%%%%%%%%%%%%%5
\subsection{Frozen coupling features}

Let us first consider the features of $G(Y;t_1,t_2)$ in the limit of frozen
coupling $\asb=\asb(k_0^2)$, i.e.\ $b=0$. In such a case the kernel $\CK_\om$
becomes scale invariant, but the solution to Eq.~(\ref{eq:bfklequation}) is
still non-trivial, due to the $\om$-dependence which complicates the
$Y$-evolution, it no longer being purely diffusive.  In fact, the characteristic
function becomes
\begin{equation}
\label{eq:chifix}
\asb \chi_{\om}(\gamma,\asb) = \asb(\chi^\om_0+\om \chi_{\ci}^{\om})
+\asb^2 \tilde{\chi}_1^{\om} \;,
\end{equation}
and the important $\om$ values, corresponding to the pole of the resolvent, are
defined by
\begin{equation}
\label{eq:omdefchi}
\om = \asb \chi_\om(\gamma,\asb)\;,
\end{equation}
whose solution at fixed $\gamma$ we denote by
\begin{equation}
\om = \asb \chi^{(0)}_{\eff}(\gamma,\asb) \;,
\label{eq:chieff}
\end{equation}
the superscript (0) referring to the $b=0$ limit.  The effective characteristic
function (\ref{eq:chieff}) so defined has the interpretation of a BFKL-type
eigenvalue reproducing the pole~(\ref{eq:omdefchi}). As such, it can be
compared, at least for frozen coupling, to the analogous quantity defined in the
``duality'' approach of Ref.~\cite{ABF2000}. It provides information about the
hard Pomeron exponent and the diffusion coefficient $D = \chi_m''/ 2 \chi_m$.
%%%%%%%%%%%%%%%%%%%%%%%%%%%%%%%%%%%%%%%%%%%%%%%%%%%%%%%%%%%%%%%%%%%%
\begin{figure}[t]
\centering{
 \includegraphics[width=0.7\textwidth]{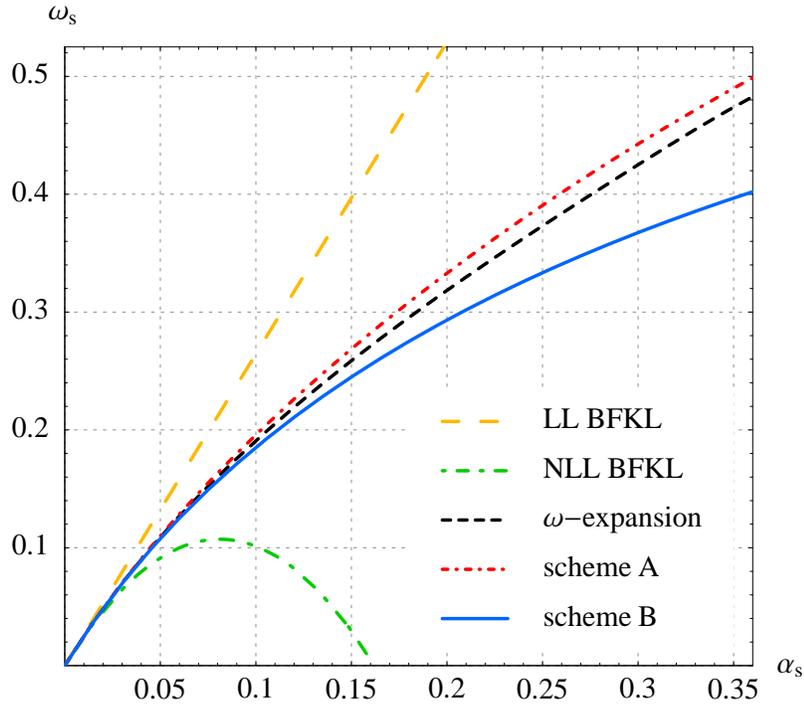}\\}
\caption{\it  $\oms$ as a function of $\as$ for different subtraction
schemes together with the original result for the $\om$-expansion. The
calculation is done in the fixed coupling case.
\label{f:oms}}
\end{figure}
%%%%%%%%%%%%%%%%%%%%%%%%%%%%%%%%%%%%%%%%%%%%%%%%%%%%%%%%%%%%%%%%%%%%%
In Fig.~\ref{f:oms} we compare the results for the exponent $\om_s$ as a
function of $\as$ calculated in the case of fixed coupling for schemes
$A,B$ and the original $\om$-expansion method presented in~\cite{CC,CCS1}. The
critical exponent is obtained by evaluating the effective kernel eigenvalue at
the minimum
\begin{equation}
\label{eq:omsad0}
\om_s^{(0)} =  \asb \chi^{(0)}_{\eff}(\gamma_m,\asb)\;.
\end{equation}
All resummed results for the intercept are significantly reduced in
comparison with the LL result and they all give stable predictions
even for large values of $\asb$. As we see from the plot, the changes
of resummation procedure as well as subtraction scheme do not
significantly influence the values of $\om_s$. They give at most $20\%
$ change at the highest $\as \simeq 0.35$.
%%%%%%%%%%%%%%%%%%%%%%%%%%%%%%%%%%%%%%%%%%%%%%%%%%%%%%%%%%%%%%%%%%%%%
\begin{figure}[t]
\noindent
 \includegraphics[width=0.48\textwidth]{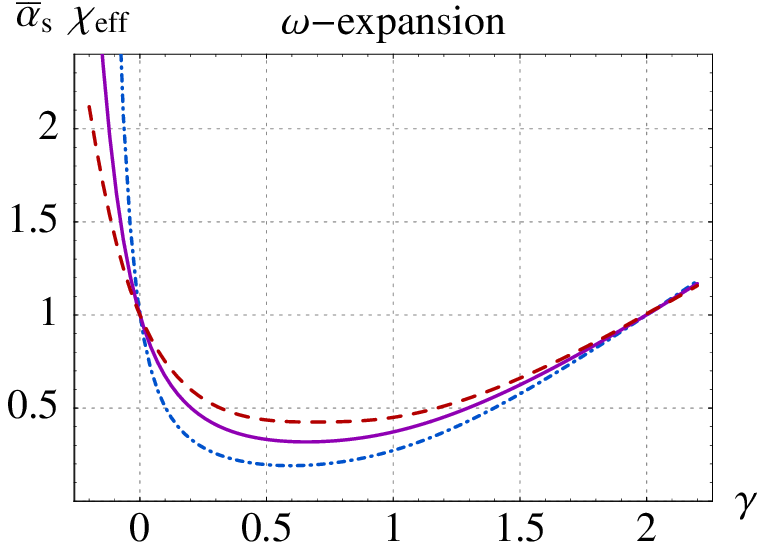}\\[5mm]
 \includegraphics[width=0.48\textwidth]{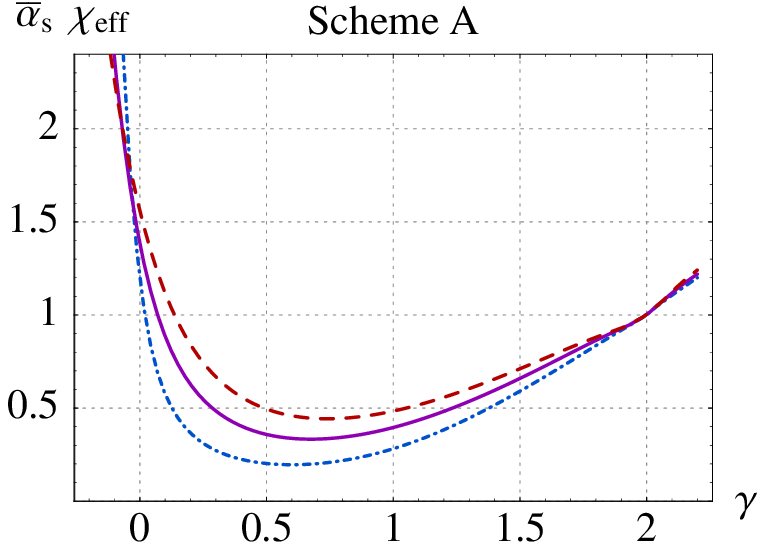} \hfill
 \includegraphics[width=0.48\textwidth]{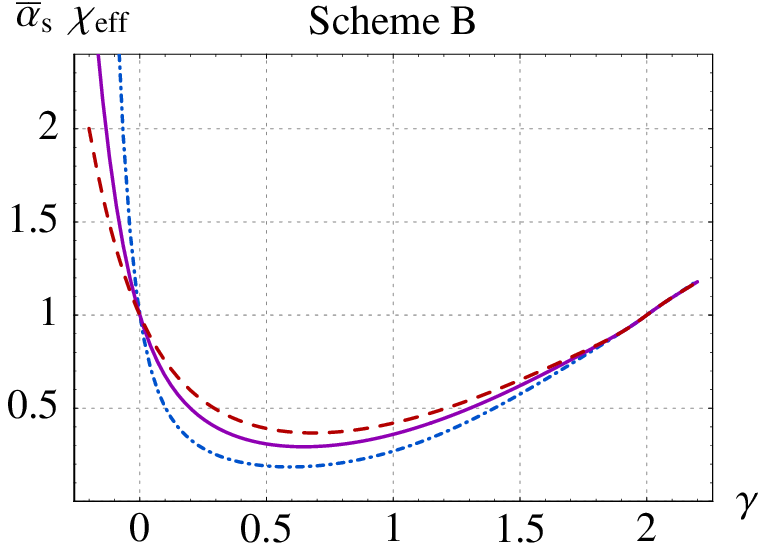}\\
\caption{
\it  $\asb \chi_{\eff}(\gamma,\asb)$ as a function of $\gamma$ in different
schemes for different values of $\as$: $\as=0.1$ (dash-dotted line), $\as=0.2$
(solid line), $\as=0.3$ (dashed line). The calculation is done in the fixed
coupling case.
\label{f:chi_eff}}
\end{figure}
%%%%%%%%%%%%%%%%%%%%%%%%%%%%%%%%%%%%%%%%%%%%%%%%%%%%%%%%%%%%%%%%%%%%%%
In Fig.~\ref{f:chi_eff} we show the effective kernel eigenvalue
as a function of $\gamma$. We have considered here the asymmetric $\om$-shift,
which corresponds to the upper energy scale choice $\nu_0=k^2$.  In this case it
is easy to show that close to $\gamma=0$ the effective eigenvalues from scheme
B and the original $\om$-expansion~\cite{CCS1} satisfy the momentum sum
rule. This is illustrated in Fig.~\ref{f:chi_eff} by the fact that
$\asb \chi_{\eff}(\gamma=0,\asb) = 1$ for all values of $\asb$ in these schemes.
This can be seen by expanding around $\gamma=0$, where we have
\begin{equation}\label{gammaAround0}
\chi_{\omega}(\gamma,\asb) \propto \frac{1+\om A_1(\om)}{\gamma}
\end{equation}
which for $\gamma=0$ gives $\om A_1(\om)=-1$, which has the solution
$\om=1$.  Note that a second fixed intersection point of curves with
different $\as$ occurs at $\gamma = 2$. This is expected from
energy-momentum conservation%
\footnote{
    Such an intersection occurs in scheme A also (where
    momentum conservation is not satisfied) as an artefact of the
    collision of the 
    shifted pole at $\gamma = 1+\om$ with the unshifted one at $\gamma = 2$.}
in the collinear regime $Q_0^2 \gg Q^2$, because of a behavior similar to
Eq.~(\ref{gammaAround0}) around the shifted pole $1+\om-\gamma = 0$.  This
intersection has no counterpart in the approach of Ref.~\cite{ABF2000}.

\begin{figure}[t]
  %\vspace*{0.0cm}
        \centerline{ \epsfig{figure=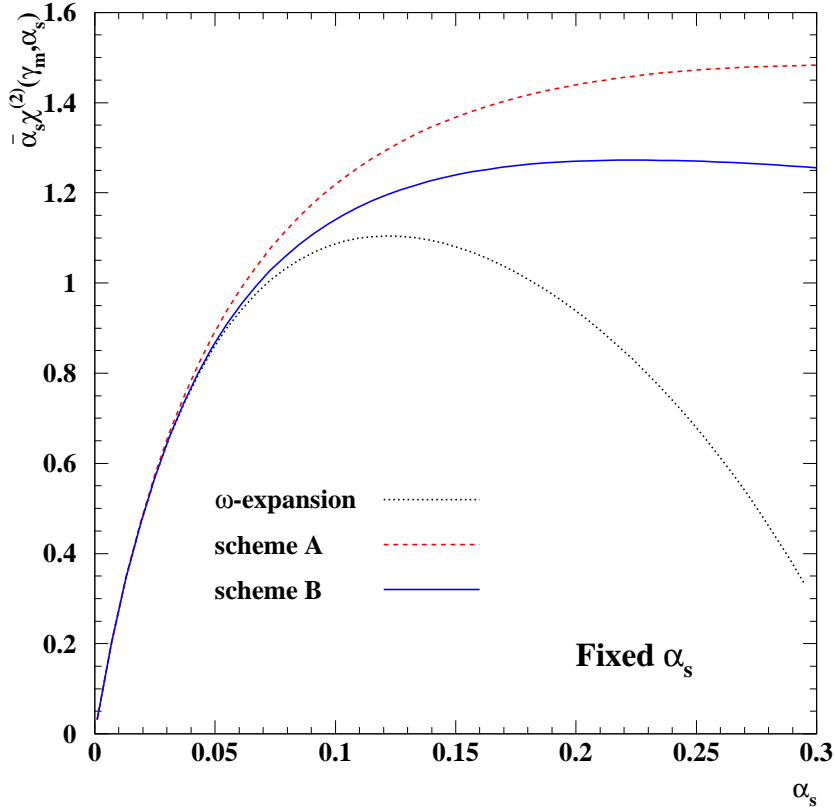,width=12cm} }
\vspace*{0.5cm}
\caption{\it  $\chi''(\gamma_m,\asb)$  as a function of $\as$ for two
  different subtraction models and the $\om$-expansion scheme.
\label{f:chi_dp}}
\end{figure}

We also examine the second derivative $\chi_{\eff}''(\gamma,\asb)$ which
controls the diffusion properties of the small-$x$ equation, Fig.~\ref{f:chi_dp}.
As we see from the plot, the second derivative is more model-dependent than the
intercept $\om_s$, though the two models A and B presented in this paper
give quite similar answers. The value of the second derivative will influence
the diffusion corrections to the hard Pomeron, as we shall see in
Sec.~\ref{s:eidc}, and also the transition of the solution to the
non-perturbative regime.
%%%%%%%%%%%%%%%%%%%%%%%%%%%%%%%%%%%%%%%%%%%%%%%%%%%%%%%%%%%%%%%%%%%%

%%%%%%%%%%%%%%%%%%%%%%%%%%%%%%%%%%%%%%%%%%%%%%%%%%%%%%%%%%%%%%%%%%%%
\subsection{Numerical methods for solution}
\label{sec:numSoln}

In this section we are going to investigate in detail the shape of the
solutions to the integral equation\footnote{An interesting iterative
  method of solution to the NLL BFKL equation has been recently
  proposed \cite{AnSaVe03}. By using this method it is possible to
  solve the equation directly in $(x,k)$ space and keep the full
  angular dependence.} with the resummed kernel given in
sections~\ref{s:fkxks} and \ref{s:cos}.  To this aim we solve
numerically
the following integral equation%
\footnote{Here we change slightly the notation in the first argument
  of $\CK$, writing $\log\frac1{z} = Y-y$ instead of $z$.}
\begin{equation}
 G(Y;k,k_0) = G^{(0)}(k,k_0) \Theta(Y) +
 \int_0^Y dy  \int_{k_{min}}^{k_{max}} dk'^2\;  \CK(Y-y;k,k') G(y;k',k_0)\\
\label{eq:inteq}
\end{equation}
with (so as to have the same normalization as in
Eq.~\eqref{eq:bfklequation}),
\begin{equation}
 2\pi k_0^2 G^{(0)}(k,k_0) = \delta(\ln\frac{k}{k_0})\,.
\nonumber
\end{equation}
We use the method of iterations and discretized kernel similar to that
introduced in 
\cite{BoMaSaSc97}.  
More precisely in our problem (see (\ref{eq:kernelxk}))
we can rewrite the kernel in the following way:
\begin{equation}
 \CK(Y-y;k,k') = \sum_{\alpha} \CK_{\alpha} (Y-y;k,k') =
 \sum_{\alpha} K_{\alpha}(k,k') \, P_{\alpha}(Y-y) \,
 \Theta_R\Big(Y-y-{\rm max}(\ln\frac{k}{k'},\ln\frac{k'}{k})\Big) \; ,
\label{eq:kersplit}
\end{equation}
where the index $\alpha$ enumerates different terms in the equation
(\ref{eq:kernelxk}) (that is LL BFKL, LL DGLAP, and the different
components of NLL BFKL with subtractions), each of which factorize
into transverse and longitudinal parts.  The $P_{\alpha}$ are the
singular and non-singular pieces of the splitting function as well as
the subtraction terms $S_2(x)$. The additional $\Theta_R$ stands for
the kinematical constraint, applied to all terms that in Mellin-space
have an $\om$-shift.

To find the solution numerically one introduces a grid in rapidity $Y$
and logarithm of momentum, $\tau =\ln k/k_0$, with small spacings,
$\Delta Y$ and $\Delta \tau$ respectively.  The solution is then
calculated at the grid points.  Linear interpolation gives the values
of the solution in the points between the nodes of the grid
\begin{equation}
G(Y;k,k_0) = \sum_{i}\sum_{j}\, \phi_i(Y) \psi_j(k) \,G(Y_i;k_j,k_0)
\end{equation}
where $\phi_i(Y)$ and $\psi_j(k)$ are the appropriate basis functions
for linear interpolation.  To find the solution for $G$ the equation
(\ref{eq:inteq}) is solved by a method of evolution in rapidity.  In a
first step one takes $G(Y_0=0;k_m,k_0)=G^{(0)}(k_m,k_0)$ and estimates
$G(Y_1;k_m,k_0)$ at the next point of the grid, $Y_1= \Delta Y$, using
the integral equation (\ref{eq:inteq}).  This gives a first
approximated value for $G(Y_1;k_m,k_0)$.  This function is then again
used in equation (\ref{eq:inteq}) to calculate the next approximation.
Usually a few iterations are sufficient to find an accurate answer
(typically $5-8$). After obtaining $G(Y_1;k_m,k_0)$ with
the desired accuracy one proceeds to calculate the solution on the
next point of the grid $Y_2 = 2 \Delta Y$ and so on.  The procedure is
then repeated for all points of the grid in rapidity $Y_n=n\Delta Y$.

The procedure presented above requires numerous evaluations of the
right hand side of equation (\ref{eq:inteq}). Given the fact that we
have two convolutions in $y$ and $k'$, with the complicated kernel
$\CK$, such a procedure can be quite time consuming.

In order to speed up the calculation one can discretize in $k'$ the
kernels $K_{\alpha}$ and in $y$ the functions $P_{\alpha}$ using the
basis functions in the following form
\begin{eqnarray}
K^{(\alpha)}_{m,i}  & = &\int dk'^2 \,  K_{\alpha}(k_m,k') \, \psi_i(k') \nonumber \\
P^{(\alpha)}_{n-j} & = & \int dy \,  P_{\alpha}(Y_n-y) \,  \phi_j(y) \; ,
\end{eqnarray}
where we have used the fact that the functions $P_{\alpha}$ depend
only on the difference $Y_n-y$ which --- together with the linear
interpolation --- results in a one-dimensional vector $P$ instead of a
matrix.  One can simplify the treatment of the $\Theta_R$ function in
(\ref{eq:kersplit}) by using the same grid spacing in $y$ and in $\ln
k$, $\Delta y = \Delta \tau$ (or for energy-scale choice $\nu_0=k^2$,
$\Delta y = 2\Delta \tau$). After the discretization procedure, the
convolution on the right hand side in equation (\ref{eq:inteq}) (and
using (\ref{eq:kersplit})) can be then represented as a multiplication
as follows
\begin{equation}
 \int dy \int dk'^2 \; \CK_{\alpha}(Y_n-y;k_m,k') \, G(y;k',k_0) =
 \sum_{i=0}^{i_{max}} \, \sum_{j=0}^{n-|m-i|} \,
 P^{(\alpha)}_{n-j-|m-i|} \, K^{(\alpha)}_{m,i} \, G(y_j;k_i,k_0) \;,
\end{equation}
so that in practice all the integrations present in equations
(\ref{eq:inteq}) are performed once before the evolution, and then
only the multiplications of kernel matrices and gluon Green's function
vectors are done during the iterations.

\begin{figure}[t]
\centering\includegraphics[width=0.5\textwidth]{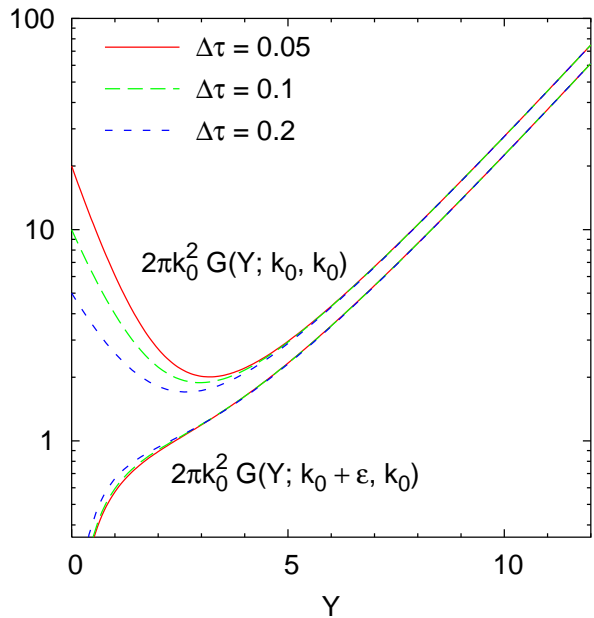}
\caption{\it Gluon Green's function as a function of rapidity
  $Y$ for three different grid spacings $\Delta \tau=0.05,0.1,0.2$. LL
  evolution is used with a fixed coupling, $\asb = 0.2$; $\epsilon
  \simeq 0.2k_0$.}
\label{f:LL_dy_var}
\end{figure}

Of course, in a numerical analysis one is not able to use exact
distributions --- in particular for the delta function in $k$ as an
initial condition, see 
Eq.~(\ref{eq:bfklequation}). In practice what is done is to set to
$1/\Delta \tau$ one point on the fine grid i.e.
$2\pi k_0^2 G^{(0)}(k_m,k_0)=\frac{1}{\Delta \tau}\delta_{m0}$, where $\Delta
\tau$ is the grid spacing in $\ln k$. The resulting Green's function
will be finite in the $Y=0$ limit but dependent on the size of the
grid spacing.  We illustrate this effect in the upper set of curves of
Fig.~\ref{f:LL_dy_var}, where we have solved the equation
(\ref{eq:inteq}) with the kernel in LL approximation with $3$
different grid spacings $\Delta \tau =0.05,\; 0.1,\; 0.2$. One might
be worried by the apparently substantial dependence on the choice of
the grid spacing $\Delta \tau$. However this is just a consequence of
the grid-dependent discretization of the initial $\delta$-function and
disappears when one convolutes the gluon Green's function with some
smooth impact factor. We will therefore consider from now on a
slightly asymmetric choice of scales, $G(Y;t_0,t_0+\epsilon)$ with
$\epsilon=0.2$. In the lower set of curves of Fig.~\ref{f:LL_dy_var}
one sees that the dependence on the grid spacing in this case is
relatively small. For the remaining plots in this paper we have used
$\Delta \tau = 0.1$ or smaller.

%----------------------------------------------------------------------
\subsection{Basic features of the Green's function}
\label{sec:basicGreenFeatures}

Let us now discuss the properties of the gluon Green's function
obtained with the method discussed above. We shall use a one-loop
coupling with $n_f=4$, normalized such that
$\asb(9\GeV^2)=0.244$.\footnote{As one obtains, roughly, by running
  $\as(M_z^2) = 0.118$ down to $9\GeV^2$, taking into account flavor
  thresholds and the two-loop $\beta$-function.}  The coupling is
cut off at scale $\kbar = 0.74\GeV$ --- a detailed analysis of the
sensitivity to this regularization is postponed to
section~\ref{sec:NPGreen}.
In the kernel, for the time being we consider $n_f=0$, since our
single-channel RGI approach does not properly account for the quark
sector (however we will see below that simply varying $n_f$ in the
kernel has only a small effect).

Results will be given given for: LL evolution (with $\asb(q^2)$); our
two resummation schemes, A and B; and two variants of `pure' NLL
evolution: one, labeled `NLL $\as(q^2)$' where the kernel is
$\asb(q^2)K_0 + \asb^2(k_>^2) K_1^{b=0}$, with $K_1^{b=0}$
corresponding to eq.~\eqref{eq:nllorg} without the first term in
square brackets; and another, labeled `NLL $\as(k^2)$', where the
kernel is $\asb(k^2)K_0 + \asb^2(k^2) K_1$, and $K_1$ corresponds to
eq.~\eqref{eq:nllorg} in full.

\begin{figure}[t]
\includegraphics[width=0.49\textwidth]{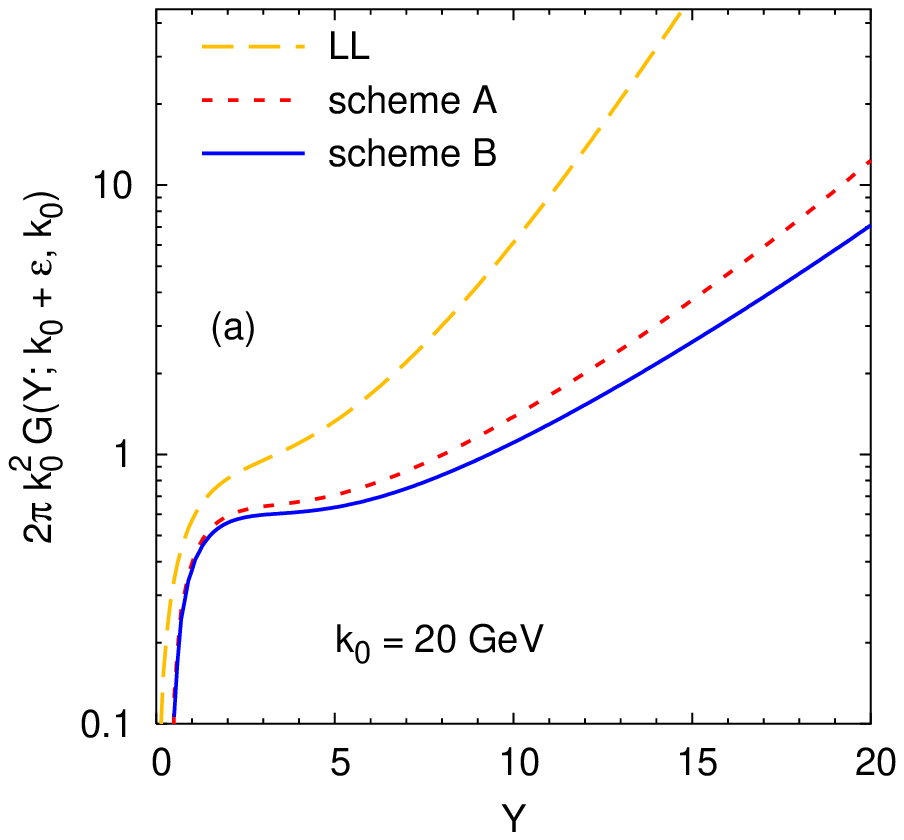}\hfill
\includegraphics[width=0.49\textwidth]{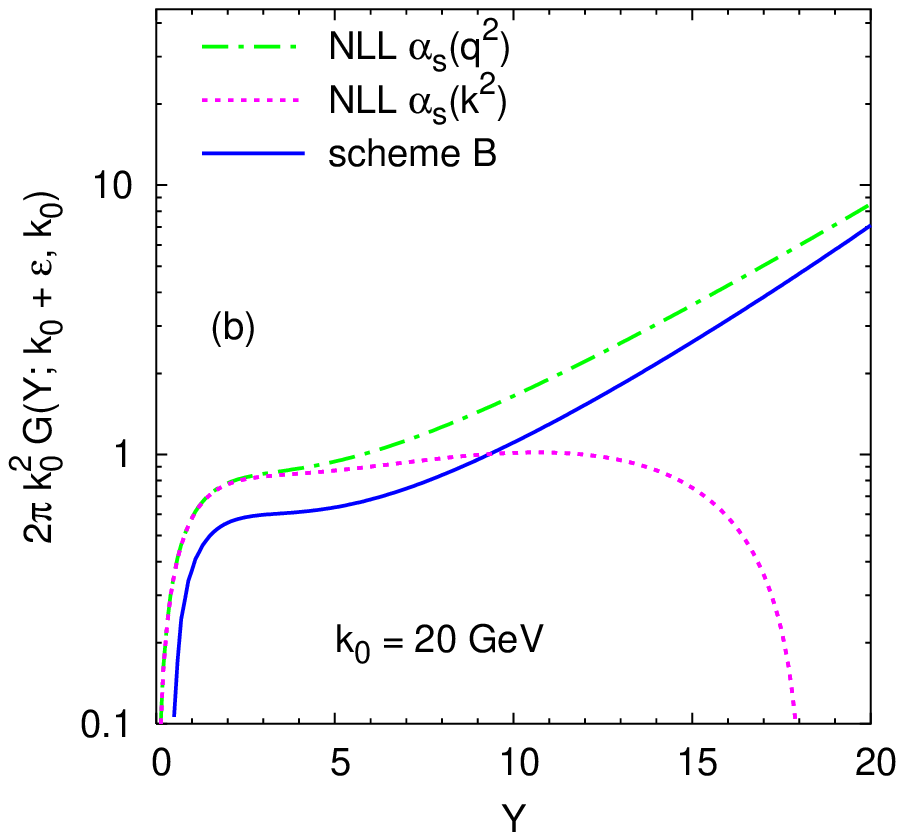}
\caption{\it Gluon Green's function $G(Y;k_0+\epsilon,k_0)$ as a function of
  rapidity $Y$: (a) for LL and the two RGI schemes A and B; (b) for
  scheme B and two variants of pure NLL evolution. The parameters are
  $k_0=20 \GeV$ and $\epsilon\simeq 0.2 k_0$.}
\label{f:gfy_ab_nf0}
\end{figure}

\begin{figure}[t]
\includegraphics[width=0.49\textwidth]{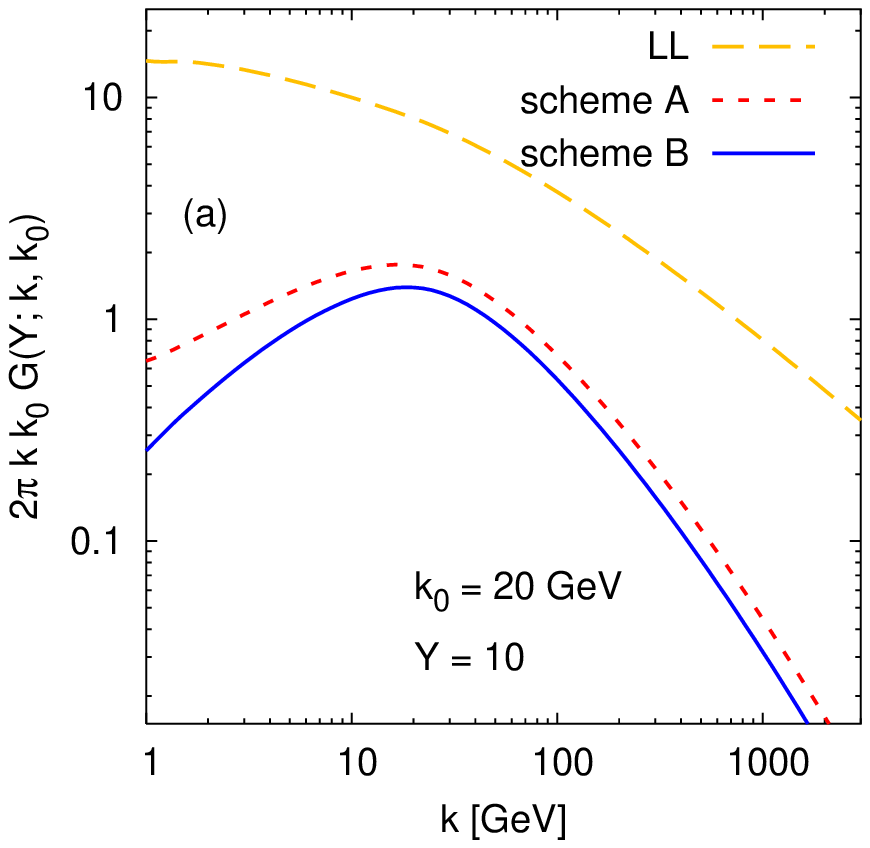}\hfill
\includegraphics[width=0.49\textwidth]{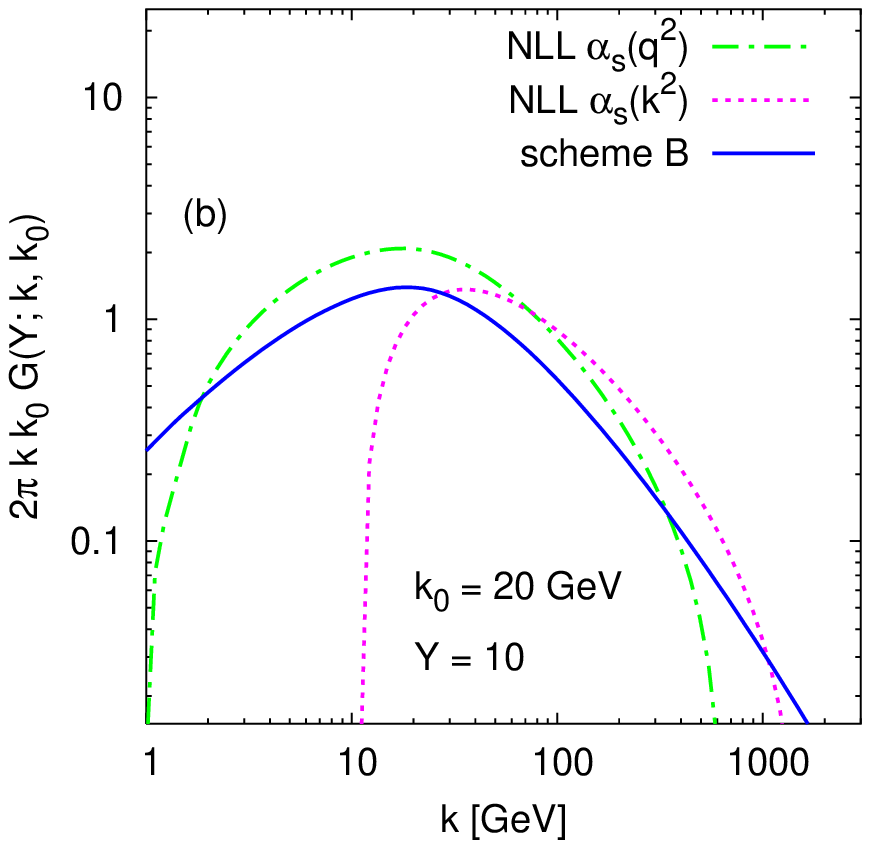}
\caption{\it Gluon Green's function $2\pi k k_0 G(Y;k,k_0)$ at rapidity $Y=10$
  as a function of the transverse scale $k$. The sets of kernels used
  in plots (a) and (b) are the same as in figure~\ref{f:gfy_ab_nf0}.}
\label{f:gfk_ab_nf0}
\end{figure}

Fig.~\ref{f:gfy_ab_nf0} shows Green's functions
$G(Y;k_0+\epsilon,k_0)$ as a function of rapidity $Y$, and
fig.~\ref{f:gfk_ab_nf0} shows $k k_0G(Y;k,k_0)$ as a function of $k$
for $Y=10$.  To aid legibility, each figure has been separated into
two plots, the left-hand one (a) showing LL and schemes A and B, while
the right-hand one (b) shows the two pure NLL curves and scheme B.  We
choose a moderately high value for the initial transverse scale,
$k_0=20\GeV$, $\asb(k_0)\simeq 0.15$, so as to be able to focus on the
perturbative aspects of the problem (non-perturbative effects are
formally suppressed by powers of $\Lambda^2/k_0^2$).  Such a scale has
been used for BFKL dijet studies at the Tevatron \cite{D0Dijet}.

A number of features of fig.~\ref{f:gfy_ab_nf0}a are worth commenting.
Most noticeable is the significant reduction in the high-energy growth
of the Green's function when going from LL evolution to our resummed
schemes A and B. This is as expected from the discussion of
high-energy exponents, fig.~\ref{f:oms}. Also important is the fact
that for the RGI schemes the high-energy growth does not start until a
rapidity of about $4$. Both of these observations are relevant to the
problem of trying to reconcile theoretical predictions with the lack
of experimental evidence for a strong high-energy growth of cross
sections at today's energies. The small difference between the two RGI
resummation schemes, A and B, is in accord with their slightly
different $\oms$ values (cf.\ fig.~\ref{f:oms}).

As regards the transverse momentum dependence of the Green's function,
fig.~\ref{f:gfk_ab_nf0}a there are a number of further differences
between the LL and RGI results. The higher overall normalization for LL
evolution is just a consequence of a larger $\oms$ value. But one also
sees that the large-$k$ tails in $k$ for the resummed models are
substantially steeper than in the LL case. This can be understood by
comparing the diffusion coefficients in these models: the RGI models
are characterized by a smaller $\chi''_{\eff}$ and, as a consequence,
they have less diffusion than in the LL case. As was the case for the
$Y$ dependence, the two RGI schemes give very similar results, here
differing essentially only in the normalization.

Some comments are due concerning the structure at low $k$: there,
there is a component of the
evolution that is sensitive to the larger coupling,
$\as(1\GeV^2)\simeq 0.4$. For the LL case the 
resulting stronger evolution (than at $k_0^2$) over-compensates the
suppression due to the large 
ratio of scales $k_0/k$, leading to the absence of a decreasing
low-$k$ tail. For 
the RGI schemes the difference between $\oms$ values at $1\GeV$ and
$k_0$ is not sufficient to bring about this overcompensation for
$Y=10$, so there still is a decreasing tail for small $k$. However the
results are sensitive to the fact that at large $\as$ the difference
between $\oms$ values for the two schemes becomes non-negligible. This
is what causes the low-$k$ Green's function to be almost three times
larger for scheme A than scheme B.  It should of course be kept in
mind that all the properties at
low $k$ are strongly dependent on the particular choice of infrared
regularization of the coupling.

Let us now examine the right-hand plots of figures~\ref{f:gfy_ab_nf0}
and \ref{f:gfk_ab_nf0}, which show results with pure NLL evolution. We
recall that the original motivation for introducing RGI resummation
schemes was the large size of the NLL corrections, and in particular
the fact that for moderate values of the coupling the NLL terms change
the sign of $\chi(\gamma)$ and its second derivative around
$\gamma=1/2$, with the situation being even worse in the collinear
region. Nevertheless, as was pointed out by Ross \cite{ROSS98},
because of the change of sign of $\chi''(\frac12)$, the usual saddle
point at $\gamma=\frac12$ is replaced by two saddle points off the
real axis, at $\gamma = 1/2 + i\nu_0$ and $1/2 - i\nu_0^*$, and it is
the value of $\chi$ at these new saddle points that determines the
high-energy behavior of the (fixed-coupling) NLL Green's function:
\begin{equation}
  \pi k k_0 G(Y;k,k_0) = \int\frac{d\gamma}{2\pi i} e^{\asb Y
    \chi(\gamma)}
  \left(\frac{k^2}{k_0^2}\right)^{\gamma-\frac12} \sim 
  e^{\asb Y
    \chi(\frac12 + i\nu_0)} \left(\frac{k^2}{k_0^2}\right)^{
  i\nu_0} + 
  e^{\asb Y
    \chi(\frac12 - i\nu_0^*)} \left(\frac{k^2}{k_0^2}\right)^{-
  i\nu_0^*}\,.
\end{equation}
Since $\chi(\gamma) = \chi^*(\gamma^*)$ this gives
\begin{equation}
  \label{eq:GNLL}
  \pi k k_0 G(Y;k,k_0) \sim e^{\asb Y \Re[\chi(\frac12 + i\nu_0)]
   - \Im[\nu_0] \ln \frac{k^2}{k_0^2}} 
  \cos \left( \Re[\nu_0] \ln \frac{k^2}{k_0^2} + \asb Y \,\Im[\chi(\frac12 + i\nu_0)]
  \right)\,.
\end{equation}
When $\chi_1(\gamma)$ is symmetric in $\gamma \leftrightarrow
1-\gamma$, as is the case if we use $\as(q^2)$ in the LL term (or as
can be achieved with the modified Mellin transform suggested in
\cite{NLLFL} and used in \cite{ROSS98}), then
$\asb\chi(\frac12+\nu_0)$ is real, having a value of about $0.2$. One
therefore expects to find a high-energy growth of the Green's function
that numerically is not so different from that with out RGI resummed
schemes. This is precisely what is observed in fig.~\ref{f:gfy_ab_nf0}b
for the `NLL~$\as(q^2)$' result.

On the other hand if $\chi_1(\gamma)$ is not symmetric in $\gamma
\leftrightarrow 1-\gamma$ then $\chi$ will be complex at the saddle
points. This is the case for the `NLL~$\as(k^2)$' kernel and the
change in sign of the Green's function around $Y=18$ can be understood
as a direct consequence of a zero of eq.~\eqref{eq:GNLL} when $\asb Y
\,\Im[\chi(\frac12 + i\nu_0)] = \pi/2$.

The oscillatory behavior of eq.~\eqref{eq:GNLL} also becomes an issue
when $k\neq k_0$, as is visible in fig.~\ref{f:gfk_ab_nf0}b. For NLL
evolution with $\as(q^2)$ the change of sign intervenes only for
ratios of $k/k_0$ that are fairly small or large from a
phenomenological point of view (at least for Mueller-Navelet or
$\gamma^*\gamma^*$ type processes). For evolution with $\as(k^2)$ the
situation is more dramatic because of the sum of terms in the argument
of the cosine of eq.~\eqref{eq:GNLL}.

So our overall conclusions regarding NLL evolution is that, while in
certain instances it may give results that are not too different
from those with RGI methods, in general it offers only limited
predictive power, because of the strong sensitivity to the details of the
formulation. Though here we have just discussed renormalisation scale
sensitivity, we note that changing the energy scale $\nu_0$ from say
$k k_0$ to $k^2$ also leads to a Green's function that oscillates as a
function of $Y$, since once again the characteristic function is
asymmetric.

A final point relating to figures~\ref{f:gfy_ab_nf0} and
\ref{f:gfk_ab_nf0} concerns the overall normalization of the results.
One sees that at low $Y$ the LL and NLL results all have similar
normalizations, while the RGI results are slightly lower. This is
because the $\om$-dependence is associated with an implicit
NLO impact factor. This of course has to be taken into account should
one wish to use the RGI Green's function in conjunction with any NLO
impact factor calculation, as is discussed in detail in
section~\ref{s:iif}.
To close this section we present brief results on $\nf$ and
renormalisation scale dependence for the RGI schemes.

\begin{figure}[t]
\centering\includegraphics[width=0.8\textwidth]{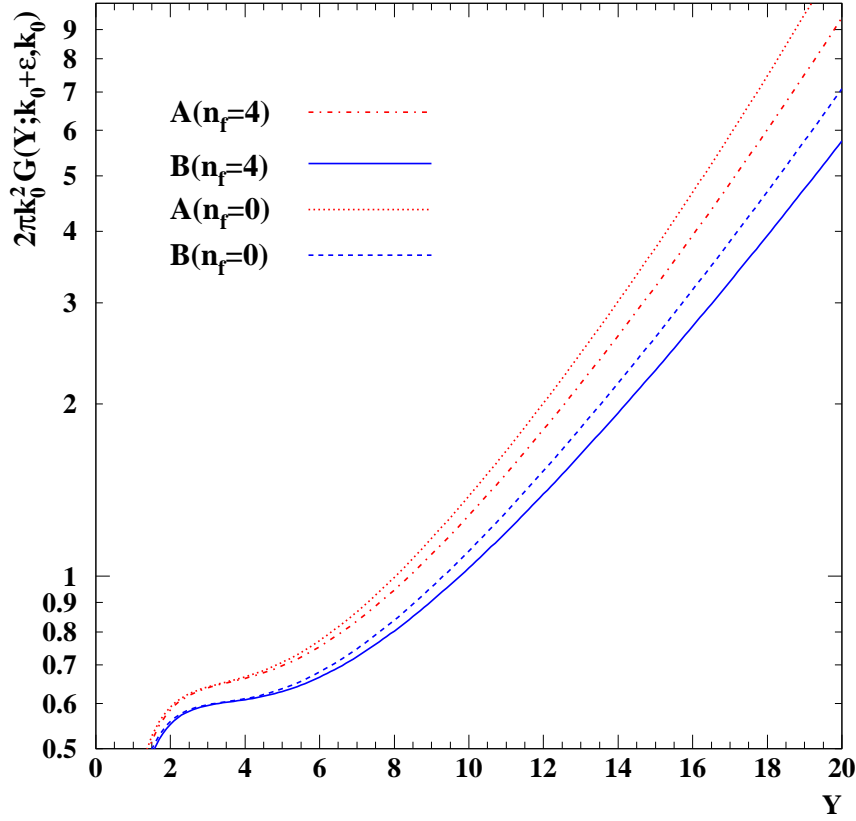}
\caption{\it Gluon Green's function $G(Y;k_0,k_0+\epsilon)$
  as a function of rapidity $Y$ for two different resummation schemes
  $A,B$ in the $n_f=0,4$ cases. Parameter $\epsilon\simeq 0.2 k_0$}
\label{f:gfy_ab_nf4}
\end{figure}

Our RGI approach has been constructed for a purely gluonic channel and only
scheme B satisfies the momentum sum rule in this case.  For
phenomenological purposes one would wish to include quarks and in
Fig.~\ref{f:gfy_ab_nf4} we present the two schemes in the cases when
$n_f=0$ and $n_f=4$ in the NLL kernel. As is clear from the plot,
having $n_f \neq 0$ does not change the result in a significant way.
We note that the full inclusion of quarks in a RG-consistent manner is
a non-trivial operation in this framework especially if one is to
construct resummed quark anomalous dimensions that satisfy the
momentum sum rules. 

\begin{figure}[t]
\centering\includegraphics[width=0.8\textwidth]{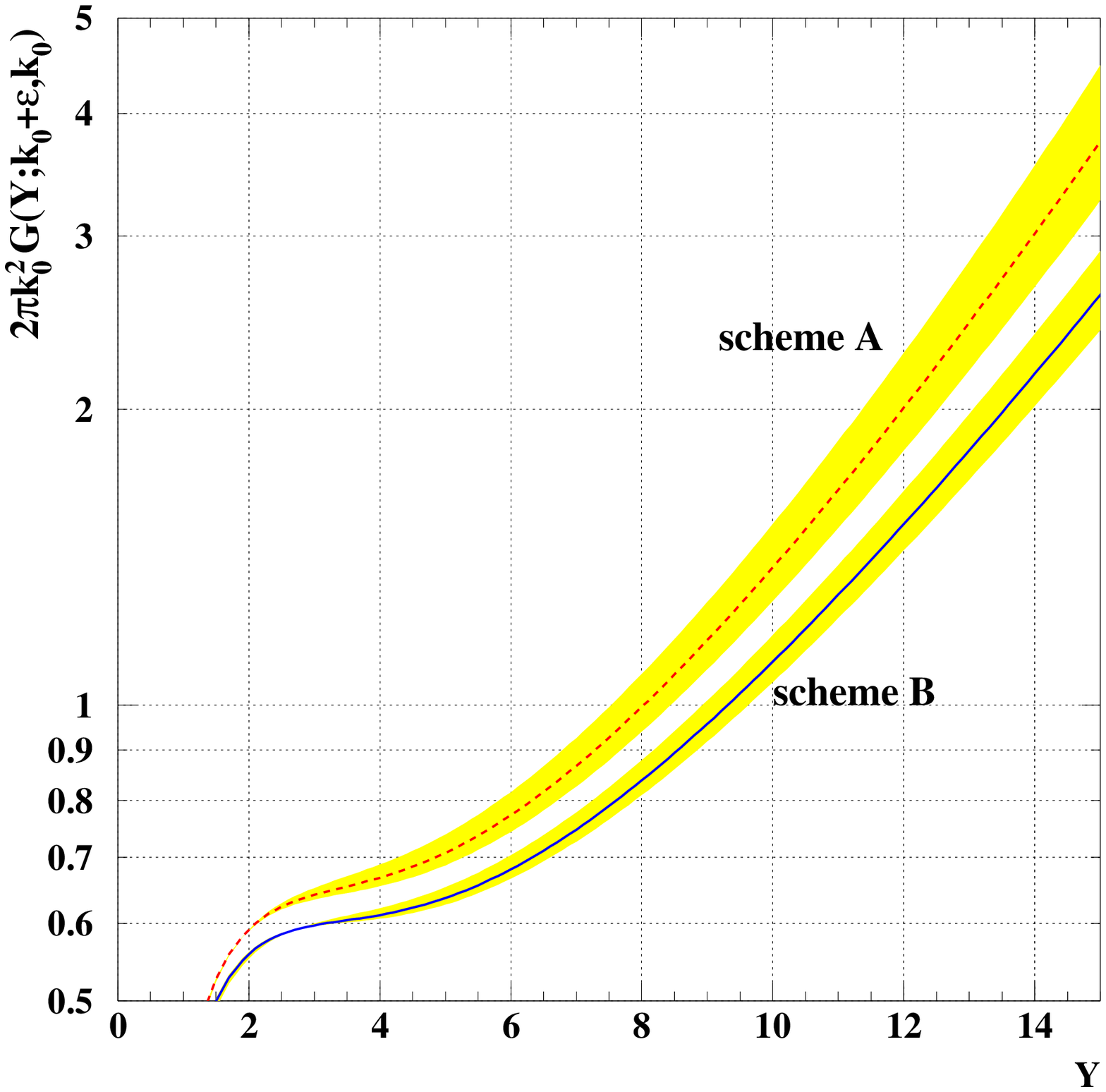}
\caption{\it Gluon Green's function $G(Y;k_0,k_0+\epsilon)$
  as a function of rapidity $Y$ for two different resummation models
  $A,B$. The bands represent the uncertainty due to a variation of the
  renormalisation scale in the range $1/2 < x_{\mu}^2<2$. Parameter
  $\epsilon\simeq 0.2 k_0$}
\label{f:gfy_ab_xmu}
\end{figure}

Finally, we show the dependence of the gluon Green's function on the
renormalisation scale choice eq.~(\ref{xmukernel}). We have varied the
scale $x_{\mu}$ in the range $1/2 < x^2_{\mu}  < 2 $. The results
of the calculation are presented in Fig.~\ref{f:gfy_ab_xmu} where the
yellow bands correspond to the renormalisation scale variation for two
resummation schemes.

%%%%%%%%%%%%%%%%%%%%%%%%%%%%%%%%%%%
\subsection{$\boldsymbol{b}$-expansion of intercept and of diffusion
 coefficient\label{s:eidc}}

In order to properly evaluate the hard Pomeron intercept $\om_s$ in
the case with running coupling it is necessary to control the
corrections with respect to the frozen coupling limit. To this end we
shall apply the $b$-expansion method presented in~\cite{CCSS1}.

According to this method, we use the formal limit $b \to 0$ (with
$\as(t_0)$ kept fixed) in order to suppress the non-perturbative
Pomeron.  The left-over perturbative Green's function can then be
expanded in $b$ in the form
\begin{equation}\label{d:oms1}
 G(Y;t_0,t_0) = G^{(0)}(Y;t_0,t_0) \exp \left[ b \oms^{(1)} Y +
 \ord(b^2 \as^5 Y^3) \right] \left( 1+\ord(b^2 \as^4 Y^2) \right) \;,
\end{equation}
which shows a shift of $\oms$ of order $b \as^2$, as well as
diffusion corrections of order $(b\as)^2 (\om_s Y)^2$ and $(b\as)^2
(\om_s Y)^3$. The purpose of this subsection is to compute
$\oms^{(1)}$ [defined by Eq.~(\ref{d:oms1})] and the $Y^3$ terms both
analytically and numerically. Further corrections to $\oms$ of order
$b^2 \as^3$ appear as subleading contributions in this expansion and
are probably not really meaningful, given the complex $Y$-dependence
of the exponent involving the parameter $b \as^2 Y$~\cite{CCSS1}.

We start by expanding the $\as$-dependence of the kernel around the
frozen-coupling limit up to $\ord(b^2)$ by setting, for instance at
scale $q^2$,
\begin{equation}\label{svilAlphaq}
 \as(q^2) - \as(k_0^2) = -b \az^2
 \left( \log\frac{q^2}{k^2} + (t-t_0) \right) + b^2 \az^3
 \left( \log^2\frac{q^2}{k^2} +2 (t-t_0) \log\frac{q^2}{k^2} + (t-t_0)^2
 \right) \;,
\end{equation}
where $\az \equiv \asb(k_0^2)$ throughout this section.  We then
define the kernel with frozen-coupling $\CK_\om^{(0)} \equiv \left.
  \CK_\om \right|_{\asb \to \az}$ and the correction kernel $\Delta$
as
\begin{align}\label{defDelta}
 \Delta(t,t') &\equiv \CK_\om - \CK_\om^{(0)} =
 \CK_\om - \az \left( K_0^\om + \om K_\ci^\om \right) - \az^2 \tilde{K}_1^\om
 \\ \label{svilDelta}
 &= \Delta_0(t-t') + (t_0-t) \Delta_1(t-t') + (t_0-t)^2 \Delta_2(t-t')\;,
\end{align}
where the $\Delta_i$'s are scale-invariant, and are obtained from the
definition~(\ref{eq:kernelxk}) by picking up the relevant terms in the
running coupling expansions of type~(\ref{svilAlphaq}). We obtain:
\begin{subequations}\label{DeltaCoeff}
\begin{align}\label{Delta0}
 \Delta_0 &= - b \az^2 \left[ \log\frac{q^2}{k^2} K_0^\om +
  \log\frac{k_>^2}{k^2} (\om K_\ci^\om + 2 \az \tilde{K}_1^\om ) \right]
  +\ord( b^2 \az^3 ) \\ \nonumber
 \Delta_1 &= b \az^2 \left[ K_0^\om + \om K_\ci^\om + 2 \az \tilde{K}_1^\om
  \right] + \ord( b^2 \az^3 ) \\ \label{Delta1}
 &= b \az^2 \frac{\partial}{\partial\az} K_\om^{(0)} (\az; k^2, k_0^2 )
  + \ord( b^2 \az^3 ) \\ \label{Delta2}
 \Delta_2 &= b^2 \az^3 \left[ (K_0^\om + \om K_\ci^\om) + 3 \az \tilde{K}_1^\om\
  \right] \;.
\end{align}
\end{subequations}
Now we evaluate the Green's function $G_\om(t,t_0)$ in $\om$-space up
to second
order in $b$, with the purpose of deriving the leading diffusion terms%
\footnote{In principle all diffusion correction terms can be derived
  using this method.}  $\sim b^2 Y^3$ and the intercept shift at
$\ord( b )$; to this purpose, expansion~(\ref{DeltaCoeff}) is
sufficient.  We have
\begin{equation}\label{GExpan}
 G = G^{(0)} + G^{(1)} + G^{(2)} + \cdots
\end{equation}
with
\begin{subequations}\label{diffCorr}
\begin{align}\label{G0}
 G^{(0)} &= \big[ \om - K_\om^{(0)} \big]^{-1} \\ \label{G1}
 G^{(1)} &= G^{(0)} \Delta\; G^{(0)} = \int \frac{d\gamma}{2\pi i} \;
  e^{(\gamma-\half )(t-t_0)} G^{(0)}(\gamma) \left[ \Delta_0 G^{(0)} +
  \Big( \Delta_1 G^{(0)} \Big)' + \Big( \Delta_2 G^{(0)} \Big)'' \right](\gamma)
  \\ \label{G2}
 G^{(2)} &= G^{(0)} \Delta\; G^{(1)} = \int \frac{d\gamma}{2\pi i} \;
  e^{(\gamma-\half )(t-t_0)} G^{(0)}(\gamma) \left[ \Delta_0 G^{(1)} +
  \Big( \Delta_1 G^{(1)} \Big)' + \Big(\Delta_2 G^{(1)} \Big)'' \right](\gamma)
\end{align}
\end{subequations}
where, inside the integrals, we have used the same notation for the
kernels and their $\gamma$-space eigenvalues.  We restrict our
attention to $t=t_0$ and perform partial integrations to obtain
\begin{align}
 G^{(1)}_\om(t_0,t_0) &= \int \frac{d\gamma}{2\pi i} \; \left[
  \Big( \Delta_0+\frac12 \Delta_1' \Big) G^{(0)2} + \Delta_2
  \left( \frac{1}{2} \Big( G^{(0)2} \Big)'' - \Big( G^{(0)}{}' \Big)^2 \right)
 \right] \nonumber \\
 &= \int \frac{d\gamma}{2\pi i} \; \left[
  \frac{\Delta_0 + \frac12 \Delta_1' + \frac13 \Delta_2''}{
  \big[ \om - \chi_\om^{(0)} \big]^2}
  +\frac13 \, \frac{\Delta_2 \chi_\om^{(0)}{}''}{
  \big[ \om - \chi_\om^{(0)} \big]^3} \right] \;,
 \label{valueG1}
\end{align}
where the $\gamma$-variable dependence is understood in the
$\Delta$'s, $G$'s and $\chi$'s.  Up to this order, the maximal energy
dependence comes from the cubic pole, which yields a $\sim b^2 \az^3
Y^2$ dependence. The double pole yields instead terms $\sim b \az^2 Y$
which provide the $\ord(b)$ correction to $\oms$. By noting that
\begin{align}\label{valueG0}
 G^{(0)}(Y;t_0,t_0) &= \int \frac{d\gamma}{2\pi i} \frac{d\om}{2\pi i} \;
  \frac{e^{\om Y}}{\om - \chi_\om^{(0)}(\gamma)}
  \simeq \frac{J e^{\oms^{(0)} Y}}{\sqrt{4 \pi D \oms^{(0)} Y}} \\
 \label{jacobian}
 J &= \left[ 1 - \partial_{\om} \chi_\om^{(0)}(\half)
  \right]^{-1}_{\om = \oms^{(0)}} \;,
\end{align}
and that a squared Jacobian factor $J^2$ occurs in $G^{(1)}$, we
obtain the $\ord(b)$ correction
\begin{equation}\label{ordbCorr}
 b \oms^{(1)} =
 \left[ \Delta_0(\half) + \frac12 \Delta_1'(\half) \right] J \;,
\end{equation}
where actually $\Delta_1'(\half) = 0$, because $\Delta_1(\gamma)$ is
symmetric for $\gamma \leftrightarrow 1-\gamma$.

The eigenvalue function $\Delta_0(\gamma)$ is found from the
definition in Eq.~(\ref{Delta0}) by noting that the generalized
regularized kernel
\begin{equation}\label{K0Regularised}
 \frac1{\pi \qt^2} \left(\frac{\qt^2}{\kt^2}\right)^\lambda
 \left(\frac{k_<}{k_>}\right)^\om - \frac1{\lambda} \delta^2(\qt) \;,
\end{equation}
has characteristic function
$\chi_L^{[\lambda]}(\gamma+\omhalf) + \chi_R^{[\lambda]}(1-\gamma+\omhalf)$,
where
\begin{equation}\label{chiRegularised}
 \chi^{[\lambda]}(\gamma) = \frac1{\lambda} \left[
 \exp\left( \lambda \chi_0(\gamma) + \frac12 \lambda^2 \chi_0'(\gamma)
 + \ord(\lambda^3) \right) -1 \right] \;,
\end{equation}
and the subscript $L$ ($R$) refers to the projection with left-hand
(right-hand) poles. By proper expansion in $\lambda$ we obtain (See
App.~\ref{a:krn})
\begin{equation}\label{expanDelta01}
 \Delta_0(\gamma) + \frac12 \Delta_1'(\gamma) = - \frac{b \az^2}{2}
 \left\{ [\chi_0^2]_L(\gamma+\omhalf) -\om\chi_{\ci L}'(\gamma+\omhalf)
 -2\az\tilde{\chi}_{1L}^\om{}'(\gamma) +
 [ \gamma \leftrightarrow 1-\gamma ] \right\} \;,
\end{equation}
and finally
\begin{equation}\label{DeltaOms}
 \oms^{(1)} = - \az^2 \Big[
 [\chi_0^2]_L({\textstyle\frac{1+\om}{2}})
 - \om \chi_{\ci L}'({\textstyle\frac{1+\om}{2}})
 -2\az \tilde{\chi}_{1L}^\om{}'(\half) \Big] J + \cdots \;.
\end{equation}
We note that the expressions of the left projections are (See App.~\ref{a:krn})
\begin{align}\label{chi0L}
  \chi_{0L}(\gamma) &= \psi(1) - \psi(\gamma) \;, \\
 [\chi_{0}^2]_L(\gamma) &= 2 [\chi_{0L}(\gamma)]^2 - \psi'(\gamma)
  + \frac{\pi^2}{2}   \label{chi0squareL} \\
  \chi_{\ci L}(\gamma) &= \frac{A_1(\om)}{\gamma} \;,   \label{chiColL}
\end{align}
and $\tilde{\chi}_{1L}(\gamma)$, depending on the resummation scheme,
is quoted in App.~\ref{a:krn}.

While the $\sim b^2 Y^2$ terms exponentiate $\Delta\oms$ and provide a
further normalization correction~\cite{CCSS1}, the $\sim b^2 Y^3$
terms provide the leading diffusion corrections and occur in
$G^{(2)}$. Considering Eq.~(\ref{G2}) for $t=t_0$ and performing
partial integrations, we obtain at $\ord(b^2)$
\begin{align}\nonumber
 G_\om^{(2)}(t_0,t_0) &\simeq \int \frac{d\gamma}{2\pi i} \;
  \left[ \Delta_0 G^{(0)} - \Delta_1 G^{(0)}{}' \right] G^{(0)}
  \left[ \Delta_0 G^{(0)} + \Big( \Delta_1 G^{(0)} \Big)' \right] \\
 \label{G2integral}
  &= \int \frac{d\gamma}{2\pi i} \; \left[ \Big( \Delta_0 G^{(0)} \Big)^2
  - \Delta_1 G^{(0)}{}' \Big( \Delta_1 G^{(0)} \Big)' \right] G^{(0)} \;.
\end{align}
This result contains up to a fifth order pole, which can be reduced to
a quartic one by partial integration, to yield
\begin{align}\nonumber
 G^{(2)}(Y;t_0,t_0) &\simeq \int \frac{d\gamma}{2\pi i} \frac{d\om}{2\pi i} \;
  e^{\om Y} \frac14 \,
 \frac{\Delta_1^2 \chi_\om^{(0)}{}''}{\big[ \om - \chi_\om^{(0)} \big]^4}
 \\ \nonumber
 &\simeq  \int \frac{d\gamma}{2\pi i} \; e^{\asb \chi_\eff(\gamma) Y}
  \frac{Y^3}{24} \,
  \frac{\chi_\om^{(0)}{}''(\gamma) \left[\partial_{\log k_0^2}
  \chi_\om^{(0)}(\gamma,\as(k_0^2))
  \right]^2}{\big[ 1 - \partial_\om\chi_\om^{(0)}(\gamma) \big]^4} \\
 \label{OmLeading}
 &\simeq G^{(0)}(Y;t_0,t_0) \frac{Y^3}{24}
 \left[ \partial_{t_0}\oms^{(0)}(t_0) \right]^2 \asb \chi_\eff ''(\half) \;.
\end{align}
The last factor provides the leading diffusion exponent we were
looking for. Note that the Jacobian factor $J^3$ has been reabsorbed
in the $t_0$-derivative of $\oms$ and in the curvature of the
effective characteristic function:
$\chi_\eff ''(\half) = J \left.\chi_\om''(\half)\right|_{\om=\as\chi_\eff}$.
This particular form for the generalization of the LO $Y^3$ diffusion term is
quite natural when one considers the physical mechanism at play: diffusion
causes a symmetric spread over a logarithmic range of transverse scales
of order $\sqrt{\asb\chi _\eff '' Y}$. The exponent of the evolution at a
scale $t'$ is given by
$\om_s^{(0)}(t_0)+(t'-t_0)\partial_{t_0}\oms^{(0)}(t_0)$.  In a
first-order expansion of the evolution there is a cancellation between
components above and below $t_0$. But in a second order expansion of
the evolution, there are corrections from above and below $t_0$ that
enter with the same sign,
$\sim [\pm \sqrt{\asb\chi _\eff '' Y}\partial_{t_0}\oms^{(0)}(t_0)Y]^2$. This is
precisely the form of~\eqref{OmLeading}.

The analytical treatment given above has its counterpart in the
numerical extraction of the running-coupling diffusion coefficients
presented in~\cite{CCSS1}.  We illustrate here that the
method can also be applied to a more general case with an
$\om$-dependent resummed NLL BFKL kernel.

Formally we write the logarithm of the Green's function as a power
series in $b$:
\begin{equation}
  \label{eq:formalbexpGT}
  \ln G(Y;t;t_0) = \sum_{i=0} b^i \,[\ln G(Y;t;t_0)]_i\,,
\end{equation}
where the expansion is defined such that $\as(t_0)$ (or optionally
some other scale) is kept independent of $b$. We can then write the
effective exponent
\begin{equation}
  \om_{\eff}(Y;t_0) = \frac{d}{dY} \ln G(Y;t_0;t_0)\,,
\end{equation}
also as a series in $b$:
\begin{equation}
  \label{eq:omeffbexp}
  \om_{\eff}(Y;t_0) = \sum_{i=0} b^i \om_{\eff,i} = \sum_{i=0} b^i \frac{d}{dY}
  [\ln G(Y;t_0;t_0)]_i\,.
\end{equation}
In practice the power series is determined numerically by carrying out
the evolution with a generalized $b$-dependent coupling $\asb^{[b]}(k^2)$,
\begin{equation}
\asb^{[b]}(k^2) = \frac{\asb(k_0^2)}{1+(t-t_0)b\asb(k_0^2)}.
\end{equation}
using several values of $b = i\delta b$ (typically $\delta b = 0.01$
and $i$ ranges from $-3$ to $3$). In the formal limit of small $\delta
b$, the knowledge of $\ln G(Y;t_0;t_0)$ for $n$ values of $b$ allows
one to determine the power series up to order $b^{n-1}$.

\begin{figure}[t]
  %\vspace*{0.0cm}
  \centerline{ \epsfig{figure=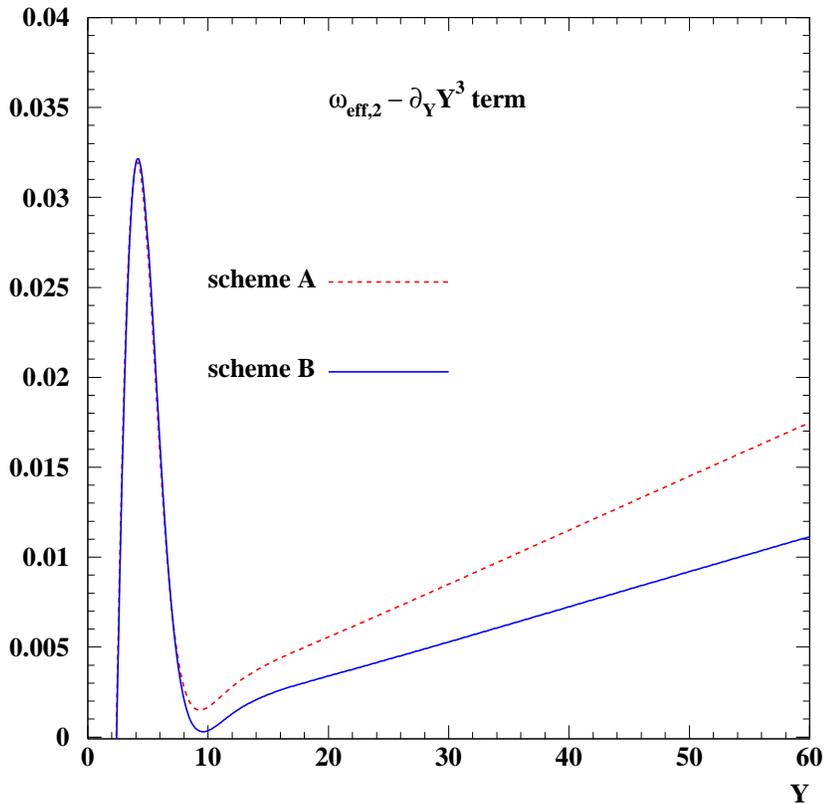,width=12cm} } \vspace*{0.5cm}
\caption{\it The difference between the $\om_{\eff,2}$ coefficient
  as from (\ref{eq:omeffbexp}) and the leading diffusion term
  calculated from equation (\ref{OmLeading}) for resummed kernel in
  schemes A (dashed line) and B(solid line).  The $b$-expansion has
  been performed around the fixed coupling value $\asb(k_0^2)=0.1$. }
\label{fig:bexp1}
\end{figure}

In Fig.~\ref{fig:bexp1} we test the analytical prediction for the
leading diffusion term $\sim Y^3$ as given by Eq.~(\ref{OmLeading}).
We show on this plot the term $\om_{\eff,2}$ from
expansion~(\ref{eq:omeffbexp}) with the subtracted $\partial_Y Y^3$ term
calculated for schemes A and B with scale $\asb(q^2)$ as a function of
rapidity $Y$.  We clearly see that after the subtraction there is only
a linear dependence left, which signals presence of the subleading
$\partial_Y Y^2$ terms 
The numerical value of the diffusion terms is much lower in the
resummed models than in the LL BFKL equation. For example the
coefficient of the leading $\sim Y^3$ term, see (\ref{OmLeading}) in
the LL BFKL case is about $8$ times larger than the one in the
resummed models. As a consequence the regime in which the
solution is perturbative is much broader in the case of the NLL BFKL.
One can see this by studying the contour plots in
Fig.~(\ref{fig:ValRegions}), as will be discussed in more detail in
the next section. In particular one finds,
Fig.~(\ref{fig:ValRegions}a), that the
region where the LL solution is insensitive to non-perturbative
results is much smaller than in Figs.~(\ref{fig:ValRegions}b,c,d) with
the resummed evolution. This result is quite encouraging as far as the
phenomenological predictions for high energy processes with two hard
scales are concerned.

In principle, one could extend our procedure to extract the $Y^2$
terms too, as has been done in ref.~\cite{CCSS1} for the case of the
LL BFKL with running coupling.  However, the analytical calculation
here would be quite involved, since these terms originate from a
number of different sources, i.e. they come both from (\ref{valueG1})
and (\ref{G2integral}), and moreover they mix with the terms coming
from the normalization.  In practice these $Y^2$ terms are expected to
be rather small and not as relevant for phenomenology as the leading
$Y^3$ terms.

\begin{figure}[t]
  %\vspace*{0.0cm}
        \centerline{ \epsfig{figure=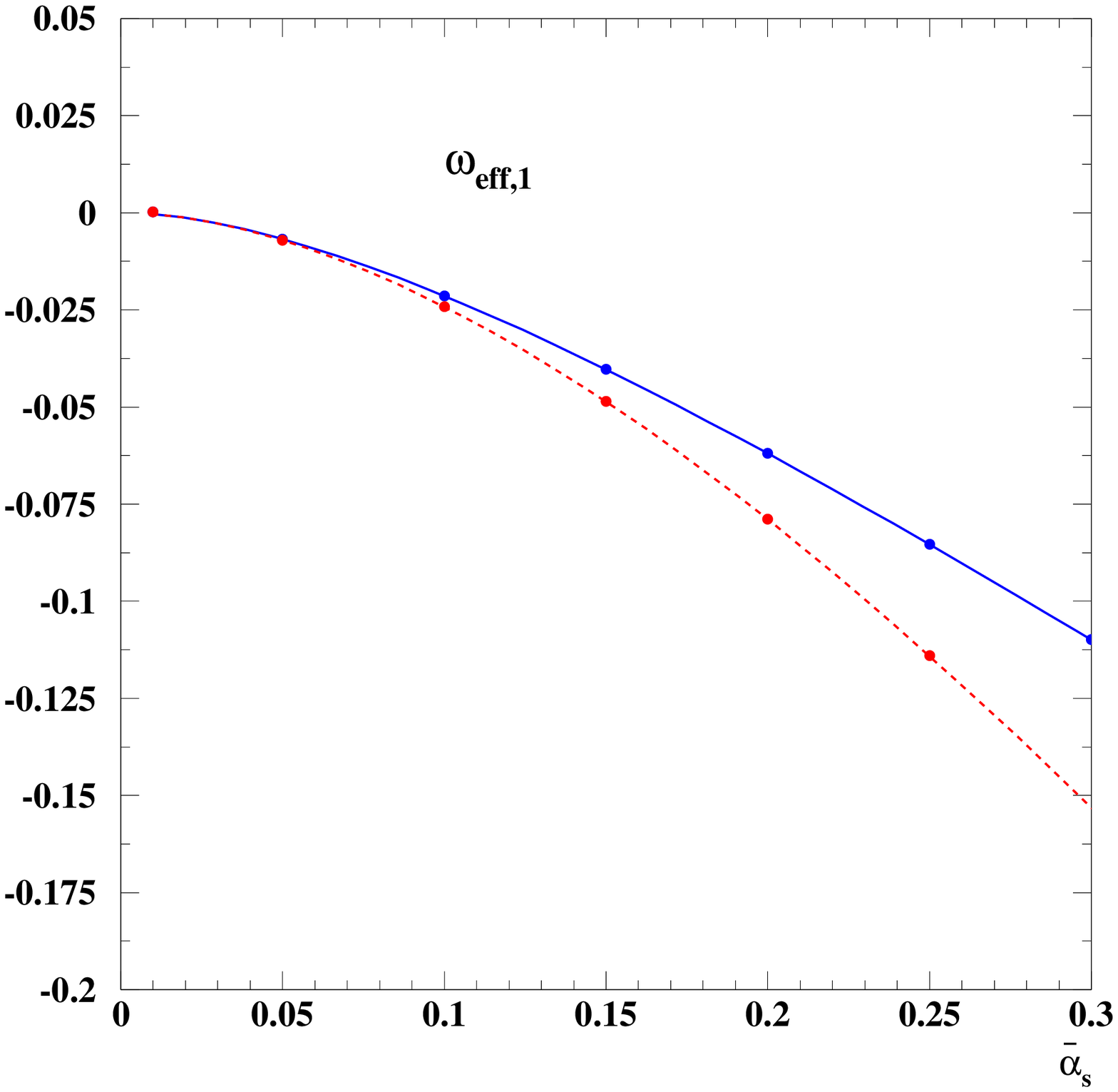,width=12cm}
           }
\vspace*{0.5cm}
\caption{\it $\om_{{\rm eff},1}$ from Eq.~(\ref{eq:omeffbexp}) as a function of
  the coupling $\asb$ calculated in schemes A(dashed) and B(solid).
  Lines represent analytical evaluation based on Eq.~(\ref{DeltaOms}),
  the points correspond to the numerical extraction. }
\label{fig:bexp2}
\end{figure}

We restrict therefore ourselves to showing only the ${\cal O}(b)$
shift to $\oms$ given by the analytical expression Eq.~(\ref{DeltaOms})
and compared with the numerical calculation, see Fig.~\ref{fig:bexp2}.
There is clearly a perfect agreement between the two methods,
exhibiting the leading $\alpha_0^2$ behavior of $\oms^{(1)}$.

\begin{figure}[t]
  %\vspace*{0.0cm}
  \centerline{ \epsfig{figure=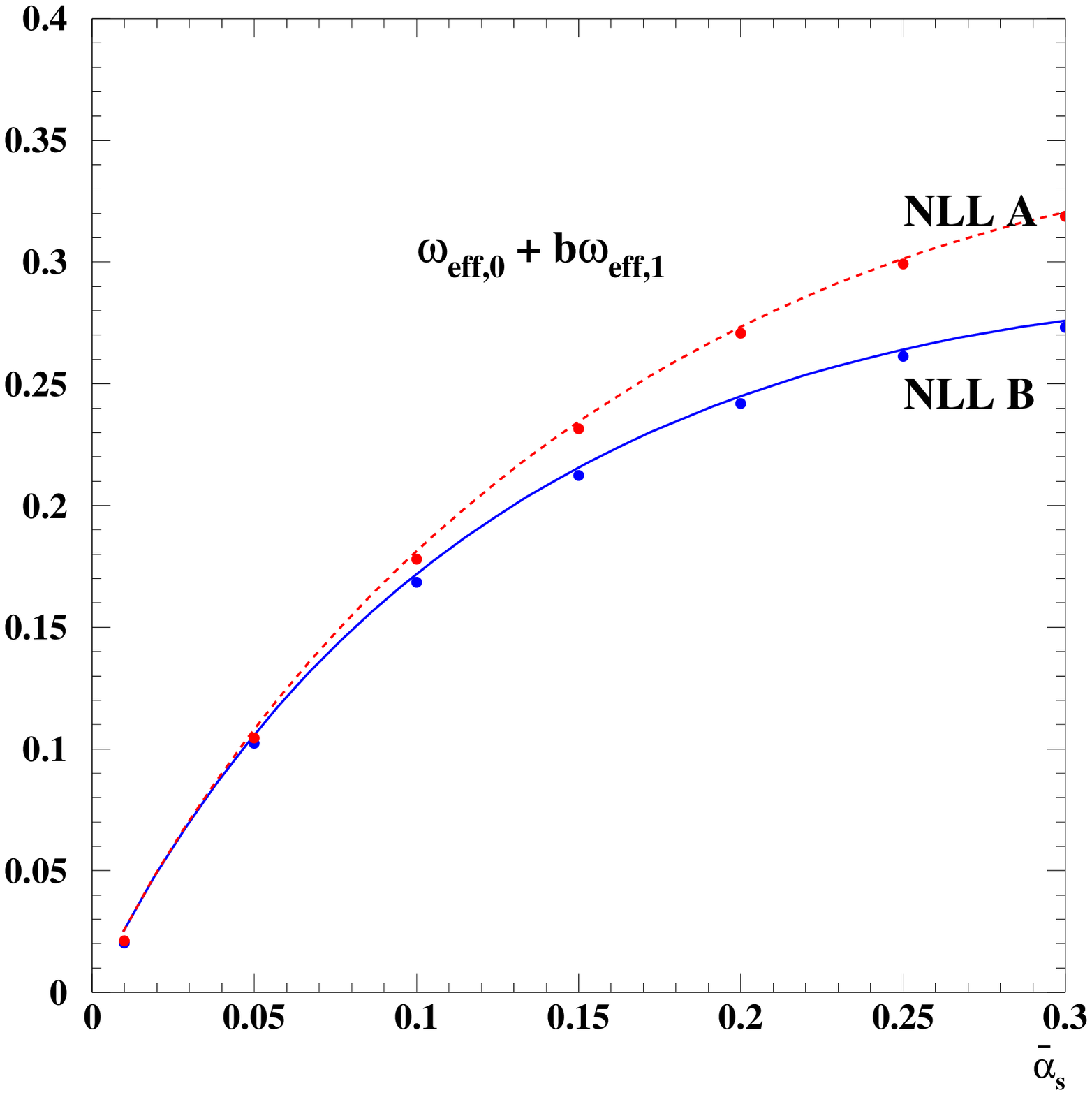,width=12cm} } \vspace*{0.5cm}
\caption{\it Sum of the first two terms  from Eq.~(\ref{eq:omeffbexp})
  as a function of the coupling $\asb$ calculated in schemes A(dashed)
  and B(solid).  }
\label{fig:bexp3}
\end{figure}

Finally, we show in Fig.~\ref{fig:bexp3} our numerical evaluation of
the sum of the first two terms of $\om_{\rm eff}$,
Eq.(\ref{eq:omeffbexp}), that is $\om^{(0)}+b\om^{(1)}$, as a function
of the coupling constant $\asb$. The correction due to the running of
the coupling reduces somewhat the value of the intercept, as
compared with the fixed coupling case ($b=0$), which is shown in
Fig.~\ref{f:oms}.
The plot in Fig.~\ref{fig:bexp3} summarizes our present understanding
of $\oms$, because the higher order terms $\sim b^2 \as^3, \cdots$ are
beyond our present level of accuracy, and are perhaps not really
meaningful, given the complex $Y$-dependence of (\ref{d:oms1}).

Note that we do not compare directly with our earlier results for
$\oms$~\cite{CCS1}, because they are based on a different definition
(the saddle-point of an effective characteristic function), which is
less directly related to the Green's function. Nevertheless, the
present results are consistent with previous ones to within NNLL
uncertainties.

%----------------------------------------------------------------------
\subsection{NP uncertainties on Green's function}
\label{sec:NPGreen}

It is well appreciated nowadays that, even with two hard scales, the
ultra-high energy behavior of the BFKL Green's function is entirely
determined by non-perturbative physics. It is only in an intermediate
high-energy regime that one is able to make reliable perturbative
predictions~\cite{JKCOL,Lipatov86,CC97,CCSS2}.

Traditionally one estimates non-perturbative uncertainties on BFKL evolution by
examining the sensitivity to variations of the infrared regularization of the
coupling. More recently we showed that a purely perturbative answer can be
defined in the context of the $b$-expansion \cite{CCSS1}, with the highest
perturbatively accessible rapidity being determined by the breakdown of
convergence of this expansion. In this section we shall examine both %of these
approaches.

Let us consider a variety of infrared (IR) regularizations of the
coupling. Mostly we shall use cutoff regularizations,
\begin{equation}
  \asb(q^2) \equiv \asb^{\mathrm{PT}}(q^2) \Theta(q - \kbar)\,,
\end{equation}
with three different values of $\kbar$. It will also be instructive to examine
a `freezing' regularization,
\begin{equation}
  \asb(q^2) \equiv \asb^{\mathrm{PT}}(\max(q^2,\kbar^2)) \,.
\end{equation}
We believe this freezing regularization to be somewhat less physical, since
it allows diffusion to arbitrarily low scales in the infrared, in
contradiction with confinement.  However
for the purposes of our general discussion it will be helpful to have it too at
our disposal.

In all cases $\asb^{\mathrm{PT}}$ is the perturbative one-loop coupling with
$\nf=4$, chosen such that $\asb(9\GeV^2) = 0.244$,
and no cutoff is placed on exchanged gluon virtualities. The complete set of IR
regularizations is summarized in table~\ref{tab:omp}, together with the resulting
Pomeron properties, both for LL and resummation scheme B (\NLLB) evolution.

\begin{table}[htbp]
  \centering
  \begin{tabular}{|c|c|c|c|c|} \hline
    $\kbar$ (GeV)   & $\asb(\kbar^2)$  & asymptotic growth & $\omp$ (LL)& $\omp$
    (\NLLB)\\ \hline
    1.00 (cutoff) &   0.39      & $\exp(\omp Y)$        & 0.44 & 0.32 \\ \hline
    0.74 (cutoff) &   0.46      & $\exp(\omp Y)$        & 0.49 & 0.35 \\ \hline
    0.50 (cutoff) &   0.62      & $\exp(\omp Y)$        & 0.58 & 0.41 \\ \hline\hline
    0.74 (frozen) &   0.46      & $Y^{-3/2}\exp(\omp Y)$& 1.28 & 0.46 \\ \hline
  \end{tabular}
  \caption{\it Our set of infrared regularizations of the coupling,
    together with the resulting asymptotic `Pomeron' behavior and
    $\omp$ values for LL with running coupling $\as(q)$ and \NLLB evolution.}
  \label{tab:omp}
\end{table}

The two main Pomeron features that one may wish to study are its
analytical structure and the power, $\omp$ of asymptotic growth, both
shown in table~\ref{tab:omp}. It is well known that with a cutoff one
expects the Pomeron to be a pole, while for a frozen coupling one
expects a branch cut, giving a $Y^{-3/2}\exp(\omp Y)$ growth. Though
these properties are most easily derived for LL BFKL and a coupling
that runs as $\as(k^2)$, they apply quite generally.

As regards the $\omp$ values, a first point to note concerns the results for
LL evolution, which with cutoffs on $\asb$, are much smaller than the
naive expectation of $\asb(\kbar^2)\chi_0(1/2)$ --- the difference stems
from large $\asb^{5/3}$ (and higher) contributions, originally noticed
by Hancock and Ross \cite{hr92} (discussed also in \cite{FRBook}).

For \NLLB evolution the difference between the cutoff and frozen
coupling evolutions is less dramatic because of the smaller value of
the `raw' $\omega_s$ value (Figs.~\ref{f:oms} and \ref{fig:bexp3}). As
a result the uncertainty on the properties of the `Pomeron' are
somewhat reduced.  It is interesting to note that these values for the
Pomeron intercept are not too different from those found for the hard
Pomeron in `two-Pomeron' fits to data in \cite{DL2pom}. It is not
clear however to what extent this can be considered significant, since
on one hand non-perturbative aspects of small-$x$ evolution are likely
to be extensively modified by the true non-perturbative physics,
including saturation effects; and on the other hand because the
two-Pomeron fits involve rather strong simplifying assumptions.

\begin{figure}[tbp]
  \centering
  \includegraphics[width=0.49\textwidth]{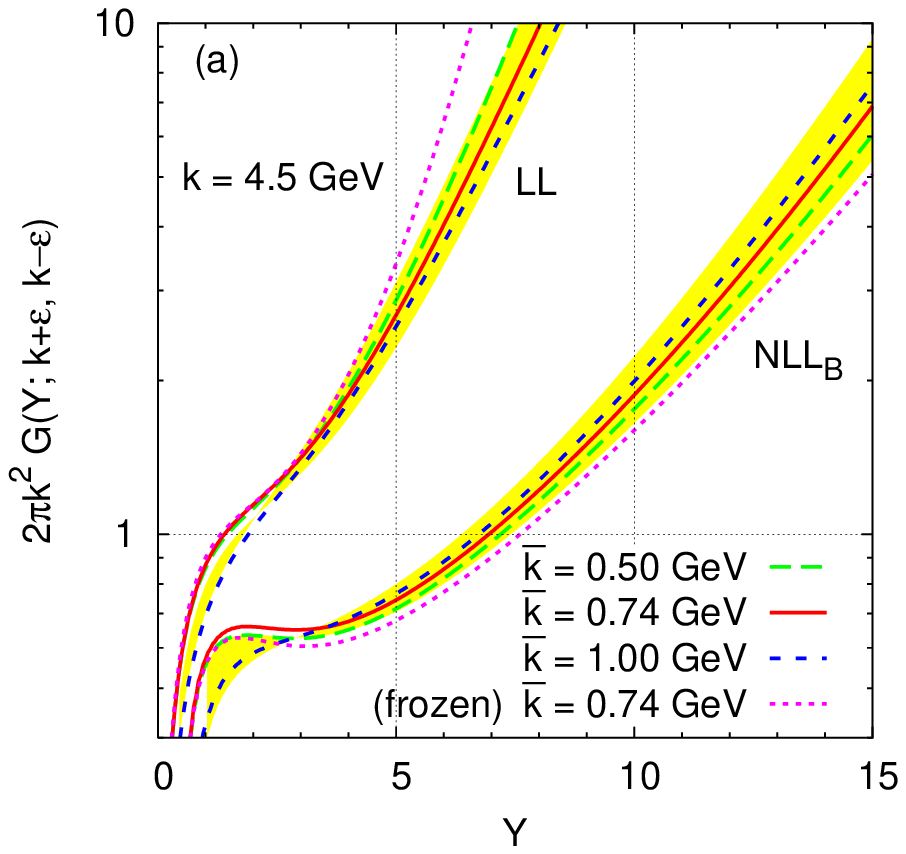}
  \hfill
  \includegraphics[width=0.49\textwidth]{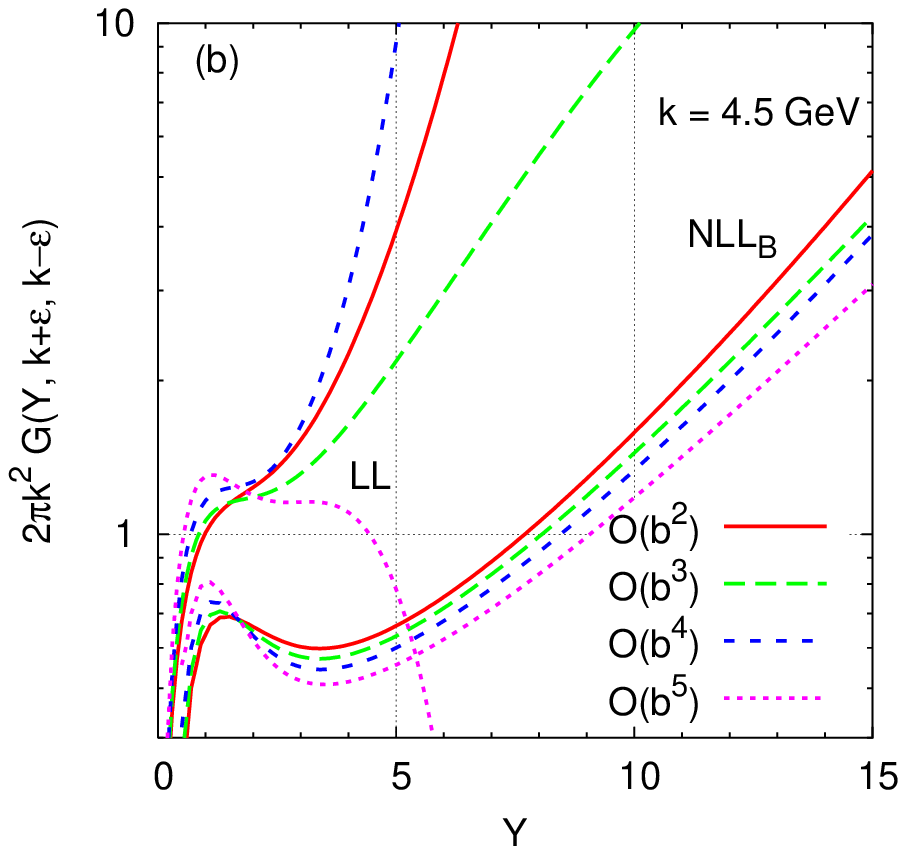}\medskip\\
  \includegraphics[width=0.49\textwidth]{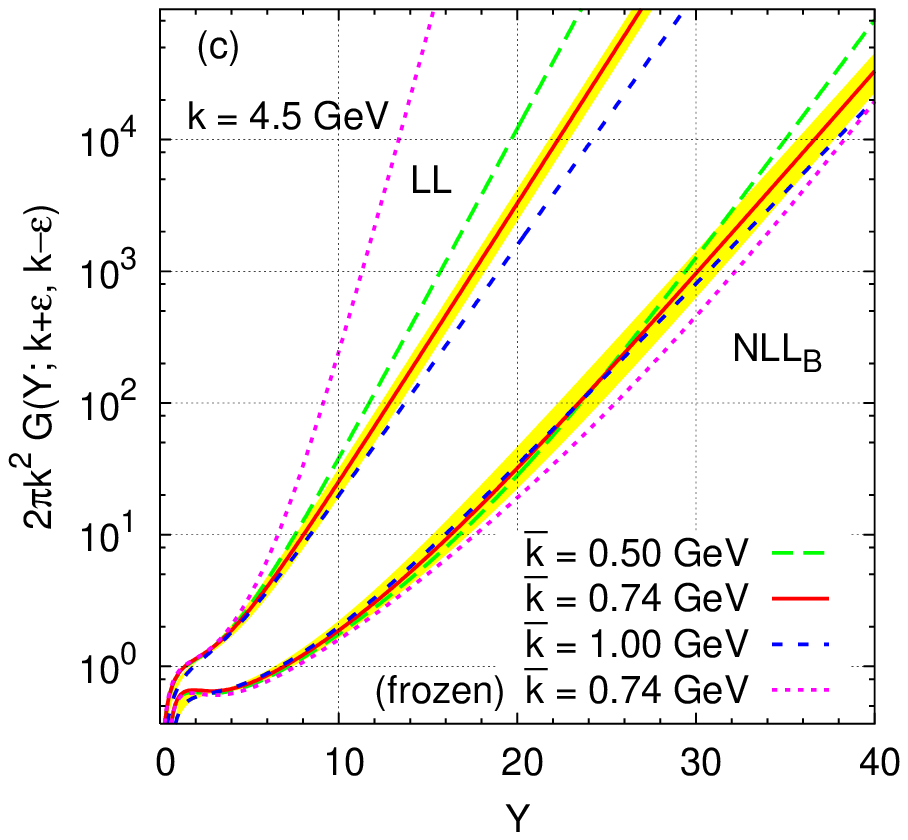}
  \hfill
  \includegraphics[width=0.49\textwidth]{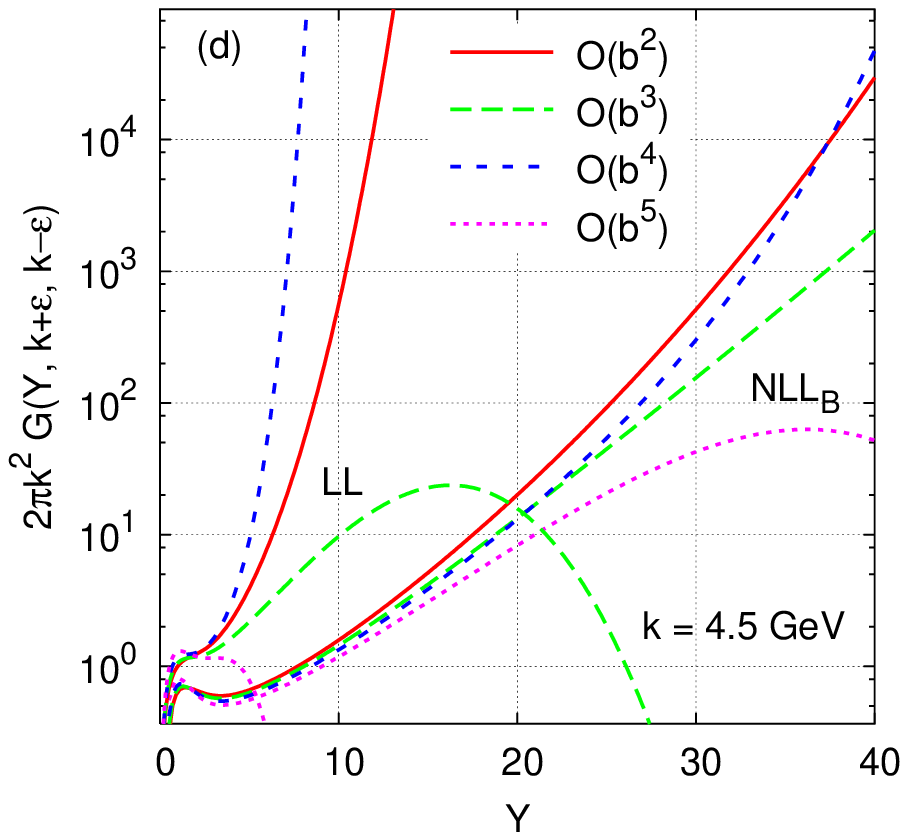}
  \caption{\it (a) Green's function calculated with four different
    infrared regularizations of the coupling, with a
    renormalisation-scale band ($1/2<x_\mu^2<2$) included for
    reference for the $\kbar = 0.74\GeV$ curve; (b) Green's function
    calculated in the $b$-expansion, up to and including second,
    third, fourth and fifth orders in $b$; (c) and (d) are the same as
    (a) and (b) respectively but on a different scale. In all cases
    $\epsilon \simeq 0.1k$.}
  \label{fig:manyRegBeta}
\end{figure}

Having examined the asymptotic properties of the various infrared
regularizations, we can now move on to examine the IR sensitivity of
`perturbative' Green's functions. The left hand plots of
Fig.~\ref{fig:manyRegBeta} ((a) and (c) simply have different rapidity
ranges) show $G(Y,k-\epsilon,k+\epsilon)$ for the four infrared
coupling regularizations of Tab.~\ref{tab:omp}. The transverse
momentum $k=4.5\GeV$ is chosen lower than in the plots of
section~\ref{sec:basicGreenFeatures} in order enhance the sensitivity to the IR
region. For reference we also include the uncertainty band due to
renormalisation scale uncertainty. The discussion that follows will
concentrate on the \NLLB results, however all the plots of
Fig.~\ref{fig:manyRegBeta} include also LL results, so as to
illustrate the dramatically different IR sensitivity between LL and
\NLLB evolution.

So let us first consider the three cutoff regularizations (\NLLB). 
One sees that up to $Y\simeq30$ they give
very similar results. Beyond this point, tunneling occurs (for the
lowest cutoff), and the three curves start to diverge, indicating that
according to this prescription the Green's function is no longer under
perturbative control.

When instead one examines the curve with an infrared-frozen coupling,
one finds a result that at first sight appears paradoxical: the Green's
function is somewhat lower than with a cutoff regularization,
over a wide range of $Y$ in which the cutoff regularization looks relatively
insensitive to NP effects. Naively one might have expected to see
little difference until the tunneling point. Our understanding of the
observed behavior is that it is connected with the use of $\asb(q^2)$
in Eq.~\eqref{eq:LOBFKLkc}, which causes the regularization of the
coupling to affect, among other things, the virtual corrections of the
BFKL equation. Having a larger infrared coupling increases the size of
the (negative) virtual corrections. In situations where the Green's
function has a substantially negative second derivative (as it does
over a wide range of $Y$) there is an incomplete cancellation with the
real contributions (of order $1/Q^2$), which means that a larger
infrared coupling leads to \emph{smaller} preasymptotic growth of the
Green's function.\footnote{One cross-check of this understanding comes
  from the fact that when evolving with a scale $\asb(k^2)$ in the
  kernel, differences between cutoff and freezing IR regularizations
  appear only in the asymptotic Y dependence.} This also explains why
the curves with a cutoff IR coupling initially evolve more slowly for
smaller values of $\kbar$.

One could also have imagined more sophisticated IR regularization
schemes. For example, while maintaining an infrared-frozen coupling,
one could have placed an IR cutoff on the exchanged transverse momentum
$k$. We expect that this would give curves whose initial evolution is
very similar to that of the IR-frozen coupling case, but whose
asymptotic NP behavior is a pole, as in the cases with a cutoff on
the coupling.

This confusion arising from this wide range of regularization options
was in part the motivation for introducing the $b$-expansion in
\cite{CCSS1}. The $b$-expansion allows one to define a perturbative
prediction in close analogy with the prescription that is implicitly
contained in standard fixed-order perturbative predictions. There, one
never has to specify any IR regularization. Rather, momentum integrals
are implicitly carried out over a perturbative fixed-order expansion
of the coupling, which is well behaved down to zero momentum.
Sensitivity to non-perturbative effects then manifests itself through
the appearance of renormalons (see for example the review by Beneke
\cite{renormalon}), \ie factorially divergent coefficients in the
series expansion for one's observable.

In a small-$x$ resummation, a pure fixed-order expansion would defeat
the purpose of the resummation in the first place. However it was
shown in \cite{CCSS1} that one can expand $\ln G$ in powers of the
$\beta$-function coefficient $b$, and that a truncation of the
resulting series maintains the advantages of small-$x$ resummation,
while providing a prescription for defining purely `perturbative'
predictions. This is in addition to its usefulness for studying
analytical properties of the running-coupling dependence of the
Green's function, as has already been exploited in Sec.~\ref{s:eidc}.

Figs.~\ref{fig:manyRegBeta}b and \ref{fig:manyRegBeta}d show the same
Green's function as in Figs.~\ref{fig:manyRegBeta}a and
\ref{fig:manyRegBeta}c, but in truncations of the $b$-expansion
ranging from orders $b^2$ to $b^5$. One sees how all different
truncations give fairly similar answers at low $Y$. But at large $Y$,
the presence of the terms in $\ln G$ involving additional factors of
$b^2 \as^4 Y^2$ leads to the splaying out of the different
truncations, signaling the fundamental limit of the $b$-expansion.
In certain models (\eg \cite{CMT}) this is associated with the
appearance of non-analyticity in $b$. It is to be noted that this
large-$Y$ breakdown of the $b$-expansion is not of the renormalon type
that is expected in normal perturbative series.

\begin{figure}[tbp]
  \centering
  \includegraphics[width=0.48\textwidth]{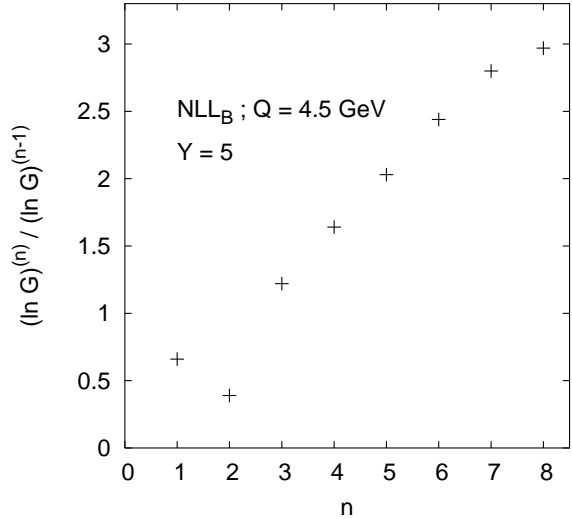}
  \caption{\it Ratios of successive coefficients of $b^n$ in the
    $b$-expansion of $\ln G(Y,k-\epsilon, k+\epsilon)$ for $Y=5$.}
  \label{fig:renormalon}
\end{figure}
A detailed study of the figure also reveals that even at low $Y$ the
expansion is not entirely well-behaved. Indeed successive coefficients
of the $b$-expansion are all of the same sign and grow quite rapidly,
in a way that \emph{is} suggestive of an infrared renormalon. Infrared
renormalons are a factorially divergent behavior of the perturbative
series whereby the $n^\mathrm{th} \gg 1$ order term is proportional to
$(\asb b /p)^n n!$ (in simple cases). When interpreted in the language
of asymptotic series, this translates to an uncertainty on the sum of
the perturbative series of order $(\Lambda^2/Q^2)^p$.

To establish whether it is renormalon behavior that we are seeing, in
Fig.~\ref{fig:renormalon} we show ratios of successive coefficients of
$b^n$ in the expansion of $\ln G$. The fact that, over a significant
range of $n$, one
sees a large-$n$ behavior consistent%
\footnote{Except for the last point --- indeed while it is the largest
  values of $n$ that are the hardest to determine accurately with our
  numerical methods, we have not been able to determine with certainty
  that the value obtained for $n=8$ is truly unreliable. Accordingly
  we have chosen
  to show the point despite our limited confidence in it.} %
with $(\ln G)^{(n)}/(\ln G)^{(n-1)}\simeq c n$, implies that it is
renormalon behavior.  Furthermore by examining a second value of $Q$
one can establish that $c$ itself is roughly proportional to $\asb$,
$c\simeq 1.9\asb$.
However the constant of proportionality, corresponding to a value of
$p = \frac{asb}{c}\simeq 0.53$, is somewhat surprising, because it
implies power 
corrections of order $(\Lambda/Q)^{2p}$, \ie roughly $\Lambda/Q$.
Naively one would have expected $p=1$ (see also~\cite{Hautmann}). This
difference has yet to be understood, though it should be kept mind
that significant enhancements of naively expected power-suppressed
effects are known to be possible due to certain classes of resummation
effects~\cite{DMWEEC}. It is interesting additionally to note that the
formally higher-twist non-perturbative effects that we expect for
splitting functions in Sec.~\ref{s:resanomdim} will also turn out to
scale roughly as $\Lambda/Q$ rather than $\Lambda^2/Q^2$.

Regardless of the precise reason for the unexpected scaling, it can be
quite straightforwardly established that the renormalon behavior is
directly connected with the use of $\asb(q^2)$ in the LL part of the
kernel; \ie it has the same origin as the preasymptotic effects that
arise when modifying the IR regularization of the coupling,
Fig.~\ref{fig:manyRegBeta}a.

These preasymptotic effects are a feature of BFKL evolution that to
the best of our knowledge have not been observed before. Given that
they are strictly connected to the use of $\asb(q^2)$, they are somewhat
model dependent. However the motivations for using $\asb(q^2)$ are quite
strong. In particular, as we have mentioned above, this is the scale
that is explicitly suggested by the form of the NLO corrections;
furthermore the appearance of the transverse momentum of the emitted
gluon as the scale of the coupling is a phenomenon that is
well-motivated in many other contexts of QCD \cite{ktscale}.

\begin{figure}[tbp]
  \newcommand{\figfrac}{0.48}
  \centering
  \includegraphics[height=\figfrac\textwidth]{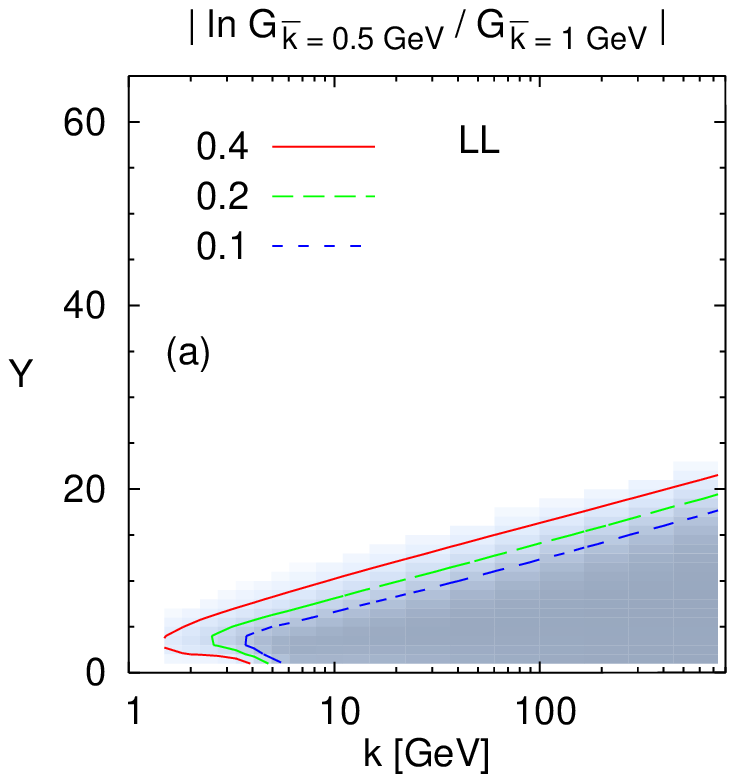}
  \hfill
  \includegraphics[height=\figfrac\textwidth]{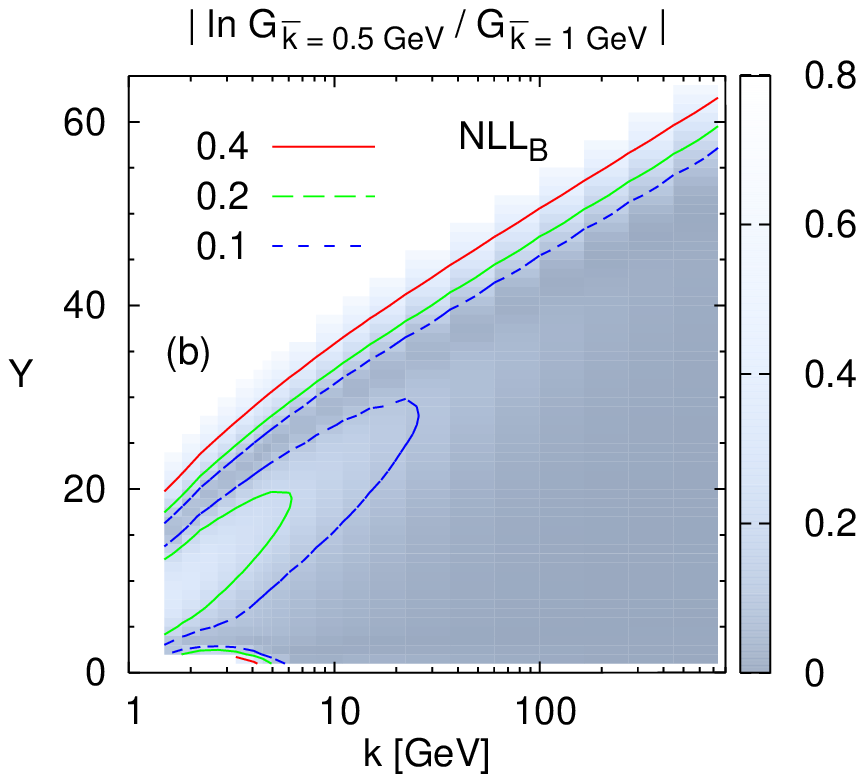}\bigskip\\
  \includegraphics[height=\figfrac\textwidth]{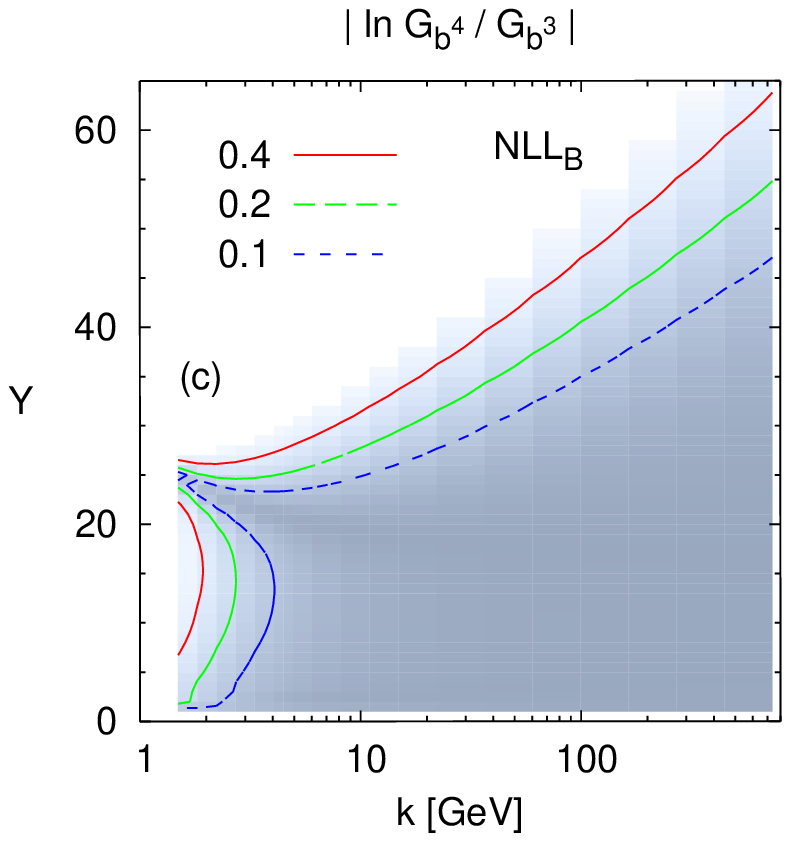}
  \hfill
  \includegraphics[height=\figfrac\textwidth]{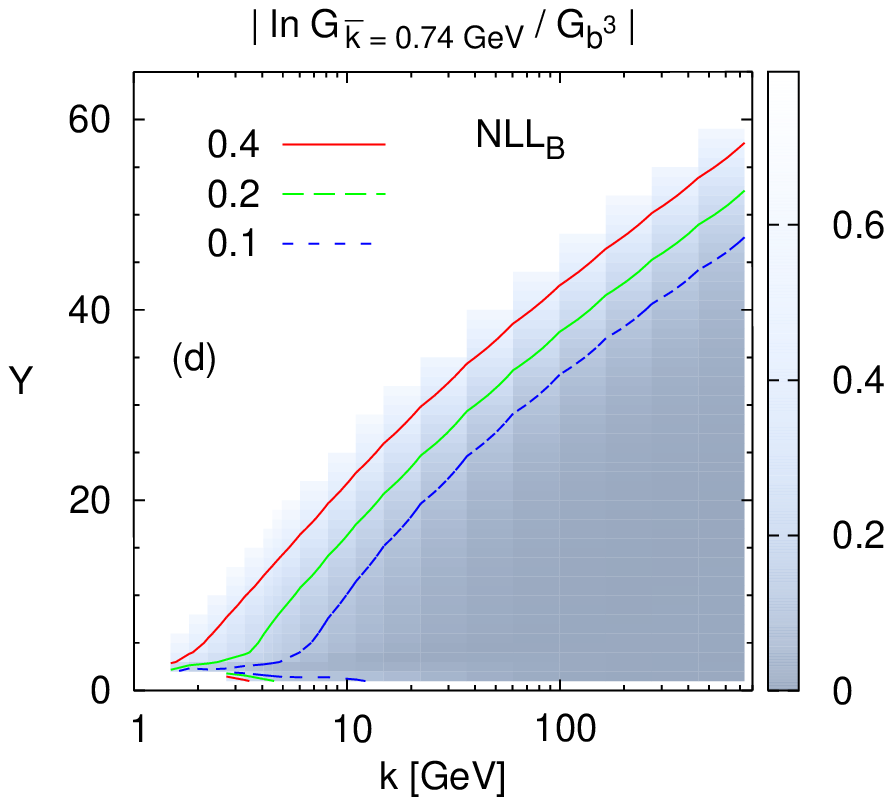}
  \caption{\it Contour plots showing the sensitivity of
    $G(Y,k-\epsilon, k+\epsilon)$ to the choice of non-perturbative
    regularization, obtained by examining the absolute value of the
    logarithm of the ratio of pairs of regularizations. Darker shades
    indicate insensitivity to the NP regularization, and contours have
    been drawn where the logarithm of the ratio is equal to $0.1$,
    $0.2$ and $0.4$. Plot (a) shows the result for LL evolution (with
    $\as(q)$) and two cutoff regularizations ($\kbar = 0.5\GeV$ and
    $\kbar= 1.0\GeV$); (b) shows \NLLB evolution with the same pair of
    cutoffs; (c) shows \NLLB evolution with truncations of the
    $b$-expansion at orders $b^3$ and $b^4$; and (d) shows \NLLB
    evolution, comparing a cutoff regularization ($\kbar = 0.74\GeV$)
    with a $b$-expansion truncation (at order $b^3$).}
  \label{fig:ValRegions}
\end{figure}

The appearance of significant preasymptotic NP effects complicates
somewhat any attempt to give a compact summary of NP limits in BFKL
evolution. In their absence one might have parameterised NP effects at
a given transverse scale $k$, by the rapidity at which one loses
predictability for the Green's function (\eg \cite{KOVMUELLER,CCSS1,CCSS2}).
Instead, we examine contour plots, Fig.~\ref{fig:ValRegions}, of
\begin{equation}
  \left|\ln
    \frac{G_{a}(Y;k-\epsilon,k+\epsilon)}{G_{b}(Y;k-\epsilon,k+\epsilon)}
  \right|,
\end{equation}
where the subscripts $a$ and $b$ indicate the different
non-perturbative treatments in the two evaluations of the Green's
function. Darker shades indicate good agreement between the two
evaluations, while lighter shades indicate disagreement. Additionally,
to guide the eye, we have added explicit contours where the (absolute
value of the) log of the ratio is equal to $0.1$, $0.2$ and $0.4$,
which for brevity we shall refer to as the $10\%$, $20\%$ and $40\%$
contours respectively.

The first plot, Fig.~\ref{fig:ValRegions}a, given for reference, shows
results for LL evolution with two different IR cutoffs on the coupling
($0.5\GeV$ and $1\GeV$). Preasymptotic effects are fairly irrelevant
here, in part because the asymptotic NP contributions set in quite
quickly. The contours indicate a linear relation between the maximum
perturbatively accessible $Y$ value, $Y_{\max}$, and $\ln k$, as would
be expected if this limit is due to tunneling in the Green's function
with the lower cutoff. From the simplified version of the tunneling
formula \cite{CCS2,CCSS2},
\begin{equation}
  \label{eq:YtunnelRecall}
  Y_\mathrm{tunnel}(k^2) \simeq \frac{\ln k^2/\kbar^2}{\omp - \om_s(k^2)}\,,
\end{equation}
we expect that for asymptotically large $k$, we should see
$dY_{\max}/d\ln k \simeq 2/\omp \simeq 3.45$. In practice, the slope
that is measured (for $k$ between $10^3$ and $10^4\GeV$) is about 2.7;
given that the measurement region is not truly asymptotic,
the $20\%$ disagreement between the two numbers is not unreasonable.

Fig.~\ref{fig:ValRegions}b uses the same pair of NP regularizations,
but with \NLLB evolution. A first striking difference is the
significant region (lower left-hand quadrant) in which there are
preasymptotic NP effects at the $20\%$ level. This is connected with
the preasymptotic effects (due to $\asb(q^2)$) mentioned earlier in this
section. The second important observation is that the rapidity where
asymptotic non-perturbative effects become important, $Y_{\max}$, is
significantly larger than for LL. But as before it is roughly
consistent with a manifestation of tunneling in the Green's function
with the lower cutoff:\footnote{The linear dependence of $Y_{\max}$ on
  $\ln k$ only becomes convincingly evident at very large $k$; we have
  limited the scale to only moderately large $k$ in order to maintain
  the visibility of the phenomenologically relevant region of $k$.}
this time the tunneling formula differs slightly from that in
\cite{CCS2,CCSS2}, because of the presence of the kinematical
constraint in the evolution, giving
\begin{equation}
  \label{eq:YtunnelKC}
  Y_\mathrm{tunnel,\,k.c.}(k^2) \simeq \frac{(1+\omp)\ln k^2/\kbar^2}{\omp -
    \om_s(k^2)}\,.
\end{equation}
At very large $k$ one would therefore expect a slope $dY_{\max}/\ln k
\simeq 2(1 + 1/\omp) \simeq 6.9$. The measured slope (same $k$ range
as above) is roughly $6.1$.  As for LL evolution, these two results
are not perfectly compatible, but given that the $k$-region is not
formally asymptotic, the disagreement is not unreasonable.

As is discussed above, using different infrared regularizations is
not the only way of gauging non-perturbative effects.
Fig.~\ref{fig:ValRegions}c shows what happens if instead we consider
two truncations of the $b$-expansion, at orders $b^3$ and $b^4$. Once
again, for smaller values of $k$ there are significant preasymptotic
NP effects, though the range of $k$ for which they matter is more
limited. The upper (`asymptotic') limit on $Y$ due to NP uncertainties
also behaves differently with the $b$-expansion. As was shown in
\cite{CCSS1}, the $b$-expansion allows one to reach rapidities of the
order of the fundamental perturbative limit
\cite{KOVMUELLER,ABB,LEVIN,CMT}, $Y_{\max} \sim \ln^2 k^2/\Lambda^2$.
This different parametric behavior of $Y_{\max}$, though not directly
relevant for phenomenological parameter ranges, is evident from the
large-$k$ curvature of the contours, and becomes even more so when
going to yet larger $k$.

The plots so far have shown comparisons of pairs of IR
regularizations, or pairs of $b$-expansion truncations. However if we
look once again at Fig.~\ref{fig:manyRegBeta}, we see that the largest
preasymptotic `NP' differences are to be seen when comparing an IR
cutoff with the $b$-expansion. Accordingly in
Fig.~\ref{fig:ValRegions}d, we show contours for the ratio of Green's
functions where one is evolved with a central IR cutoff ($\kbar =
0.74\GeV$) and the other is determined by a $b^3$ truncation of the
$b$-expansion.  This is to be considered as a conservative estimate of
the impact of non-perturbative effects.

In this comparison, preasymptotic NP effects are so important at lower
$k$ values (below a few GeV), that one loses the ability to
distinguish them clearly from asymptotic NP effects associated with
tunneling or diffusion. Only for $k \gtrsim 6\GeV$ is one able to
calculate the Green's function over a reasonable range of rapidity (at
least up to $Y=10$) with better than $20\%$ accuracy. One comes to a
similar conclusion if one compares the cutoff and frozen IR coupling
regularizations, as was illustrated in \cite{CCSSletter}.

%%%%%%%%%%%%%%%%%%%%%%%%%%%%%%%%%%%%%%%%%%%%%%%%%%%%%
\section{Resummed anomalous dimension and splitting function
\label{s:resanomdim}}

So far, we have investigated the gluon Green's function in the hard
Pomeron regime, in which the hard scales $k^2, k_0^2$ are of the same
order, and --- by the $b$-expansion method --- we have isolated
diffusion and running coupling effects from the non-perturbative
Pomeron behavior.  In the complementary regime $k^2\gg k_0^2$ (or
$k_0^2\gg k^2$), the collinear properties become dominant, and the
Green's function is characterized by scaling violations and by the
corresponding anomalous dimensions.  The relation to non-perturbative
physics changes also, because of the validity of the RG factorization
property.  By arguments based on the double
$\gamma$-representation~\cite{JKCOL,ABF2001,CC1} or on truncated
models~\cite{Lipatov86,CCS2,CCS3} we can state that, for $t\gg t_0$,
\begin{equation}
\CGO(k,k_0) = \CFO(k) \tilde{\CF}_{\om}(k_0) + \rm higher \; twists \;,
\label{eq:fact}
\end{equation}
where $\CFO$ ($\tilde{\CF}_{\om}$) is a solution of the homogeneous
equation (\ref{eq:homo}) which is regular for $t\rightarrow \infty$
($t_0\rightarrow -\infty$).  While the $t$-dependence, because of its
boundary conditions, is expected to be perturbatively calculable, the
$t_0$-dependence is sensitive to the strong-coupling region and to
non-perturbative physics, but is factorized so that the standard
approach of DGLAP evolution~\cite{DGLAP} can apply. We are thus
entitled to define
\begin{equation}
 \gres(\om,t) = {\CFO(t) \over g_{\om}(t)} \;, \qquad
 g_{\om}(t) = \int\frac{d\gamma}{2\pi i \gamma} \;
 e^{\gamma t} f_{\om}(\gamma) \;,
\label{eq:resanomdim}
\end{equation}
where $f_{\om}(\gamma)$ represents $\CFO$ in $\gamma$-space.

%----------------------------------------------------------------------
\subsection{Resummation by $\boldsymbol{\om}$-expansion\label{s:roe}}

The analytical form of the resummed eigenfunction $f_{\om}$ was found
in~\cite{CCS1} on the basis of the $\om$-expansion --- summarized in
Sec.~\ref{s:omegaexp} --- which provides the solution
\begin{equation}\label{omexpSolution}
 f_{\om}(\gamma) = \exp\left( -\frac1{b\om} X_{\om}(\gamma) \right) \;,
 \qquad X'_{\om}(\gamma) \equiv \partial_{\gamma} X_{\om}(\gamma)
 \equiv \chi_{\om}(\gamma) \;, 
\end{equation}
in terms of the eigenvalue function $\chi_{\om}(\gamma)$ in
Eqs.~(\ref{eq:nonlinear}) and (\ref{eq:omexp}). Furthermore, in the
``semiclassical'' regime when $bt > 1 / \om \gg 1$, the behavior of
$\CFO(t)$ can be found from the saddle point estimate
\begin{equation}
\label{eq:sadpointestimate}
 b \om t = \chi_{\om}(\bar{\gamma}_{\om}(t))
 =X'_{\om}(\bar{\gamma}_{\om}(t)) \;,
\end{equation}
and the solution is then given by
\begin{equation}
\label{eq:solsad}
\CF_\om(t)  = k^2 \CFO(k) \sim
{1 \over \sqrt{-2 \pi \chi'_{\om}(\bar{\gamma}_{\om}(t))} }
\exp\left[ \int^t d \tau\; \bar{\gamma}_{\om}(\tau) \right] \;,
\end{equation}
where the function $\bar{\gamma}_{\om}(t)$ satisfies the following identity
\begin{equation}
\bar{\gamma}_{\om}(t) t -{1\over b\om} X_{\om}(\bar{\gamma})
= \int^t \bar{\gamma}_{\om}(\tau)d \tau \;.
\end{equation}
The corresponding gluon anomalous dimension is given by~\cite{CC}
\begin{equation}
\gres(\om,t) =
 \bar{\gamma}_{\om}(t) - \frac{b \om}{\chi_{\om}'(\bar{\gamma})}
 \left[ \frac{1}{\bar{\gamma}}
 + \frac{1}{2}\frac{\chi_{\om}''(\bar{\gamma}) }{\chi_{\om}'(\bar{\gamma})}
 + \dots \right ]\;.
\label{eq:sadanom}
\end{equation}

Recall, however, that Eq.~(\ref{eq:sadanom}) is an acceptable approximation only
away from the turning point
\begin{equation}
\label{eq:turning}
 \chi_{\omsb}' \big( \bar\gamma(\omsb,t) \big) = 0\;,
\end{equation}
which is a singularity of~(\ref{eq:sadanom}) with infinite fluctuations, and
defines the exponent $\om = \omsb(t)$ at anomalous dimension level.

Therefore, when $\om$ approaches $\omsb(t)$, one can only rely on the
$\gamma$-representation~(\ref{eq:resanomdim}) in order to define the anomalous
dimension past the turning point. This was the method followed in~\cite{CCS1}
(with the choice of scheme in Eq.~(\ref{eq:chiomnll})) in order to provide the
resummed anomalous dimension and its exponent $\omc$. In the following, we refer
to this this calculation as the ``$\om$-expansion'' result.

%----------------------------------------------------------------------
\subsection{Practical determination of splitting functions\label{s:pdsf}}

Here we are more interested in providing the resummed gluon splitting function
directly in $x$-space, by using the resummation scheme defined by the kernel
$\CK_\om$ and by the corresponding Green's function.  Two methods are available
to this purpose. One can exploit the $\gamma$-representation for the
$t$-dependence on the gluon distribution, and define an anomalous dimension in
$\omega$-space as given in
Eqs.~(\ref{eq:resanomdim},\ref{omexpSolution}). To obtain a result in 
$x$-space, it is then necessary to take the inverse Mellin transform of
$\gres(\omega,t)$.  However our formalism for calculating the Green's function
involves a kernel with higher-order terms in $\as$ and this cannot be
straightforwardly represented with a $\gamma$-representation, so in order to
obtain a splitting function within the same `model' as the Green's function we
shall need to resort to $x$-space deconvolution directly from the Green's
function, using the numerical method presented in \cite{CCS3}. This involves
calculating the Green's function $G(y,t,t_0)$ and a corresponding integrated
gluon density
\begin{equation}
  x g(x,Q^2) \equiv \int^{Q} d^2k \; G^{(\nu_0=k^2)}(\ln 1/x, k, k_0) \;, 
\end{equation}
and then solving numerically the following equation for the effective
splitting function $P_\eff(z,Q^2)$,
\begin{equation}\label{eq:deconv}
  \frac{d g(x,Q^2)}{d \ln Q^2} = \int \frac{dz}{z}\, P_\eff(z,Q^2)\,
   g\left(\frac{x}{z},Q^2\right)\;.
\end{equation}
In the limit of $Q^2 \gg \max \{k_0^2,\Lambda^2 \}$, $P_\eff(z,Q^2)$
should be independent of the particular choice of $k_0$ and of
regularization of the coupling, modulo higher-twist corrections. That
this is true in practice is an important verification of
factorization, and provides complementarity to analytical
`proofs' based on simplified models.

As a first step it is interesting to check that the two methods for
obtaining splitting functions are equivalent. We do this for a
`LL$+$DGLAP' model (which includes the kinematical constraint), namely
\begin{equation}
  \label{eq:LLDGLAP}
  \chi_\om(\gamma) = 2\psi(1) - \psi(\gamma) - \psi(1 - \gamma + \om) + \om
  A_1(\om) \left(\frac{1}{\gamma} + \frac{1}{1-\gamma+\om}\right) \;,
\end{equation}
where the running coupling is evaluated at scale $k$. Such a model is
of interest because it can be fully represented in both the
$\gamma$-representation, since it has no higher-order terms in $\as$,
and the Green's function approach, since it is straightforwardly
expressed in $k$-space. It also contains some of the typical sources
of potential numerical instability (e.g.\ the $1/(1-z)_+$ term),
making it a powerful `test-case'.

\begin{figure}[t]
\centering\includegraphics[width=0.8\textwidth]{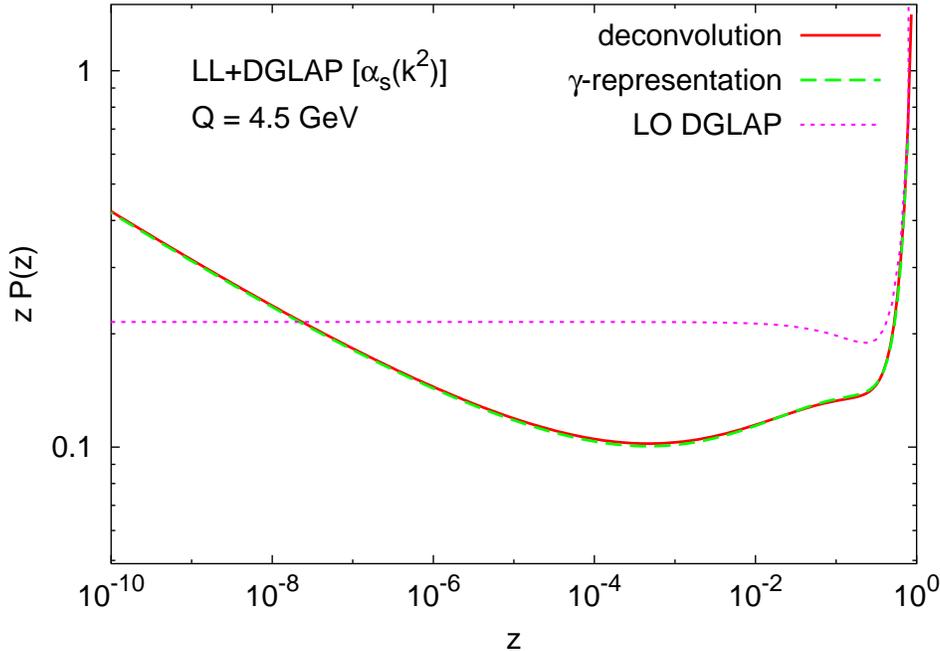}
\caption{\it Small-$z$ splitting function determined by two
  complementary numerical methods ($\gamma$-representation and
  deconvolution) for the test case of the LL+DGLAP model. For
  reference the pure DGLAP splitting function is also shown.}
\label{f:LLDGLsplit}
\end{figure}

Fig.~\ref{f:LLDGLsplit} shows that the effective splitting functions
obtained with the two methods are nearly identical. The difference
between them is of the same order as the higher-twist effects that
come from varying the regularization of the coupling in the
deconvolution method (not shown). Also plotted is the 1-loop (LO) pure DGLAP
splitting function for comparison. We note that at large $z$ one sees
the standard $1/(1-z)$ behavior in all three curves.

\begin{figure}[t]
\centering\includegraphics[width=0.8\textwidth]{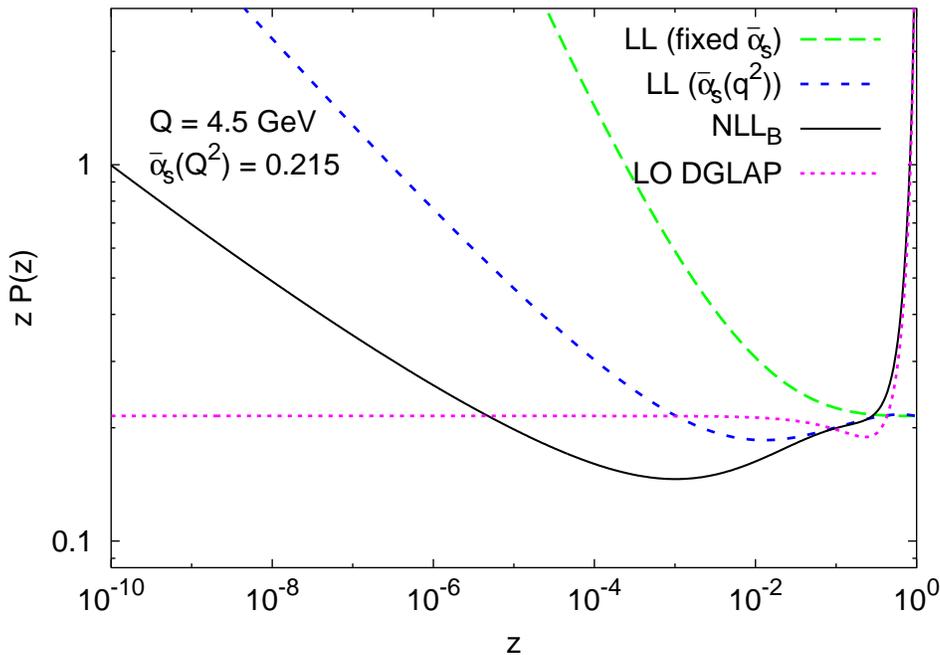}
\caption{\it Small-$z$ resummed splitting function from resummation scheme B,
  compared to the pure 1-loop DGLAP and BFKL splitting functions (the
  latter with fixed and running couplings).}
\label{f:OldNewsplit}
\end{figure}

Having established the validity of the deconvolution approach, one can
examine the effective splitting functions in the context of the full
resummed kernel. We restrict our attention to scheme B, given that scheme A
is not expected to obey the momentum sum rule. Since we are
determining a purely 
gluonic splitting function we take $n_f=0$ in the subtracted NLL
kernel, though we keep $n_f=4$ in occurrences of the $\beta$ function,
so as to maintain a realistic running of the coupling. Switching to
$n_f=4$ in the kernel as well has a relatively small effect, cf.\ 
Fig.~\ref{f:gfy_ab_nf4}.

Fig.~\ref{f:OldNewsplit} shows the effective splitting function for
$Q=4.5\GeV$. It is compared to the 1-loop DGLAP splitting function,
and to BFKL splitting functions obtained in the pure LL approximation
with fixed ($\asb \equiv \asb(Q^2) \simeq 0.215$) and running
($\asb(q^2)$) couplings.  

It is perhaps of interest to discuss first the two LL curves.  As can
be seen from the figure (and as has been discussed extensively
elsewhere \cite{CCS1,CCS3,THORNE,ABF2001,ABF2003}), running coupling
effects alone give strong modifications relative to the fixed-coupling
LL splitting function. There is a taming of the asymptotic behavior:
the cut at $\omega_s = 4\ln 2 \asb \simeq 0.60$ is converted to a
series of poles, the leading one being at $\omega_c \simeq 0.25$, with
the difference $\omega_s-\omega_c$ formally of order $\asb^{5/3}$
\cite{hr92,FRBook,CCS1}. The running of the coupling also leads to
preasymptotic effects, in particular it is associated with a dip at
moderately small $z$.  Similar features have been discussed by
other authors as well, though the details differ: in \cite{THORNE} the
running as $\asb(q^2)$ is fully implemented only through to NLL order.
In \cite{ABF2001,ABF2003} the coupling runs as $\asb(k^2)$ (a NLL
difference) and furthermore the use of the Airy approximation in the
evaluation of the expressions analogous to our
Eqs.~(\ref{eq:resanomdim},\ref{omexpSolution}) means that their
results do not quite correspond to an exact solution of
Eqs.~(\ref{eq:gluongreen}) and (\ref{eq:deconv}).

From the discussion in section~\ref{s:resGGF} for the Green's
function, one expects a further strong suppression of the asymptotic
growth when going from LL to \NLLB --- for example (for $b(n_f=4)$)
$\oms$ goes from $0.60$ to $0.27$. However because of non-linearities
(and the compensation of some double counting), the correction to the
splitting function from the combination of running coupling and \NLLB
effects is weaker than would be expected from a simple linear
combination of the two separate effects.  Indeed the final
running-coupling, \NLLB result for $\omc$ with $\asb(Q^2) = 0.215$ is
$\omc \simeq 0.18$. The preasymptotic dip, to which we return below, is
also modified in the \NLLB resummation, becoming somewhat deeper
(about $30\%$ of $\asb$) and moving to smaller $z$ ($\sim 10^{-3}$).

Other important characteristics of the splitting function extracted in
scheme B are the large-$z$ behavior, which coincides with the
expected LO DGLAP result and the value of its first moment ($\om=1$):
the scheme has been constructed such that for fixed coupling, the
effective characteristic function satisfies
$\asb \chi_\eff(\gamma=0,\asb) = 1$. At fixed-coupling, duality
arguments~\cite{ABF2000} then
automatically lead to the splitting function having a zero first
moment (to within higher-twist corrections), i.e.\ validity of the
momentum sum rule.  More generally, for running coupling we expect the
momentum sum-rule to hold because at $\om=1$ the kernel is free of
leading-twist poles.  It is therefore interesting to observe that
after full inclusion of the IR regularized running coupling, and our
rather sophisticated deconvolution approach, the numerically derived
splitting function of Fig.~\ref{f:OldNewsplit} does indeed have a
first moment which is zero, to within a few parts in $10^4$.  (We have
not so far succeeded in establishing the detailed origin of this small
departure from zero, though it may well be a higher-twist
contribution).

\begin{figure}[t]
\centering\includegraphics[width=0.8\textwidth]{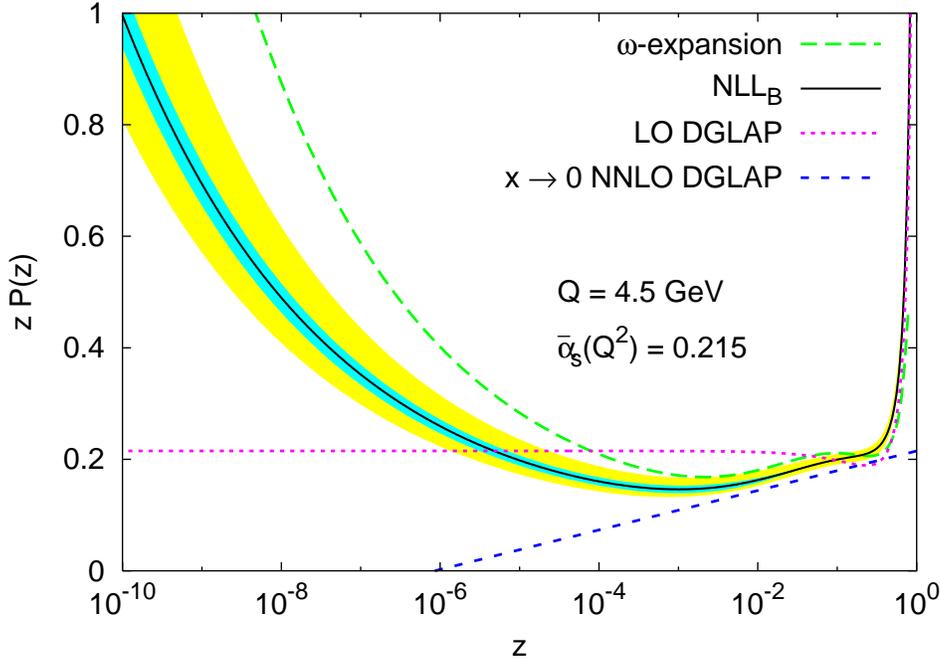}
\caption{\it The small-$z$ resummed splitting function for $\kbar=0.74\GeV$
  and $x_\mu=1$ together with renormalisation scale and
  IR-regularization uncertainties; the inner band is due to the
  variation of $\kbar$ between $0.5\GeV$ and $1.0\GeV$, while the
  outer band comes from varying the renormalisation scale in the range
  $1/2<x_\mu^2 <2$. Also shown are the splitting function obtained
  with the $\om$-expansion \cite{CCS1} (calculated with a $\beta_0$
  corresponding to $n_f=4$), the LO DGLAP $P_{gg}$ and the
  known small-$x$ parts of the NNLO DGLAP $P_{gg}$.}
\label{f:SplitUncert}
\end{figure}

Given the large difference between the original fixed-coupling LL
splitting function and the running coupling scheme B result, it is
important to establish the order of magnitude of potential
higher-order and non-perturbative uncertainties. This question is
addressed in figure~\ref{f:SplitUncert}, where the scheme B splitting
function is shown together with two uncertainty bands. The inner band
is that associated with the variation of the infrared cutoff $\kbar$
between $0.5$ and $1\GeV$, indicating a modest non-perturbative
uncertainty.\footnote{A more conservative NP uncertainty estimate
  would consider also an IR frozen coupling. Unfortunately this leads
  to numerical instabilities and we are only able to study the case of
  a coupling frozen down to some moderately low scale (below which it
  is cutoff). From these studies we deduce that including the results
  from a full IR-frozen coupling would roughly double the size of the
  NP uncertainty band.}

The outer band shows the effect of varying $x_\mu$ in the range $0.5 <
x_\mu^2 < 2$ (a range commonly used for fully inclusive quantities).
This should give an estimate of the importance of potential
higher-order corrections. One sees that the main features of the
splitting function are stable, though at small $z$ the uncertainty
grows because different renormalisation scales lead to slightly
different $\om_c$ powers. Another way of investigating higher-order
uncertainties is to consider the $\om$-expansion of \cite{CCS1} ---
here recalculated with the same $n_f$ convention as used for scheme B
($n_f=4$ in the $\beta$-function and $n_f=0$ in the rest of the
kernel) and transformed to $z$-space. We recall that the
$\om$-expansion is based on the same assumptions as scheme B, namely
LL+NLL BFKL and the requirement of correct LO DGLAP limits. From
fig.~\ref{f:SplitUncert} one sees that down to $z\sim 10^{-3}$ it
agrees with scheme B to within the renormalisation scale
uncertainties. Below, the \NLLB and $\om$-expansion curves move
further apart, essentially because their $\omc$ values ($0.18$ and
$0.20$ respectively) differ by more than would be expected based on
the $x_\mu$ variation. This suggests that for future phenomenological
purposes, in the very small-$z$ region one might wish to consider the
effects of a larger range of $x_\mu$ variation.

An aspect of the splitting function that deserves more comment is the
dip at moderately small $z$. A priori one may wonder about its origin
and indeed whether it might not be some form of artefact of our
resummation procedure. To help resolve the issue
fig.~\ref{f:SplitUncert} also shows the known small-$z$ part of the
NNL DGLAP splitting function (for $n_f=0$):
\begin{equation}
  \label{eq:NNLDGLAP}
  z P_{gg}(z) = \asb + B \asb^3 \ln \frac1z\,, \qquad B = -\frac{395}{108}
  + \frac{\zeta(3)}2 + \frac{11 \pi^2}{72} \simeq -1.549\,.
\end{equation}
One sees that the initial decrease of the scheme-B splitting function
corresponds closely to the decrease of the pure NNL DGLAP splitting
function, associated with the $B\asb^3 \ln 1/z$ term. At a certain
point however small-$z$ resummation effects set in and the scheme-B
structure function starts to rise, giving the characteristic dip
structure. The fact that the initial decrease of the full $P_{gg}$ is
correlated with an exactly determined (NLL$x$) piece of the NNL DGLAP
splitting function suggests that the dip structure is a true feature
of the small-$z$ splitting function. This belief is reinforced by the
observed robustness of the dip structure under renormalisation scale
and resummation scheme changes (though the depth of the dip is subject
to some degree of uncertainty).\footnote{It is worth noting that the
  dips observed in figs.~\ref{f:LLDGLsplit} and \ref{f:OldNewsplit}
  for the running-coupling LL+DGLAP and LL models have different
  compared to that of scheme~B. This is at least in part
  because the NLL$x$ terms 
  of their low-order expansions differ substantially from the true
  NLL$x$ terms contained in scheme~B.}

An interesting question concerns the impact of the dip on fits to
parton distributions.  Calculations in a (partially) RGI LL
model~\cite{KMS1997} whose effective splitting function also has a
dip, suggest that it is not incompatible with the available structure
function data. Ref.~\cite{THORNE} mentions work in progress on fits
involving a resummed splitting function with a dip (actually
considerably deeper than ours), but detailed results have yet to be
presented. It should of course be kept in mind that so far we have
only presented results for purely gluonic problems --- phenomenological
studies will additionally require a treatment of the quark sector.

%----------------------------------------------------------------------
\begin{figure}[t]
\centering
\includegraphics[height=0.55\textwidth]{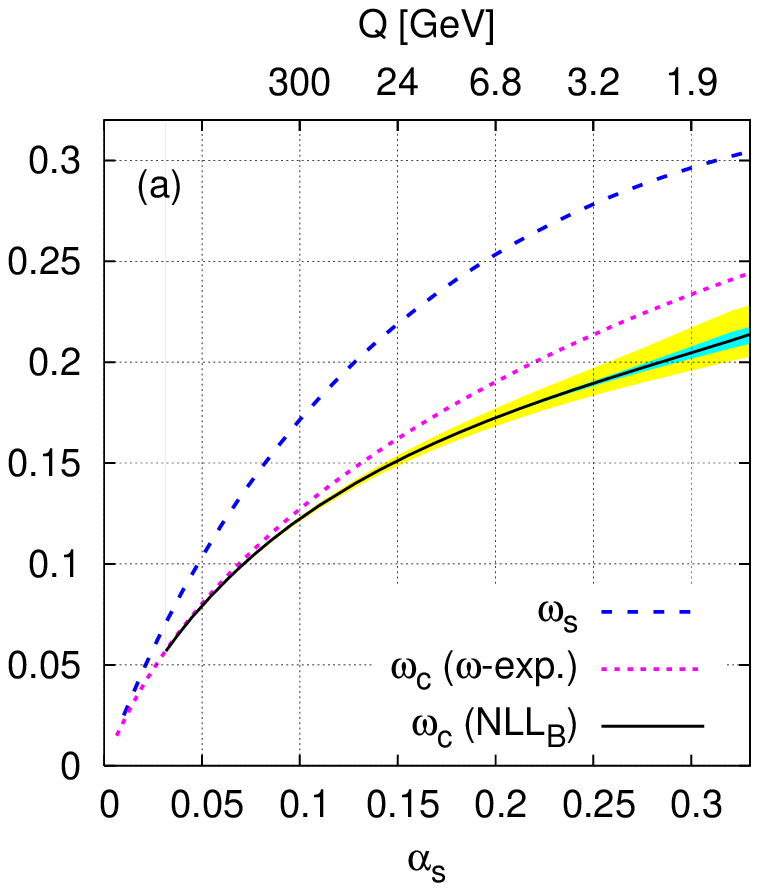}\;
\includegraphics[height=0.55\textwidth]{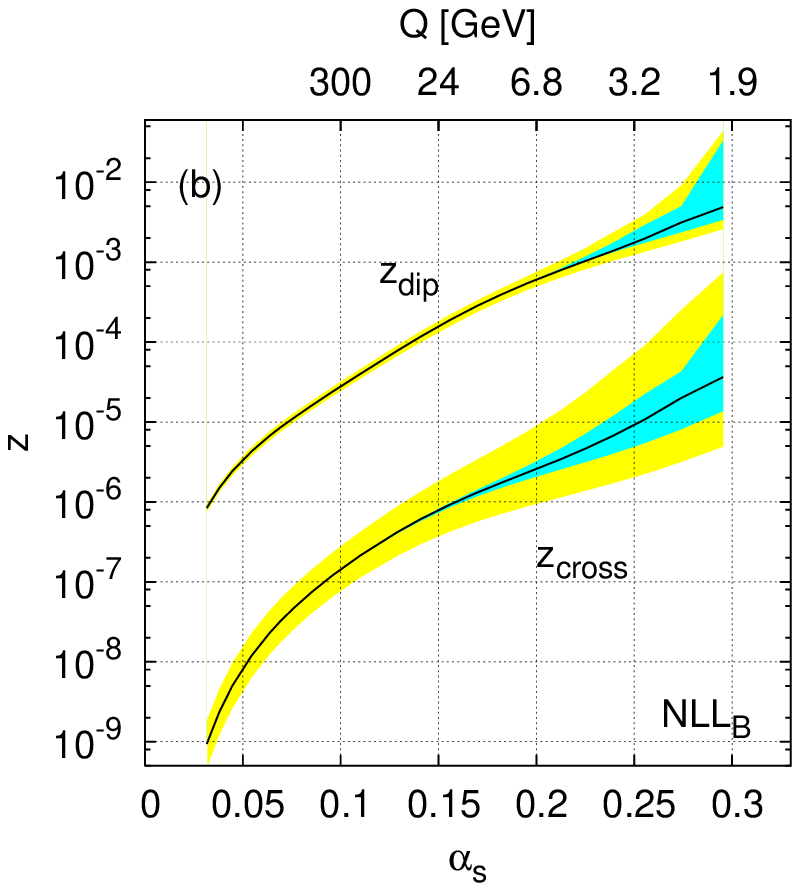}
\caption{\it (a) the small-$z$ exponent, $\omega_c$ of the effective BFKL
  splitting function in resummation scheme B, compared to the previous
  $\om$-expansion result, and to the Green's function $\om_s$ as
  determined in section~\ref{s:eidc} (\NLLB); (b) the position
  ($z_\mathrm{dip}$) of the small-$z$ minimum of the splitting
  function and the point ($z_\mathrm{cross}$) below which the resummed
  splitting function becomes larger than the 1-loop DGLAP $P_{gg}$.
  The inner and outer bands have the same meaning as in
  Fig.~\ref{f:SplitUncert}.}
\label{f:omegac}
\end{figure}

To close off this section, we examine how certain properties of the
effective splitting function depend on the coupling $\as$,
Fig.~\ref{f:omegac}. One quantity of interest is the formal small-$z$
exponent, $\omega_c$, shown in the left-hand plot, together with
uncertainty bands from varying the IR regularization and the
renormalisation scale. One sees that at small $\as$, regularization
uncertainties very quickly become negligible, in accord with their
expected higher-twist nature,\footnote{Actually the regularization
  uncertainties seem to decrease roughly as $1/Q$ whereas one would
  have expected a $1/Q^2$ behavior --- this fact (cf.~also the
  discussion of renormalons for Green's functions in
  section~\ref{s:resGGF}) has yet to be understood.} while
renormalisation-scale uncertainties decrease much more slowly with
$\as$. Also shown, for comparison, are curves for $\om_c$ in the
$\om$-expansion (quite similar to scheme B) and a reproduction of the
results of section~\ref{s:eidc} for $\om_s$ to first order in $b$
(here shown with $n_f=4$ in $b$, whereas in section~\ref{s:eidc} $n_f$
was uniformly 0).

Given the late onset of the small-$z$ power growth, other interesting
quantities are the position of the dip, $z_\mathrm{dip}$, and the
point where the effective splitting function becomes larger than the
plain 1-loop DGLAP splitting function (always defined with $x_\mu =
1$), $z_\mathrm{cross}$. Both quantities are shown as a function
of $\as$ in the right-hand plot of Fig.~\ref{f:omegac}. As one would
perhaps expect, as one decreases $\as$, one has to go to progressively
smaller values of $z$ before the BFKL increase of the splitting
function becomes visible. In this plot too we note the contrast
between regularization uncertainties which vanish rapidly with $Q$ and
renormalisation scale uncertainties which vary much more slowly with
$Q$.

%=======================================================================
\section{Inclusion of impact factors\label{s:iif}}

For a realistic calculation of a physical cross section, high energy
factorization requires that one specify also the impact factors characterizing
the external probes (Sec.~\ref{s:kf}). The impact factors are known in the LL
approximation for a variety of physical processes~\cite{KTFAC,KTSUB}, and also
in the NLL approximation for virtual photons~\cite{BaCoGiKy} and for forward jet
production~\cite{Bartels02}. However, the corresponding expressions are quite
involved and still to be implemented in numerical algorithms. Furthermore, their
accuracy stops at the first non-trivial order in $\as$.

The purpose of this section is to show how the resummed scheme for $\CGOT$ can
be extended to the corresponding impact factors $\tilde{h}$'s by incorporating
subleading corrections due to {\it 1)} phase-space and threshold effects and
{\it 2)} leading log collinear singularities. The inclusion of the exact NL
impact factor expressions~\cite{Bartels02,BaCoGiKy} is left to a future
investigation.

\subsection{Phase space and threshold effects\label{s:pste}}

\begin{figure}[t]
\centering\includegraphics[width=0.8\textwidth]{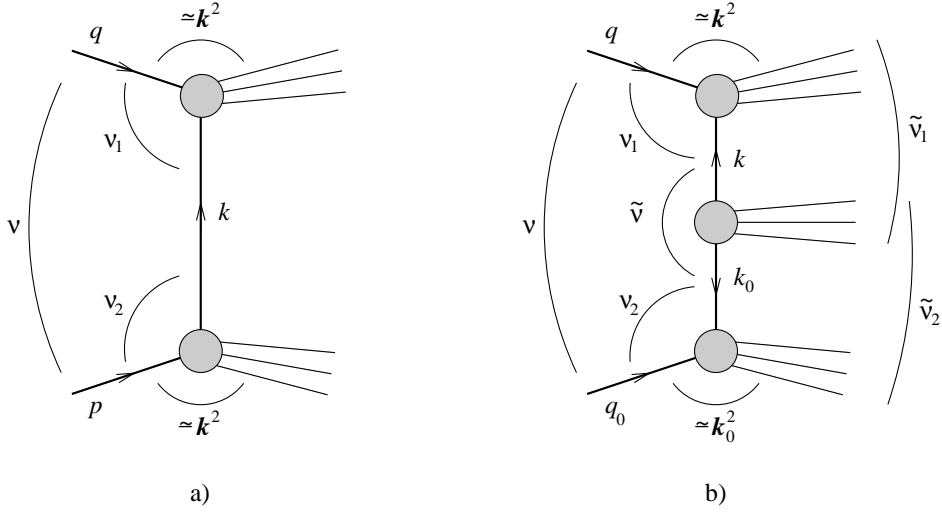}
\caption{\label{f:invariants}\it Kinematic diagrams for: a) deep inelastic scattering;
b) $\gs \gs$ cross section. The variables correspond to (2 times) the scalar
product of the corresponding momenta, e.g.,
$\nu_1 = 2 q \cdot k$,
$\nu_2 = 2 p \cdot (-k) = 2 q_0 \cdot k_0$,
$\tilde{\nu}_2 = 2 q_0 \cdot (-k)$ etc..}
\end{figure}

Let's first consider deep inelastic scattering $\gs(q) + p \to hadrons$ in the
high energy regime $\nu \equiv 2 p q \equiv Q^2 / x_B \gg Q^2$. According to the
analysis presented in Ref.~\cite{KTFAC} we can factorize the LL contribution to
the $\gs p$ cross section in the form
\begin{equation}\label{eq:sigDIS}
 \sigma^{\gs p}(\nu, Q) = \int \frac{d \nu_1}{\nu_1} \,  \frac{d \nu_2}{\nu_2}
 \, \frac{d^2 \kt}{\kt^2} \; h(\nu_1, Q, \kt) \, f(\nu_2, \kt) \,
 \Theta\left(\nu - \frac{\nu_1 \nu_2}{\kt^2} \right)\;.
\end{equation}
$h$ and $f$ represent the off-shell $\gs \pg^*$ and $\pg^* p$ cross sections in
which the virtual gluon has a particular polarization. The $\Theta$ function
indicates the threshold condition to be satisfied in the multi-Regge kinematics
(MRK) $\nu \gg \nu_1, \nu_2 \gg Q^2, \kt^2$ by the invariants defined in
Fig.~\ref{f:invariants}a.

This threshold condition expresses the fact that the longitudinal part of the
momentum transfer is small with respect to its longitudinal part (consistency
constraint~\cite{Ciaf88,LDC,KMS1997}). In fact, in a frame where the
momenta $p$ and 
$q$ have no transverse component, one has
\begin{subequations}\label{eq:param}
\begin{align}
 q &= q'- x p\\
 k &= -\bar{z} q' + z p + \kt  \;: \quad q' \cdot \kt = 0 = p \cdot \kt \;.
\end{align}
\end{subequations}
In the last equation, one has to remember the euclidean nature of $\kt$:
$-k_{\mu} k^{\mu} = z \bar{z} \nu +\kt^2$.
The relations among invariants and Sudakov parameters are given by
\begin{subequations}\label{eq:relinv}
\begin{align}
 z   &= \frac1{\nu} \left( \nu_1 - \frac{Q^2}{\nu} \nu_2 \right) \label{z_val}\\
 \bar{z}   &= \frac{\nu_2}{\nu} \\
 |k_{\mu} k^{\mu}| &= \kt^2 + \frac{\nu_2}{\nu}
  \left( \nu_1 - \frac{Q^2}{\nu} \nu_2 \right) \;. \label{kt_val}
\end{align}
\end{subequations}
MRK implies $\nu_1 \gg Q^2 \nu_2 / \nu$, hence Eqs.~(\ref{z_val},\ref{kt_val})
can be approximated by
\begin{subequations}\label{eq:relinvApp}
\begin{align}
 z &\simeq \frac{\nu_1}{\nu} \label{z_valApp}\\
 |k_{\mu} k^{\mu}| &\simeq \kt^2 + \frac{\nu_1 \nu_2}{\nu}
  \geq \frac{\nu_1 \nu_2}{\nu} \;. \label{kt_valApp}
\end{align}
\end{subequations}
Therefore, if $\kt^2$ were replaced by $|k_{\mu} k^{\mu}|$, the $\Theta$
function would represent just a phase space threshold. As it stands, it yields
the consistency condition that the virtuality of the gluon is essentially
transverse, that is
\begin{equation}\label{appThreshold}
 \kt^2 \simeq |k_{\mu} k^{\mu}| \geq  \frac{\nu_1 \nu_2}{\nu} \;,
\end{equation}
i.e., the condition in Eq.~(\ref{eq:sigDIS}).  The additional thresholds
$(q+k)^2 \simeq \nu_1 - Q^2 + \kt^2 > 0$ and $(p-k)^2 \simeq \nu_2 - \kt^2 > 0$
--- ensuring the final state particles to be in the forward light cone --- are
implicitly contained in $h$ and $f$ respectively.
According to Eq.~(\ref{z_valApp}), we can rewrite Eq.~(\ref{eq:sigDIS}) as
\begin{align}\label{eq:DISnu}
 \sigma^{\gs p}(\nu, Q) &= \int_{Q^2}^{\nu} \frac{d \nu_1}{\nu_1} \,
  \frac{d^2 \kt}{\kt^2} \; h\left( \frac{Q^2}{\nu_1}, Q, \kt \right)
  \CF\left( \frac{\nu_1}{\nu}, \kt \right) \\ \label{eq:DISz}
 &= \int_{x_B}^1 \frac{d z}{z} \frac{d^2 \kt}{\kt^2} \;
  h\left( \frac{x_B}{z}, Q, \kt \right) \CF(z, \kt)
\end{align}
in terms of the unintegrated gluon density
\begin{equation}\label{eq:ugd}
 \CF(z, \kt) = \int \frac{d \nu_2}{\nu_2} \; f(\nu_2, \kt) \,
 \Theta\left(\nu - \frac{\nu_1 \nu_2}{\kt^2} \right)\;.
\end{equation}
Eq.~(\ref{eq:DISz}) is the well known factorization formula for DIS which we
present for later convenience also as a convolution in the invariant ``energy
variable'' $\nu_1$ (Eq.~(\ref{eq:DISnu})).

Taking the Mellin transform w.r.t.\ $\nu / Q^2 = x_B^{-1}$, yields the simpler
structure
\begin{equation}\label{eq:sigDISom}
 \sigma^{\gs p}_{\om}( Q ) = \int \frac{d^2 \kt}{\kt^2} \;
 h_{\om}(Q, \kt) \CF_{\om}(\kt)\;,
\end{equation}
in terms of Mellin transforms%
\footnote{Here we define $h_{\om}$ using $Q^2$ as energy scale for $\nu_1$,
at variance with Eq.~(\ref{eq:if}), where we used $Q |\kt|$ as energy scale for
$\nu_1$. The difference is a multiplicative factor $(Q/|\kt|)^{\om}$.}
of the original factors.

We remark that the LL behavior of $\sigma^{\gs p}$ is determined by the leading
(rightmost) singularity
$\om = \omp(\as) \stackrel{\as \to 0}{\longrightarrow} 0$ of
$\CF_{\om}$ in the $\om$-plane, while
$h_{\om} = h_0 + \ord(\om)$ contributes at LL level only
through its zero-moment $h_0$. This amounts to integrating
$h(\nu_1, Q, \kt)$ in $\nu_1$ regardless of the $\nu_1$-dependence in $\CF$ and
identifying in $\CF$: $\nu_1 = Q^2$, i.e., $z = x_B$. This shows that the
details of the phase space effects, in particular those at threshold ---
evidently ignored in the approximation $h_{\om} \simeq h_0$
just mentioned --- appear only as a NL contribution. On the other hand, they
are expected to be important when the total energy $\nu$ has moderately high
values. Therefore, the $\om$-dependent formulation of impact factors is suitable to
describe subleading effects coming from the proper treatment of the phase space.

This applies also to the double $\kt$-factorization formula describing the high
energy $\gs \gs$ cross section: in the MRK
$\nu\gg\tilde{\nu_1},\tilde{\nu_2}\gg\tilde{\nu},\nu_1,\nu_2\gg Q^2,Q_0^2,\kt^2,\kt_0^2$
the threshold condition~(\ref{kt_valApp}) can be applied to any $2 \to 2$
subdiagram of Fig.~\ref{f:invariants}b:
\begin{equation}\label{other_thresholds}
 \tilde{\nu}_1 > \frac{\tilde{\nu} \nu_1}{\kt^2} \;, \qquad
 \tilde{\nu}_2 > \frac{\tilde{\nu} \nu_2}{\kt_0^2} \;, \qquad
 \nu           > \frac{\nu_1 \tilde{\nu}_2}{\kt^2} \;, \qquad
 \nu           > \frac{\nu_2 \tilde{\nu}_1}{\kt_0^2} \quad \Longrightarrow \quad
 \nu > \frac{\tilde{\nu} \nu_1 \nu_2}{\kt^2 \kt_0^2} >
 \frac{\nu_1 \nu_2}{|\kt| |\kt_0|} \;.
\end{equation}
The last inequality (obtained by using $\tilde{\nu} > |\kt||\kt_0|$) shows that
the boundary of phase space can be very simply described by the combination
$\nu_1 \nu_2 / (\nu |\kt| |\kt_0|)$. This suggests that we write the high energy
cross section for photons of polarization $A$ and $B$ ($A,B = T,L$) as
\begin{align}\label{hef_inv}
 \sigma^{AB} (\nu, Q^2 ,Q_0^2 ) =
  \int& \frac{d \nu_1}{\nu_1} \, \frac{d \nu_2}{\nu_2} \,
  \frac{d^2 \kt}{\kt^2} \, \frac{d^2 \kt_0}{\kt_0^2} \\ \nonumber
 &h^A( \nu_1, Q, \kt ) \,
  \CG \left( \frac{\nu \kt^2 \kt_0^2}{\nu_1 \nu_2}, \kt, \kt_0 \right) \,
  h^B( \nu_2, Q_0, \kt_0 ) \,
\end{align}
where $\CG$, representing the $\pg^* \pg^*$ off shell cross section integrated
in the ``invariant mass'' $\tilde{\nu}$, contains the total energy dependence
and is constrained by the last of the threshold
conditions~(\ref{other_thresholds}).

Eq.~(\ref{hef_inv}) is just equivalent to the $\kt$-factorization
formula~(\ref{eq:sigma}) in energy space, because the convolution in the energy
variables can be diagonalized by means of the following Mellin transforms:
\begin{subequations}\label{eq:defMel}
\begin{align}\label{eq:defMela}
 \sigma^{AB}_{\om}(Q, Q_0) &= \int_{Q Q_0}^{\infty} \frac{d \nu}{\nu}
  \left( \frac{Q Q_0}{\nu} \right)^{\om}  \sigma^{AB}(\nu, Q, Q_0)\\
 h_{\om}(Q,\kt) &= \int_{Q|\kt|}^{\infty} \frac{d \nu_1}{\nu_1}
  \left( \frac{Q |\kt|}{\nu_1} \right)^{\om} h(\nu_1, Q, \kt) \\
 \CGO(\kt, \kt_0) &= \int_0^1 \frac{d u}{u} \; u^{\om}
  \CG\left( \frac{|\kt| |\kt_0|}{u}, \kt, \kt_0\right) \;, \qquad u = \frac{\nu_1 \nu_2}{\nu |\kt| |\kt_0|} \;.
\end{align}
\end{subequations}
In fact, by using the equality
\begin{equation}\label{equality}
 \frac{Q Q_0}{\nu} = \frac{Q |\kt|}{\nu_1} \;
 \frac{\nu_1 \nu_2}{\nu |\kt| |\kt_0|} \; \frac{Q_0 |\kt_0|}{\nu_2}
\end{equation}
and the thresholds~(\ref{appThreshold}) and (\ref{other_thresholds}), we obtain
\begin{equation}\label{eq:hef_om}
 \sigma^{AB}_{\om}(Q, Q_0) = \int \frac{d^2 \kt}{\kt^2} \,
 \frac{d^2 \kt_0}{\kt_0^2} \; h^A_{\om}(Q, \kt) \,
 \CGO( \kt, \kt_0 ) h^B_{\om}(Q_0, \kt_0) \;.
\end{equation}
The choice of a symmetric energy scale $\nu_0 = Q Q_0$ leads naturally to
symmetric energy scales for the individual factors $h_{\om}$ and
$\CGO$, as one can see in Eqs.~(\ref{eq:defMel}).

The lesson is that even in the double high energy factorization formula, a
single Mellin variable $\om$ allows one to treat in a proper way the kinematics
of the process, in particular the threshold effects. This motivates our choice
to use $\om$-dependent impact factors and kernel.

\subsection{Collinear resummation of impact factors\label{s:crif}}

Additional subleading contributions to $\sigma^{\gs \gs}$ not taken into account
in the Green's function are higher order perturbative corrections to the impact
factors. Here we want to analyze the additional corrections which are important
in the collinear limits $Q \gg Q_0$ and $Q \ll Q_0$.
In order to keep the discussion as simple as possible, we analyze only the
fixed coupling ($b = 0$) case.

In Sec.~\ref{s:omegaexp} we have shown that we can replace the original RG
improved Green's function $\CG$ of Eq.~(\ref{eq:gluongreen}) with $\CGT$
of Eq.~(\ref{d:GGFeff}) --- the latter being defined in term of the modified
kernel~(\ref{eq:kerresum2}) --- up to NNLL differences. Correspondingly one
should define impact factors $\tilde{h}$'s which provide the same cross section
in the new scheme.

We begin by considering Eq.~(\ref{eq:hef_om}) with LO impact factors and with
the effective Green's function $\CGT$. With fixed coupling, both the
impact factor and the Green's function are scale invariant, and the cross
section can be given the integral representation
\begin{align}\label{sig_rep}
 \sigma^{AB}_{\om}(Q, Q_0) &= \frac{\pi}{Q Q_0}
  \int \frac{d \gamma}{2 \pi i} \left(\frac{Q_0^2}{Q^2}\right)^{\gamma-\frac12}
  \sigma_{\om}(\gamma) \\ \label{sig_ga}
 \sigma_{\om}(\gamma) &\simeq \tilde{h}_{\om}^A(\gamma) \, \CGOT(\gamma) \,
  \tilde{h}_{\om}^B(1 - \gamma)
\end{align}
where we have introduced the Mellin transforms
\begin{subequations}\label{d:mellin}
\begin{align}
 h_{\om}(\gamma) &\equiv \int \frac{d \kt^2}{\kt^2}
  \left( \frac{\kt^2}{Q^2} \right)^{\gamma - 1} h_{\om}(Q, \kt) \\
 \CGO(\gamma) &\equiv \int d^2 \kt_0
  \left( \frac{\kt_0^2}{\kt^2} \right)^{\gamma-1} \CGO(\kt, \kt_0 )
\end{align}
\end{subequations}
We shall now compare the collinear behavior of (\ref{sig_ga}) to that
predicted for the total cross section in order to find the NLL corrections in
$\tilde{h}^A$ at collinear level.  In $\gamma$-space the formulation of
collinear factorization becomes particularly simple in the fixed coupling case,
and can be stated as follows: the leading $\log Q^2 / Q_0^2$ contribution ($Q
\gg Q_0$) --- corresponding to the behavior at $\gamma \simeq -\omhalf$ for the
Mellin transform --- to the photon-photon cross section at order
$\alpha^2 \as^n$ is given by
\begin{equation}\label{sig_coll}
 \sigma_{\om}^{AB}(\gamma) \sim
 \frac{4 \pi \alpha V_{\om}^{A}}{\gamma+\omhalf}
 \sum_{\pa_1 \cdots \pa_{n-1} = \pq, \pg}
 \frac{\gamma_{\om}^{\pq       \pa_1}}{\gamma+\omhalf} \;
 \frac{\gamma_{\om}^{\pa_1     \pa_2}}{\gamma+\omhalf}
 \cdots
 \frac{\gamma_{\om}^{\pa_{n-1} \pq  }}{\gamma+\omhalf} \;
 \frac{\gamma_{\om}^{\pq       \gamma} V_{\om}^B}{\gamma+\omhalf}
\end{equation}
in terms of the one-loop anomalous dimensions $\gamma_{\om}^{\pb \pa}$
describing the ``probability'' of the $\pa \to \pb$ splitting, and of
additional ``photon-vertex factors'' $V_{\om}^{j}$ distinguishing the
polarization of the corresponding photon:
\begin{equation}\label{eq:pvf}
 V_{\om}^{T} = 1\;, \qquad
 V_{\om}^{L} = \left( \gamma+\omhalf \right) \big( 1+\ord(\om) \big) \;.
\end{equation}
If we restrict our analysis to gluon emission only, beside the two
$\pq \bar{\pq}$ pairs coupling to the two photons, in Eq.~(\ref{sig_coll}) all
$\pa_i$ but two are just gluons. A further approximation, valid in the high
energy limit, is to neglect collinear gluon emissions between two quarks
belonging to the same loop. In fact, this amounts to neglecting $\pq \to \pq$
splittings which are proportional to
$\gamma_{\om}^{\pq \pq} = 0 + \ord(\om)$. Therefore, $\pa_1 = \pa_{n-1} = \pq$
and $\pa_2 \cdots \pa_{n-2} = \pg$.

In order to find the collinear behavior of the resummed impact factors, we have
to compare Eq.~(\ref{sig_coll}) with the $\kt$-factorization
formula~(\ref{eq:hef_om}). The collinear behavior of the RGI
kernel~(\ref{eq:kerresumf}) at $b = 0$ is simply
\begin{equation}\label{ker_coll}
 \tilde{\chi}_{\om}(\gamma) \simeq \asb (\chi_0^{\om} + \om \chi_\ci^{\om})
 \sim \frac{\om \gamma_{\om}^{\pg \pg}}{\gamma+\omhalf}
\end{equation}
This determines the collinear behavior of the RGI Green's function
\begin{equation}\label{ggf_coll}
 \CGOT(\gamma) \sim [\om - \tilde{\chi}_{\om}(\gamma)]^{-1}
 = \frac1{\om} \sum_{n=0}^{\infty}
 \left( \frac{\gamma_{\om}^{\pg \pg}}{\gamma+\omhalf} \right)^n
\end{equation}
The collinear behavior of the LO impact factors with exact
kinematics~\cite{BiNaPe01} is
\begin{subequations}\label{imf_coll}
\begin{align}
 h_{\om}^{T(0)}(\gamma) &= h_{\om}^{T(0)}(1-\gamma) \sim
  2 \alpha \sqrt{N_c^2-1} \gamma_{\om}^{\pq \pg}
  \frac1{(\gamma + \omhalf)^2} \label{imf_collT} \\
 h_{\om}^{L(0)}(\gamma) &= h_{\om}^{L(0)}(1-\gamma) \sim
  2 \alpha \sqrt{N_c^2-1} \gamma_{\om}^{\pq \pg}
  \frac1{\gamma + \omhalf} \, \frac{1+\om}{1+\frac34\om+\frac14\om^2} \;.
  \label{imf_collL}
\end{align}
\end{subequations}
Using the relation
\begin{equation}\label{Pqboson}
 \gamma_{\om}^{\pq \gamma} = \frac{\alpha}{\as} 2 N_c \gamma_{\om}^{\pq \pg} \;,
\end{equation}
the Mellin transform of the cross section with LO impact factors assumes the
form
\begin{align}\nonumber
 \left.\sigma_{\om}^{AB}(\gamma)\right|_{\text{LO imp.fac.}} &\simeq
  h_{\om}^{A(0)}(\gamma) \CGOT(\gamma) h_{\om}^{B(0)}(1-\gamma) \\
 \label{hGh_coll} &\sim \frac{4 \pi \alpha V_{\om}^{A}}{\gamma+\omhalf} \;
 \frac{\gamma_{\om}^{\pq \pg}}{\gamma+\omhalf} \;
 \sum_{n=0}^{\infty}
 \left(\frac{\gamma_{\om}^{\pg \pg}}{\gamma+\omhalf}\right)^n \;
 \frac{\gamma_{\om}^{\pg \pq, \mathrm{sing}}}{\gamma+\omhalf} \;
 \frac{\gamma_{\om}^{\pq \gamma} V_{\om}^{B}}{\gamma+\omhalf}
\end{align}
having decomposed the anomalous dimensions relative to the $\pq \to \pg$
splitting in a singular $\propto 1 / \om$ and a non-singular part:
\begin{equation}\label{eq:gammaGQ}
 \gamma_{\om}^{\pg \pq} = \frac{C_F}{C_A}
 \left[ \frac1{\om} + B(\om) \right] \equiv
 \gamma_{\om}^{\pg \pq, \mathrm{sing}} + \gamma_{\om}^{\pg \pq, \mathrm{n.s.}}
\end{equation}
We can see from Eq.~(\ref{hGh_coll}) that the structure of Eq.~(\ref{sig_coll})
is reproduced, but in the $\pq \to \pg$ splitting we are taking into account
only the singular part of the anomalous dimension.

This is not a surprise, because the LO impact factors are by definition coupled
to the Green's function via a high energy gluon exchange, i.e., a singular
splitting. Surprising enough is the fact that, using the effective Green's
function, it suffices to use upper impact factor at LO only, in order to obtain
the correct collinear singularities on its side.  The reason is that the
additional factor
\begin{equation}
 (1-\asb K^\om_\ci)^{-1} = \sum_{n=0}^{\infty} (\asb K^\om_\ci)^n
\end{equation}
stemming from the resolvent of $\CKT_\om$ (see Eq.~(\ref{d:GGFeff}))
provides exactly the non-singular splittings needed to build up the collinear
corrections of the upper impact factor to all orders, in the collinear ordering
$Q \gg k$.

On the other hand, the full expression~(\ref{sig_coll}) contains the correction
factor $1+\om B(\om)$ w.r.t.\ Eq.~(\ref{hGh_coll}), due to the full
$\pq \to \pg$ anomalous dimension~(\ref{eq:gammaGQ}). This factor is attributed
to the lower impact factor in the collinear region $k_0 \gg Q_0$, so that we can
set, at NLL level,
\begin{align}\nonumber
 \tilde{h}_{\om}^{B}(1-\gamma) &= h_{\om}^{B(0)}(1-\gamma) +
  \om B(\om) h_{\om R}^{B(0)}(1-\gamma) \\ \label{eq:hBtilde}
 &\simeq \tilde{h}_{\om}^{B(0)}(1-\gamma)
  \left[ 1 + \asb \frac{B(\om)}{\gamma+\omhalf} \right] + \text{NNLL} \;,
\end{align}
where the additional term shows right-hand singularities only in the $1-\gamma$
variable (i.e., $\Re(1-\gamma) > 1/2$).

Analyzing the opposite ordering $Q_0 \gg Q$ --- thus the leading right-hand
singularities in the variable $\gamma$ (i.e., $\Re(\gamma) > 1/2$) --- yields
the modification of the upper impact factor
\begin{equation}\label{eq:hAtilde}
 \tilde{h}_{\om}^{A}(\gamma) = h_{\om}^{A(0)}(\gamma)
 \left[ 1 + \asb \frac{B(\om)}{1-\gamma+\omhalf} \right] + \text{NNLL} \;,
\end{equation}
which differs from Eq.~(\ref{eq:hBtilde}) by the replacements
$A \leftrightarrow B$ and $\gamma \leftrightarrow 1-\gamma$.

In conclusion, the high energy cross section can be factorized in the product of
the Green's function $\CGOT$ and impact factors $\tilde{h}$ as follows:
\begin{equation}\label{sigmaFinal}
 \sigma^{AB}_{\om}(Q, Q_0) = \int \frac{d^2 \kt}{\kt^2} \,
 \frac{d^2 \kt_0}{\kt_0^2} \; \tilde{h}^A_{\om}(Q, \kt) \,
 \CGOT( \kt, \kt_0 ) \tilde{h}^B_{\om}(Q_0, \kt_0) \;.
\end{equation}
Such a factorization formula includes the full one-loop anomalous dimensions
to all orders, the NLL contributions of the Green's function and the
NLL phase space effects.  Still missing are the running coupling
effects and the subleading collinear NLL corrections to the impact
factors. The former could be easily incorporated in the collinear
limit on the basis of a straightforward generalization of
Eq.~(\ref{hGh_coll}). The latter can be included from the known
results \cite{BaCoGiKy} on the basis of the change of scheme discussed
below.

%%%%%%%%%%%%%%%%%%%%%%%%%%%%%%%%%%%%%%%%%%%%%%%%%%%%%%%%%%%%%%%%%%%%%%
\subsection{Next-to-leading impact factors in the $\boldsymbol\om$-independent
formulation\label{s:nlif}}

Here we wish to relate the $\om$-dependent formulation of
$\kt$-factorization and of BFKL evolution used so far, to the more
conventional next-to-leading log expansion of the cross section. We
shall see that this involves a redefinition of NLL impact factors
which is somewhat ambiguous, and considerably complicates their
collinear structure. For the sake of simplicity, we shall provide the
relation in the frozen $\as$ limit.

We have already encountered the operator relation of the
$\om$-dependent Green's function to the BFKL one up to NLL order.
According to Eq.~(\ref{impact}) we have
\begin{align}\label{operators}
 \CGOT \simeq [1-\asb ( K_0^1 + K^0_\ci ) ]^{-1}
  \big[ \om - \asb (K_0 + \asb K_1 + \ord( \as^2 ) \big]^{-1} \;,
\end{align}
which differs from the pure BFKL-type expansion by the operator factor
$\ik = [1 - \asb ( K_0^1 + K^0_\ci )]^{-1}$.  The latter originates
from the $\om$-shift (expanded to first order in $\om$) and from the
collinear behavior. It was first introduced in~\cite{NLLCC} where it
was shown to provide energy-independent terms, which compensate the
symmetrical scale choice $s_0 = k k_0$, so as to provide the effective
energy scale $s_> = \max(k^2, k_0^2)$ for the Green's function. On the
other hand, the complete cross section includes the impact factors
$\tilde{h}$'s according to Eqs.~(\ref{sigmaFinal}) and
(\ref{eq:hAtilde}), and should be consistent with the NLL
parameterisation
\begin{align}\label{sigmaExpan}
 \sigma^{AB}( \nu, Q, Q_0 ) &= \int \frac{d \om}{2 \pi i}
  \left( \frac{\nu}{Q Q_0} \right)^{\om} \frac{d^2 \kt}{\kt^2} \,
  \frac{d^2 \kt_0}{\kt_0^2} \;
  \big( h^{A(0)}(Q,\kt) + \as h^{A(1)}(Q,\kt) \big) \\ \nonumber
 & \quad \times
  \langle \kt | [ \om - \asb ( K_0 + \asb K_1 ) ]^{-1} | \kt_0 \rangle
  \big( h^{B(0)}(Q_0,\kt_0) + \as h^{B(1)}(Q_0,\kt_0) \big) \;.
\end{align}
In order to compare Eq.~(\ref{sigmaFinal}) with (\ref{operators}) and
(\ref{sigmaExpan}), we see that the factor $\ik$ has to be incorporated in the
impact factors. We can do that, given the eigenvalue function
$\ik(\gamma) = \ik(1-\gamma )$, by defining some function $\ik_L(\gamma)$ such
that $\ik(\gamma) = \ik_L(\gamma) \ik_L(1-\gamma )$, and by assigning factor
$\ik_L(\gamma)$ ($\ik_L(1-\gamma)$) to impact factor $A$ ($B$). This
decomposition is not unique, however, and each solution for $\ik_L$ corresponds
to a choice of $\kt$-factorization scheme~\cite{NLLCC} in the subtraction of the
leading term at NLL level in perturbation theory. Furthermore, the
$\om$-dependence should be expanded in the $\tilde{h}$'s also. In $\gamma$-space
we have
\begin{equation}\label{imfExpan}
 \tilde{h}_{\om}^A = h^{(0)A}_{\om=0}
 + \om \left.\partial_{\om} h^{(0)A}_{\om}\right|_{\om=0}
 + \asb \tilde{h}^{(1)A}_{\om=0}
 \simeq h^{(0)A}_{\om=0} + \asb \left[
 \left.\partial_{\om} h^{(0)A}_{\om}\right|_{\om=0}
 \chi_0 + \tilde{h}^{(1)A}_{\om=0} \right] \;,
\end{equation}
where in the last line we have exploited that the factor $(\om - \asb \chi_0)$
eliminates the high energy part of the cross section.

On the other hand, at NL level the factorization of the $\ik$ factor is achieved
by setting $\ik_L(\gamma) = 1 + \asb \ikc(\gamma) + {\cal O}(\asb^2)$, with
\begin{equation}\label{ikConstraint}
 \ikc(\gamma) + \ikc(1-\gamma) = \chi_0^1(\gamma) + \chi_\ci^0(\gamma) \simeq
 -\frac12\left[ \frac1{\gamma^2} + \frac1{(1-\gamma)^2} \right]
 +A(0) \left[ \frac1{\gamma} + \frac1{(1-\gamma)} \right]
\end{equation}
Therefore, the NL contribution to the impact factor in the $\om$-independent
expansion becomes
\begin{equation}\label{hOne}
 h^{(1)A}(\gamma) = h^{(0)A}_{\om=0}(\gamma) \ikc(\gamma)
 +\left.\partial_{\om} h^{(0)A}_{\om}(\gamma)\right|_{\om=0}
 \chi_0(\gamma) + \tilde{h}^{(1)A}_{\om=0}(\gamma) \;.
\end{equation}

We see from Eq.~(\ref{hOne}) that the first two terms both generate higher order
singularities which --- e.g., for $A = T$ --- are of type $1/\gamma^4$ and
$1/(1-\gamma)^4$, and come from the $\om$-derivative and from multiplication by
the singular $\ikc$ term. In order to extract the dynamically interesting
correction $\tilde{h}^{(1)A}$ from a perturbative calculation of $h^{(1)A}$, one
has to subtract both terms from $h^{(1)A}$, by a proper choice of $\ikc$,
corresponding to a proper factorization scheme.

%%%%%%%%%%%%%%%%%%%%%%%%%%%%%%%%%%%%%%%%%%%%%%%%%%%%%%%%%%%%%%%%%%%%%%
\section{Discussion\label{s:discussion}}

In this paper we have presented a formulation of the resummed
small-$x$ equation based on the renormalisation group constraints. The
equation presented here embodies correctly the LL and NLL BFKL kernels as
well as LL DGLAP evolution. The new equation is very close to the
formulation proposed previously \cite{CCS1}, the main difference being
the treatment of the collinear terms, which are here treated as
$\om$-dependent terms of the leading kernel.  The advantage of the
small-$x$ equation proposed here is that it shows simple collinear
poles only and is defined directly in $k$ and rapidity space thus
making it easy to study the full gluon Green's function rather than
just its high-energy exponents. Therefore, after inclusion of the
impact factors, it can be used in a straightforward way for
phenomenological applications to processes with two hard scales.

In our numerical analysis we have obtained the solutions to this
equation in the case of fixed and running coupling, we have studied
the energy dependence of the Green's function both for comparable scales
and in the collinear limit and we have extracted the corresponding
splitting function.

The analysis of the Green's function has confirmed the
fact~\cite{CCS2,CMT} that the hard Pomeron exponent $\oms(\as)$
parametrizes only a transient rapidity dependence of the gluon
density, to be modified by non-linear diffusion corrections at
perturbative level and --- beyond some critical rapidity --- by the
non-perturbative Pomeron behavior. Nevertheless, subleading
resummation effects not only decrease and stabilize $\oms$ itself
(Fig.~\ref{fig:bexp3}), but basically weaken the non-perturbative
Pomeron and considerably increase the range of validity of the
perturbative behavior (Fig.~\ref{fig:ValRegions}) by 10--20 units in
rapidity compared to leading log expectations. Therefore, we are
encouraged to trust the resummed perturbative predictions for next
generation accelerators \cite{CCSSletter}.

We have provided resummed results for the gluon splitting function
also. This is a purely perturbative quantity, as has been verified
from its definition as the logarithmic derivative of the gluon
density, by checking collinear factorization for $Q \gg Q_0$. Here
resummation effects stabilize the (oscillating) $\ln s$ hierarchy, and
cause a soft departure from the DGLAP result, showing a shallow dip in
the moderate-$x$ region, followed by the expected power increase in
the very small-$x$ region, characterized by the splitting function
exponent $\omc(\as)$.

Let us now comment in a little more detail on some interesting
features of our results. 
For the high-energy exponents, this work confirms the
picture of Ref.~\cite{CCS1}. The resummed Green's function exponent
$\oms(\as)$ turns out to be numerically similar to $\as$ for relevant
values of $\as$ (Fig.~\ref{fig:bexp3}). This exponent is closely
related to the saddle point singularity discussed in~\cite{CCS1}, but
cannot really be identified with it, due to the presence of diffusion
corrections to the exponent with the same rapidity dependence. For
this reason, the extraction of $\oms$ requires the subtraction of the
leading $Y^3$ diffusion term, which turns out to be small compared to
leading log expectations. On the other hand, the splitting function
exponent $\omc(\as)$ is substantially below $\oms$ --- by about 0.1
for typical $\as$ --- due to well-known running coupling effects.

In addition, we find here interesting preasymptotic effects in the
energy dependence of the gluon density at comparable scales. In
particular, the growth of the \NLLB resummed density is delayed up to
rapidities of order $Y\simeq4$ for $\as \simeq 0.2$. It is also worth
commenting on the expectations for the onset of perturbative
non-linear (saturation) effects in the evolution. These become
relevant when the Green's function is of order $1/\as$ \cite{GLR}. For
our reference scales ($k_t\simeq 5$~GeV and $\as \simeq 0.2$), this
translates to $Y$ of order $15$, i.e.\ close to the kinematic limit of
LHC.

A special comment is needed for the splitting function's dip in the
moderate-$x$ region. It is at most a 30\% effect with respect to the
DGLAP value for $\as\simeq0.2$, spread over several orders of
magnitude in $x$. It is therefore a shallow dip, associated with
several subleading effects (notably the small-$z$ terms of the
NNLO DGLAP splitting function), and it signals a quite moderate departure
from pure LO DGLAP evolution. Considering this result and the values of
$\omc(\as)$ just mentioned, the overall picture is that our resummed
predictions are much closer to low order results than naively
expected. In turn, this may provide an explanation for the apparent
success of low order evolution to fit HERA data, despite the size of
the effective coupling $\as(Q^2) \log(1/x)$.
    
In order to compare the present results to experimental data, we need
to include the physical impact factors for two--scale processes, and
quark evolution for DIS processes. Both issues are well studied. For
quarks we can follow \cite{CC1,CCS1} and for impact factors we have
shown here how to incorporate the collinear resummation (in the frozen
coupling limit). We recall that the NL contribution to the impact
factors depends on the formulation of evolution kernel (e.g.\ 
$\om$-dependent or independent). In our $\om$-dependent approach, the
impact factors have a simple collinear behavior due to the $\om$
shift of leading poles. The relation to the more conventional
$\om$-independent calculations~\cite{BaCoGiKy} is given by the
subtraction procedure outlined in Sec.~\ref{s:nlif}.

On the whole, we have presented here a unified description of
small-$x$ deep inelastic processes, applicable to both the structure
function regime and to the $\gs \gs$ kinematics. Resummed results
push the validity of perturbative QCD towards higher energies and give
perhaps a preliminary explanation of the cross-sections' apparent
smoothness in the small-$x$ regime despite the occurrence, in their
description, of large perturbative coefficients and various
strong-coupling phenomena.

\section*{Acknowledgments}

We thank Martina Taiuti for collaboration in the early stages of this work.
This work was supported in part by the Polish Committee for Scientific Research
(KBN) grants no.\ 2P03B 05119, 5P03B 14420, and by MURST (Italy).

%AAAAAAAAAAAAAAAAAAAAAAAAAAAAAAAAAAAAAAAAAAAAAAAAAAAAAAAAAAAAAAAAAAAAAA
\appendix%                                                            A
%AAAAAAAAAAAAAAAAAAAAAAAAAAAAAAAAAAAAAAAAAAAAAAAAAAAAAAAAAAAAAAAAAAAAAA

\section{Computation of eigenvalues and projections\label{a:krn}}

We give here some details of the calculation of the kernel eigenvalues and
Mellin transforms which are needed for Secs.~\ref{s:fkxks} and \ref{s:eidc}.

Let us start from the computation of the eigenvalues of the kernel
\begin{equation}\label{defkrn}
 \frac1{\qt^2}\left(\frac{\qt^2}{\kt^2}\right)^{\lambda}
 - \frac1{\lambda} \delta^2(\qt) \equiv \krn(\kt,\kt') \;,
\end{equation}
possibly multiplied by the $\om$-shifting factor
$\left(\frac{k_<}{k_>}\right)^{\om}$, in order to obtain the kernel
of Eq.~(\ref{K0Regularised}) of the text. Such kernels are closely related to
the regularized form of the leading BFKL kernel (with or without consistency
constraint), by possibly including powers of $\log q^2$ due to the running
coupling.

Note first that, denoting by $\chi^{[\lambda]}(\gamma)$ the eigenvalue function
of $\krn$, the shifted kernel $\left(\frac{k_<}{k_>}\right)^{\om}\krn$
has eigenvalues
\begin{equation}\label{evlkrn}
 \chi_L^{[\lambda]}\Big(\gamma+\omhalf\Big) +
 \chi_R^{[\lambda]}\Big(\gamma-\omhalf\Big) \;,
\end{equation}
where the left (right) projections of $f(\gamma)$ are defined by
\begin{equation}\label{proj}
 f_{L,R}(\gamma) \equiv \pm \int_{\Re\gamma' \lessgtr \Re\gamma}
 \frac{d\gamma'}{2\pi i} \; \frac{f(\gamma')}{\gamma-\gamma'}
 = \sum_{\begin{matrix} \gamma_n \leq 0 \\ \gamma_n \geq 1 \end{matrix}}
 \frac{f(\gamma_n)}{\gamma-\gamma_n} \;,
\end{equation}
with the upper (lower) determination of signs and conditions. Note that the last
expression holds in the particular case of simple pole singularities in the left
(right) $\gamma$-plane.
In fact, such eigenvalues are found by the Fourier transform
\begin{align}\nonumber
 &\int dt'\int\frac{d\gamma'}{2\pi i} \;
 \chi^{[\lambda]}(\gamma') e^{\gamma'(t-t')}
 \left[ e^{-\omhalf(t-t')}\Theta(t-t') + e^{-\omhalf(t'-t)}\Theta(t'-t) \right]
 e^{\gamma(t'-t)} \\
 &=\int_0^\infty d\tau
 \int_{\half-i\infty}^{\half+i\infty} \frac{d\gamma'}{2\pi i} \;
 \chi^{[\lambda]}(\gamma')
 \left[ e^{(-\gamma-\omhalf+\gamma')\tau} + e^{(\gamma-\omhalf-\gamma')\tau}
 \right] \;.
\label{fourtran}
\end{align}
In this equation, the $\gamma'$ integral can be displaced to the left (right) of
$\gamma+\omhalf$ ($\gamma-\omhalf$) in the first (second) term in the r.h.s..
By then performing the $\tau$-integral in its convergence region we obtain, by
the definition~(\ref{proj}), the result~(\ref{evlkrn}).

We then compute $\chi^{[\lambda]}(\gamma)$ itself by applying the kernel $\krn$
to the test function $(\kt'{}^2)^{\gamma-1}$. By using then known integral
\begin{equation}\label{stdint}
 \int \frac{d^2 \kt'}{\pi} \;
 \frac1{(\kt'^2)^{1-\gamma} [(\kt-\kt')^2]^{1-\lambda}}
 = \frac{\Gamma(\lambda)\Gamma(\gamma)\Gamma(1-\gamma-\lambda)}{%
         \Gamma(1-\lambda)\Gamma(1-\gamma)\Gamma(\gamma+\lambda)}
\end{equation}
and the representation
\begin{equation}\label{stdrep}
 \frac{\Gamma(z+\epsilon)}{\Gamma(z)} = \exp\left[
 \epsilon \psi(z) +\frac1{2!} \epsilon^2 \psi'(z)
 + \frac1{3!} \epsilon^3 \psi''(z) + \cdots \right] \;,
\end{equation}
we obtain [$\chi_0(\gamma) \equiv 2\psi(1)-\psi(\gamma)-\psi(1-\gamma)$]
\begin{align}\label{chilambda}
 \chi^{[\lambda]}(\gamma) &= \frac1{\lambda} \left(
 \frac{\Gamma(\lambda)\Gamma(\gamma)\Gamma(1-\gamma-\lambda)}{%
       \Gamma(1-\lambda)\Gamma(1-\gamma)\Gamma(\gamma+\lambda)} -1 \right) \\
 & \simeq \frac1{\lambda}
 \left\{ \exp \left[
 \lambda \chi_0(\gamma) + \frac12 \lambda^2 \chi_0'(\gamma)
 + \frac16 \lambda^3 \left(2\psi''(1)-\psi''(\gamma)-\psi''(1-\gamma)\right)
 + \cdots \right] -1 \right\}
\nonumber
\end{align}
which proves Eq.~(\ref{chiRegularised}) of the text. Note the cancellation of
the $\lambda=0$ singularity between real emission and virtual term.

By expanding~(\ref{chilambda}) in $\lambda$ we obtain, at order $\lambda^0$,
the BFKL eigenvalue, at order $\lambda$ the eigenvalue of the running coupling
kernel of Eq.~(\ref{eq:runterm}), and so on. The corresponding kernels with
consistency constraint are obtained by shifting the left/right projections of
Eqs.~(\ref{evlkrn}) and (\ref{proj}).

The simplest left/right projections are based on the formulas
\begin{align}\label{chi0Left}
 \chi_{0L}(\gamma) &= \chi_{0R}(1-\gamma) = \psi(1) -\psi(\gamma) =
 \sum_{n=0}^\infty \frac{1-\gamma}{(1+n)(\gamma+n)} =
 \sum_{n=0}^\infty \left( \frac1{\gamma+n}-\frac1{1+n} \right) \\
\label{chi0pL}
 -\chi'_{0L}(\gamma) &= \psi'(\gamma) = \sum_{n=0}^\infty \frac1{(\gamma+n)^2}
 \;, \quad
 \psi'(\gamma) + \psi'(1-\gamma) = \frac{\pi^2}{\sin^2(\pi\gamma)} \;.
\end{align}

For the shifted running coupling kernel we need $[\chi_0^2+\chi_0']_L$ and thus
$[\chi_0^2]_L(\gamma)$. Since $\chi_0=\chi_{0L}+\chi_{0R}$ we obtain
\begin{equation}\label{chi0sqLprel}
 [\chi_0^2]_L(\gamma) = [\chi_{0L}(\gamma)]^2
 + 2[\chi_{0L}(\gamma) \chi_{0L}(1-\gamma)]_L \;.
\end{equation}
The last term has simple poles in the l.h.s., so that by Eq.~(\ref{proj}) it is
expressed by a sum over residues as
\begin{equation}\label{sumres}
 2[\chi_{0L}(\gamma) \chi_{0L}(1-\gamma)]_L = \chi_{0L}^2(\half)
 +2\sum_{n=1}^\infty \frac{\chi_{0L}(1+n)(\half-\gamma)}{(\half+n)(\gamma+n)}\;,
\end{equation}
where a subtraction at $\gamma=\half$ has been performed to make the series
convergent, and $\chi_{0L}(\half)=\half\chi_0(\half)=2\log 2$.

The series in the r.h.s.\ (with simple poles) is expressed as a
combination of $[\chi_{0L}]^2$ (which has simple and double poles) and of
$\psi'$ (with double poles only), as follows:
\begin{subequations}\label{chichi}
\begin{align}\label{chichiL}
 2[\chi_{0L}(\gamma) \chi_{0L}(1-\gamma)]_L &= [\chi_{0L}(\gamma)]^2
  - \psi'(\gamma) +\frac{\pi^2}{2} \\
\label{chichiR}
 2[\chi_{0L}(\gamma) \chi_{0L}(1-\gamma)]_R &= [\chi_{0L}(1-\gamma)]^2
  - \psi'(1-\gamma) +\frac{\pi^2}{2} \;.
\end{align}
\end{subequations}
Eqs.~(\ref{chichi}) can be proved by checking residues and values at
$\gamma=\half$ of l.h.s.\ and r.h.s.\ and coincides with Eq.~(\ref{chi0squareL})
of the text.

Finally, the left/right projections of $\tilde{\chi}_1(\gamma)$ as given in
Eq.~(\ref{eq:nllgam}) are computed on the basis of Fourier transforms of the
type~(\ref{fourtran}) by splitting the $\tau$ integration into the $]\infty,0]$
($[0,-\infty[$) intervals for the L (R) projection. The result is
\begin{align}
 \tilde{\chi}_{1L}(\gamma) &= \frac12\psi''(\gamma) + \chi_{0L}(\gamma)
 \left[\psi'(\gamma) - A_1(0)\left(\frac1{\gamma}+\frac1{1-\gamma}\right)
 + \frac{67}{36}-\frac{\pi^2}{12}-\frac{5}{18}\frac{\nf}{N_c} \right]
\label{chit1L} \\
 &\quad + \Pi(\gamma) - \Phi_{L}(\gamma)
 +\frac{\pi^2}{8} \left[\psi\Big(\frac{1+\gamma}{2}\Big)
 -\psi\Big(\frac{\gamma}{2}\Big) \right] + \frac{3}{4}\zeta(3)
\nonumber \\
 &\quad +\frac1{32} 
  \left\{
    -3\fc(\gamma) + \left(1+\frac{\nf}{N_c^3}\right)
      \left[ \frac14\left(\frac1{\gamma^2}-\frac1{(1-\gamma)^2}\right)
             -\frac12\left(\frac1{\gamma}-\frac1{1-\gamma}\right)
      \right.
  \right.
\nonumber \\
 &\qquad\qquad\qquad\qquad\qquad\qquad\qquad\;
 \left.\left. 
             +\frac1{32}\Big(\fc(\gamma+1)+\fc(\gamma-1)\Big)
             -\frac{11}{16}\fc(\gamma)
       \right]
 \right\} \;,
\nonumber
\end{align}
where
\begin{align}
\label{d:Pi}
 \Pi(\gamma) &\equiv \int_0^1 dt \; t^{\gamma-1}
 \frac{\Li(1)-\Li(t)}{1-t} = \sum_{n=0}^\infty \frac{\psi'(n+1)}{n+\gamma} \\
\label{d:PhiL}
 \Phi_{L}(\gamma) &\equiv \sum_{n=0}^\infty (-1)^n
 \frac{\psi(n+1+\gamma)-\psi(1)}{(n+\gamma)^2} \\
\label{d:fc}
 \fc(\gamma) &\equiv \frac1{\gamma-\half} \left[
 \psi'\Big(\frac{1+\gamma}{2}\Big) - \psi'\Big(\frac{\gamma}{2}\Big)
 + \psi'\Big(\frac{1}{4}\Big) -\psi'\Big(\frac{3}{4}\Big) \right] \;.
\end{align}
The corresponding expressions for $\tilde{\chi}_1$ in scheme A
(Eq.~(\ref{eq:subver1})) and in scheme B (Eq.~(\ref{eq:subver3})) are
respectively
\begin{align}
 \tilde{\chi}_{1L}^{\om,\mathbf{A}}(\gamma) &=  \tilde{\chi}_{1L}(\gamma)
 + C(0) [ \chi_{0L}(\gamma+\omhalf) - \chi_{0L}(\gamma) ]
\label{chi1tLA} \\
 \tilde{\chi}_{1L}^{\om,\mathbf{B}}(\gamma) &=  \tilde{\chi}_{1L}(\gamma)
 + C(\om) [1+\om A_1(\om)] \frac1{\gamma+\omhalf} - C(0) \frac1{\gamma} \;.
\label{chi1tLB} \\
\end{align}

%%%%%%%%%%%%%%%%%%%%%%%%%%%%%%%%%%%%%%%%%%%%%%%%%%%%%

\end{document}